\documentclass[twocolumn,traditabstract]{aa}

\makeatletter

\usepackage{pdflscape}
\makeatother

\begin{document}

\title{
Interstellar fullerene compounds and diffuse interstellar bands}

\author{Alain Omont\inst{1,2}}
\institute{UPMC Univ Paris 06, UMR7095, Institut d'Astrophysique de Paris, F-75014, Paris, France
\and  CNRS, UMR7095, Institut d'Astrophysique de Paris, F-75014, Paris, France }

\abstract{Recently, the presence of fullerenes in the interstellar medium
(ISM) has been confirmed and new findings suggest that these fullerenes
may possibly form from PAHs in the ISM. Moreover, the first confirmed
identification of two strong diffuse interstellar bands (DIBs) with the
fullerene, C$_{60}^+$, connects the long standing suggestion that various
fullerenes could be DIB carriers. These new discoveries justify reassessing
the overall importance of interstellar fullerene compounds, including fullerenes of
various sizes with endohedral or exohedral inclusions and heterofullerenes
(EEHFs). 
The phenomenology of fullerene compounds is complex.  In addition to
fullerene formation in grain shattering, fullerene formation from fully
dehydrogenated PAHs in diffuse interstellar clouds could perhaps transform
a significant percentage of the tail of low-mass PAH distribution into fullerenes
including EEHFs. But many uncertain processes make it extremely difficult to
assess their expected abundance, composition and size distribution, except
for the substantial abundance measured for C$_{60}^+$. EEHFs share many properties with pure
fullerenes, such as C$_{60}$, as regards stability, formation/destruction
and chemical processes, as well as many basic spectral features.
Because DIBs are ubiquitous in all lines of sight in the ISM, we
address several questions about the interstellar importance of various
EEHFs, especially as possible carriers of diffuse interstellar bands.
Specifically, we discuss basic interstellar properties and the likely contributions of fullerenes
of various sizes and their charged counterparts such as C$_{60}^+$, and then in turn: 
1) metallofullerenes, 2) heterofullerenes, 3) fulleranes, 4) fullerene-PAH compounds, 5) H$_2$@C$_{60}$.  From this reassessment of the literature and from combining it with
known DIB line identifications, we conclude that the general landscape
of interstellar fullerene compounds is probably much richer than heretofore
realized.  EEHFs, together with pure
fullerenes of various sizes, have many properties necessary to be suitably
carriers of DIBs: carbonaceous nature; stability and resilience in the
harsh conditions of the interstellar medium; existing with various heteroatoms and  ionization
states; relatively easy formation; few stable isomers; spectral lines in the
right spectral range; various and complex energy internal conversion; rich Jahn-Teller fine structure. This  is supported
by the first identification of a DIB carrier as C$_{60}^+$.
Unfortunately, the lack of any precise information about the complex
optical spectra of EEHFs and most pure fullerenes other than C$_{60}$ and
about their interstellar abundances still precludes definitive assessment of
the importance of fullerene compounds as DIB carriers.  Their compounds
could significantly contribute to DIBs, but it still 
seems difficult 
that they are the only important DIB carriers. Regardless, DIBs appear as
the most promising way of tracing the interstellar abundances of various
fullerene compounds if the  breakthrough in identifying C$_{60}^+$ as a
DIB carrier can be extended to more spectral features through systematic
studies of their laboratory gas-phase spectroscopy.}


\maketitle 

\section{Introduction}

Since the initial observation of the diffuse interstellar bands (DIBs)
(Heger 1922) with their confirmation and the early guesses as to their
origin (Merrill 1934, 1936; see e.g. Herbig 1975; Snow 2014, for the
history of DIB studies), identification of their carriers has
remained one of the most puzzling problems in astrophysics (see e.g.\
extensive reviews by Herbig 1995; Jenniskens \& Desert 1994; Snow \& McCall 2006; Cox 2006, 2011a, 2015;
Sarre 2006, 2008; Snow 2014, and the proceedings of two whole symposia devoted to
DIBs: Tielens \& Snow 1995; and especially Cami \& Cox 2014). The
strongest bands are amazingly conspicuous, absorbing up to 10-30\%
of the interstellar radiation at their central wavelengths in sight lines with high extinction (Fig.\ 3.2a).  They are
ubiquitous, not only along lines of sight in the Galactic disk, but also at
surprisingly high galactic latitude and  in other galaxies. With the current survey sensitivity, the number
of bands confirmed as DIBs has now increased to about 500 (see e.g.\ Hobbs et al.\ 2009, 2008 and Fig.\ 6).
This sensitivity, reaching absorption depths of less than one percent,
approaches the confusion limit in the most crowded spectral regions.

Among the most recent developments, note: 1) the extension
of DIB detection to the near-infrared (Joblin et al.\ 1990; Geballe  et al.\ 2011, 2014; Cox  et
al.\  2014a; Rawlings et al.\ 2014; Hamano et al.\ 2015; Zasowski et al.\
2015); 2) the detection of DIBs along millions of line of sight in sky surveys (e.g.\  Munari et al.\ 2008; Yuan \& Liu
2012; Kos et al.\ 2013, 2014; Lan, M\'enard \& Zhu 2015; Zasowski et al.\ 2015;
Baron et al.\ 2015a,b; Puspitarini et al.\ 2015); 3) the multiplication
of extragalactic DIB detections (see e.g.\ Cordiner 2014; Welty et al.\ 2014, and references
therein); and  4) the confirmation of the
first DIB-carrier identification as C$_{60}^+$ (Campbell et al.\ 2015; Walker et al.\ 2015).

There is more or less general agreement  that
DIB carriers are suspected of being large carbon-based molecules (with $\ga$10-100 atoms) in the gas phase,  although other possibilities remain open  (see e.g.\ Snow 1995, 2014).  
 In addition, the absence of
perfect intensity  correlation between the strongest DIBs means that most of them need
independent carriers  (e.g.\ Herbig 1995; Cami et al.\ 1997;  
Snow 2001, 2014; Tielens 2014). The three different allotropic forms
of  known interstellar carbonaceous particles are thus the most popular
candidates for DIB carriers. 

1) Linear carbon chains, which are known in dense molecular clouds with C$_n$
skeleton and n\,$\la$\,10, have been repeatedly proposed as DIB carriers (see e.\ g.\ Thaddeus 1995; Snow 1995; Maier et al.\ 2004; Zack \& Maier 2014); for example, C$_7^-$ was proposed as the carrier of several DIBs by Tulej et al.\ (1998), but rejected by McCall et al.\ (2001). Their stability is well confirmed (see e.g.\ Jin et al.\ 2009). However, one may raise questions about the survival for long of  small carbon chains  against photo-dissociation in the diffuse medium (see e.g.\ Tielens 2014). Nevertheless, long chains remain key candidates for DIB carriers. Although no completely convincing mechanism has been proposed for their gas phase synthesis in diffuse
clouds and they do not form efficiently in graphite laser ablation experiments, they could be
formed as  fragmentation products of photo-processed grain mantles after high energy cosmic ray 
bombardment or grain collisions.

2) Graphenic compounds of various size contain a substantial part of
interstellar carbon. Besides larger graphitic dust grains, polycyclic
aromatic hydrocarbons (PAHs) appear as the dominant {\it carbonaceous  particles,}
with N$_ {\rm C}$  $\sim$20-100. This is evidenced
by their strong 6-12\,$\mu$m IR emission bands which require up to 
$\sim$ 10\% of interstellar carbon.  
They have been proposed as the best potential carriers
for DIBs (Van der Zwet \& Allamandola 1985; L\'eger \& d’Hendecourt
1985; Crawford et al.\ 1985 and e.g.\ Salama 1996, 2011; Salama \& Ehrenfreund 2014); better than their slightly larger avatars of PAH-clusters (see e.g.\
Montillaud \& Joblin 2014) or flakes of hydrogenated amorphous carbon
(HAC; e.g.\ Jones 2014, 2015). But all attempts to identify a single DIB with
a known band of a defined PAH  have failed
up to now (see e.g.\ Salama et al.\ 2011; Steglich et al.\ 2011; Salama \& Ehrenfreund 2014). However, the variety
of their possible ionized and dehydrogenated forms is far from being
fully explored experimentally or theoretically. The likelihood of PAHs
as carriers of major DIBs also suffers from the very large variety of
their possible sizes, shapes and isomers (see e.g.\ Gredel et al.\ 2011), lacking the general specificity
of DIBs. However, this large variety could be mitigated by the trend of
interstellar PAHs to naturally evolve toward a few most stable forms (Tielens 2005, 2014; Andrews et al.\ 2015).

3) Fullerenes, since their discovery,  have also been advocated as very
attractive candidates for DIB carriers (e.g.\ Kroto 1988, 1989; L\'eger et
al.\ 1988b; Kroto \& Jura 1992). The identification proposed by Foing \&
Ehrenfreund (1994, 1997) of two strong DIBs with the near-IR doublet of the spectrum of C$_{60}^+$ measured by Fulara et al.\ (1993a; see also Kato et al.\ 1991; Gasyna et al.\ 1992)  (Fig.\ 3), seemed to confirm
this suggestion. Although the impact of such a breakthrough had been
tempered until very recently by the absence of gas-phase spectroscopy
of C$_{60}^+$, it is now fully confirmed by the brilliant experiment of
Campbell et al.\ (2015). With this first identification of a DIB carrier, fullerene compounds clearly appear as viable 
candidates for  
carriers of other DIBs. However, despite their many attractive features, such as
stability, spectral specificity, various forms (size, ionization state,
heteroatoms, endohedral and exohedral compounds), 
the low abundance of C$_{60}$  made it difficult
for various fullerenes to produce strong enough DIBs 
 (e.g.\ Cami (2014); §8.2).

Nevertheless,  it is clear that the first success of identifying fullerenes as
DIB carriers and the relatively  large abundance inferred for C$_{60}^+$ in
interstellar diffuse clouds (§2.4) justify a general reappraisal of various
fullerene compounds as DIB-carrier candidates. In addition to the strength of the
above arguments favouring fullerenes, this is further supported
by several recent results
including: models and laboratory demonstration of  C$_{60}$ formation
from dehydrogenated PAHs (Zhen et al.\ 2014; Bern\'e et al.\ 2015a);
and experiments confirming the easy formation of C$_{60}$ endometals
 (Dunk et al.\ 2013, 2014) and of PAH-fullerene dyads (Dunk et al.\ 2013). 

The paper is organized as follows. We first briefly review the main properties of pure fullerenes relevant for their interstellar behaviour and DIBs. Five other main classes of fullerene compounds are then discussed: 1) metallofullerenes, especially endohedral ones M@C$_{60}$, 2) heterofullerenes (mostly C$_{59}$N \& C$_{59}$Si), 3) fulleranes, 4) PAH-fullerene associations and 5) H$_2$@C$_{60}$. 
Finally we reconsider the idea that fullerene compounds might be carriers of some DIBs.

\section{Facts and questions about interstellar fullerenes}

 \begin{figure*}[htbp]
	 \begin{center}
 \includegraphics[scale=0.8]{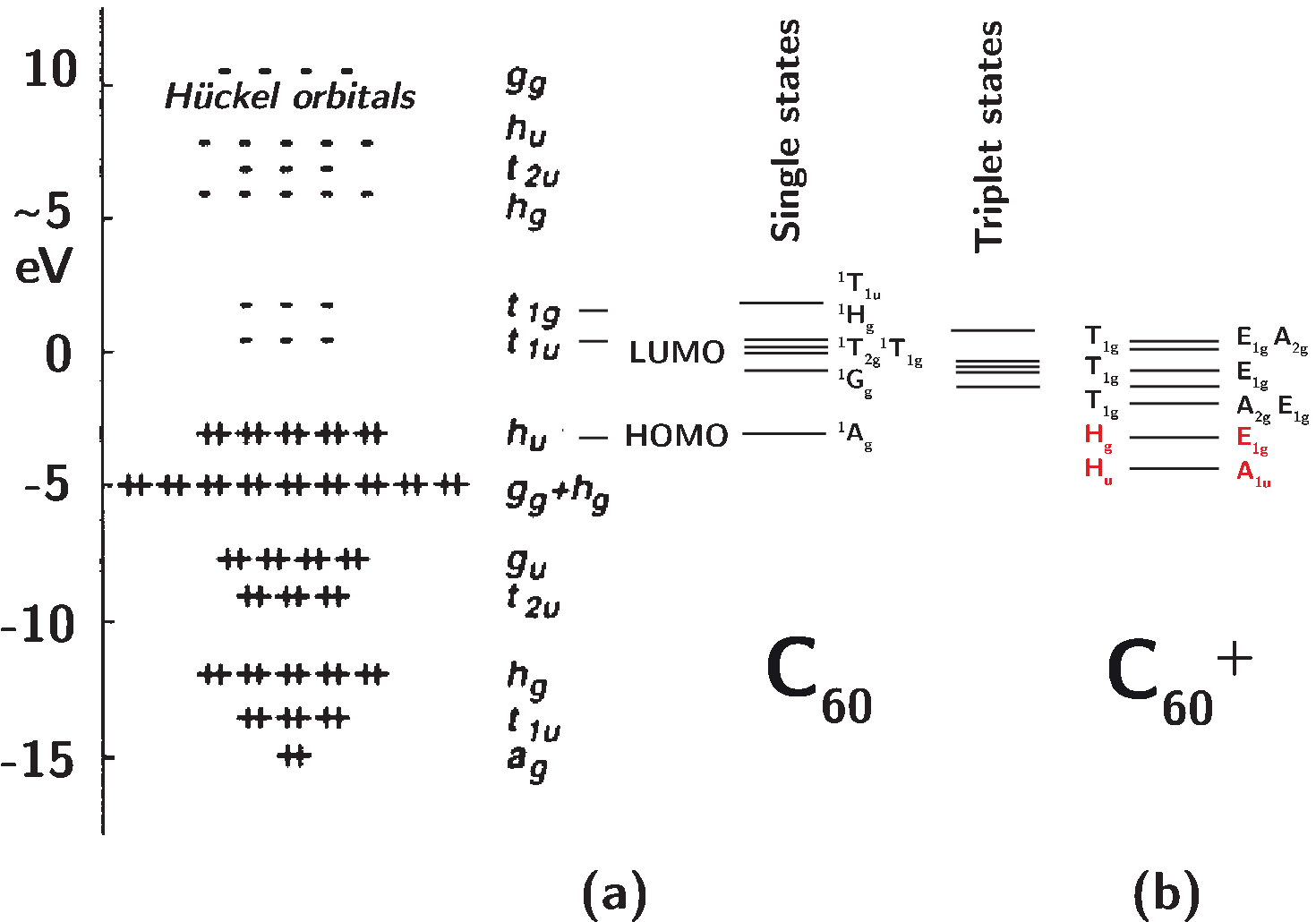}
 \caption{{\bf a)} ({\it Adapted from Fig.\ 5 of Pennington \& Stenger 1996)} Schematic diagram of basic electronic shells of C$_{60}$ in H\"{u}ckel approximation and resulting first excited states of C$_{60}$ (E\,$<\sim$\,3\,eV), singlet (see e.g.\ Sassara et al.\ 1997, 2001) and triplet. {\bf b)} {\it from Table 6 of Bendale et al.\ (1992)} Schematic diagram of the first electronic levels of C$_{60}^+$ connected to the ground state with a computed oscillator strength f\,$\ge$\,0.003. Both the I$_h$  parent geometry of C$_{60}$ and the D$_{5d}$ geometry found for C$_{60}^+$ are reported. The levels of the C$_{60}^+$ 9577/9633\,\AA\ DIBs are shown in red (the H$_g$ E$_{1g}$ excited level corresponds to the promotion of one electron of the h$_g$ HOMO-1 shell of C$_{60}$  to the h$_u$ HOMO shell).}
 \label{fig:h2o-e-level}
     \end{center}
 \end{figure*}

\subsection{Energy levels and optical properties of C$_{60}$ and its ions}
\subsubsection{C$_{60}$}

The electronic level
structure of the 60  $\pi$ electrons of C$_{60}$ in the icosahedral symmetry I$_{\rm h}$ (see e.g.\ Dresselhaus et al.\
1996) is schematized in Fig.\ 1. With its highest occupied molecular orbital, HOMO, made of
closed shells (...4g$_g^{8}$ 7h$_g^{10}$ 4h$_u^{10}$), the ground state
 is spin singlet S$_0$. There is a
significant gap between the energies of the first triplet and singlet
excited states, $\sim$1.6 and $\sim$1.9\,eV, respectively (e.g.\ Cavar
et al.\ 2005). All optical transitions from
the ground state to triplet states are forbidden, as well as
to the first five singlet excited states for symmetry reasons.  It also occurs that the first allowed transitions
at 4024\,\AA\   and 3980\,\AA\ are somewhat weak (f\,$\sim$\,0.015), so
that all strong transitions are located in the UV (Sassara et al.\ 2001;
Leach et al.\ 1992). The strength of most electronic
transitions of fullerenes are drastically reduced by screening by the $\pi$
electron shell of the cage (Westin \& Ros\'en 1993; Dresselhaus et al.\
1996). 

However, Herzberg-Teller coupling with vibrational states (plus complex Jahn-Teller effect, Chancey \& O'Brien 1997) allows spectacular weak absorption
and fluorescence emission in the forbidden transitions from the first
singlet excited states; however, its very rich energy structure is extremely difficult
to disentangle (Sassara at al.\ 1997; Orlandi \& Negri 2002; Cavar et
al.\ 2005), see Appendix E. 

The wave functions of most of all these electronic orbitals
are highly concentrated in a thin shell about the cage, as well accounted
for in the `jellium' model, for example, where the $\pi$ electrons are confined in a deep potential well (e.g.\
R\"udel et al.\ 2002; Chakraborty et al.\ 2008). However, close to the
ionization limit, there is a series of additional more extended diffuse
states studied using scanning tunnelling spectroscopy by Feng et al.\
(2008) and multi-photon photoionization by Johansson \& Campbell (2013). They are similar to, but somewhat different from atomic Rydberg states, corresponding to an
ionized cage such as C$_{60}^+$ with one  orbiting electron. Most of them have the classical binding energy close to 13.6/n$^2$\,eV with n$\ge$3,
but the most interesting ones, the lowest s and p states, have an energy
significantly lower by $\sim$1\,eV (e.g. Feng et al.\ 2008; Johansson
\& Campbell 2013). They are often called SAMOs (superatom molecular
orbitals). For pure fullerenes, their energy is too high to exhibit 
absorption at visible wavelengths.

C$_{60}$ has a large number (174) of vibrational modes. Most of them are
highly degenerate, leaving only 46 different frequencies.  Only four
transitions are infrared active with t$_{1u}$ symmetry and wavelengths equal to 7.0, 8.5, 17.4 and 18.9\,$\mu$m
(Kr\"{a}tschmer et al.\ 1990).

 \begin{figure}[htbp]
	 \begin{center}
 \includegraphics[scale=0.64]{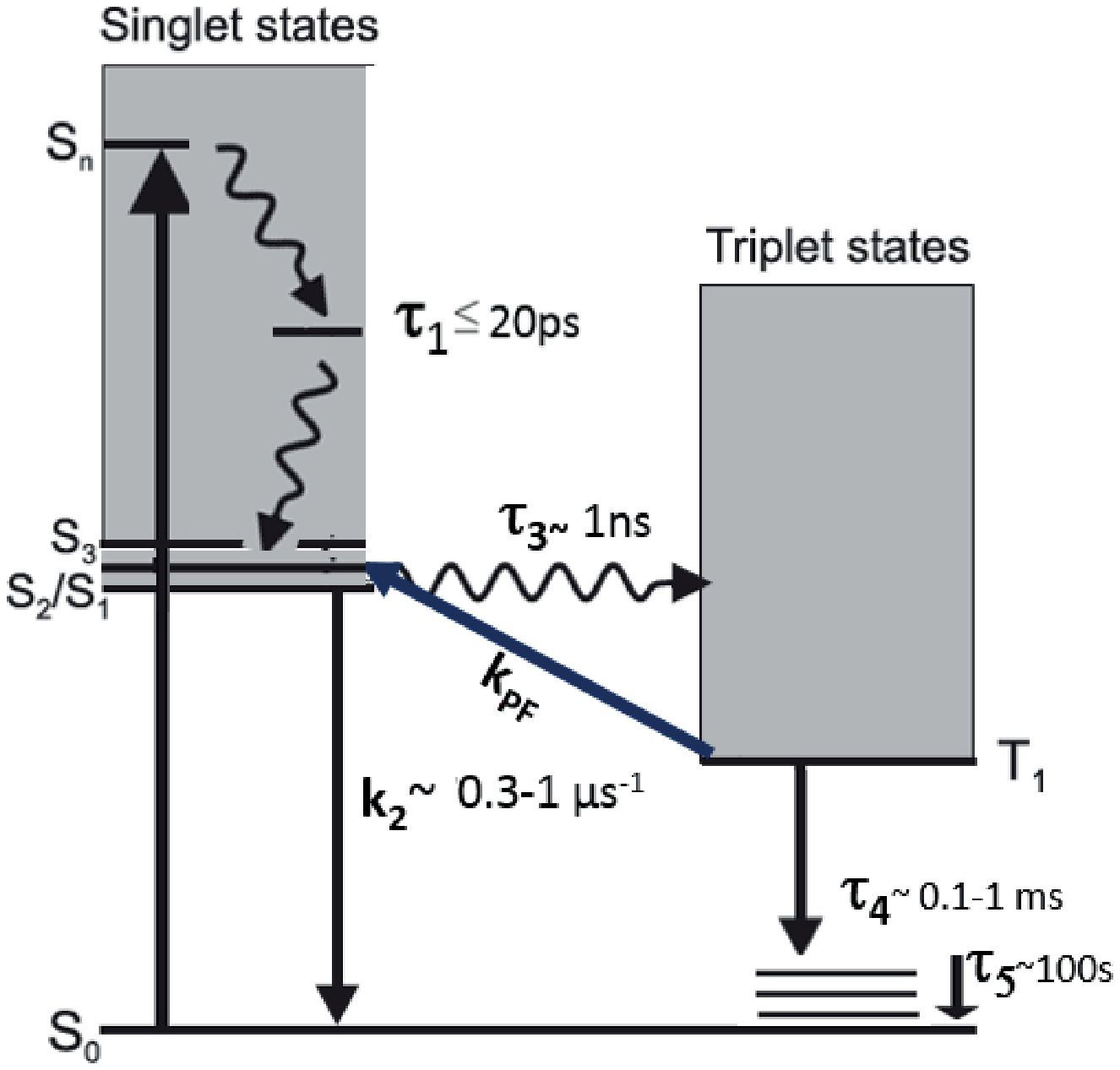}
 \caption{{(\it Adapted from Fig.\ 5 of Stepanov 2002)} Flow chart of the energy internal conversion in  C$_{60}$ with the characteristic timescales and rates:  
$\tau_1$ $<\sim$\,20\,ps is the time to reach the S$_{1,2,3}$ states from the initially excited state S$_{\rm n}$, 
by electronic to vibrational internal energy conversion; 
$\tau_{\rm 2}$ $\approx$  1ns reflects the intersystem crossing time;  
k$_3$ $\approx$  0.3--1\,$\mu$s$^{-1}$ is the slow decay rate through fluorescence directly to the ground state S$_0$; 
$\tau_4$ $\approx$ 100\,$\mu$s to 1\,ms or shorter is the radiationless  decay from the lowest triplet state to the ground state; 
k$_{\rm PF}$ is the rate of Poincar\'e fluorescence from the triplet state  discussed in Appendix G; 
and $\tau_5$, possibly $>\sim$ 100\,s, is the very slow decay of vibrational excitation through infrared emission.
Of course, as long as there is no emission of visible or infrared photons, the total energy, E(S$_{\rm n}$), is conserved; when the excitation is tranferred  to the electronic state i, the energy difference  E(S$_{\rm n}$)-E$_{\rm i}$ appears as vibrational energy.}
 \label{fig:h2o-e-level}
     \end{center}
 \end{figure}

The fluorescence and phosphorescence of C$_{60}$ have been extensively studied and are well understood (e.g. Dresselhaus et al.\ 1996; Weisman 1999;
Sassara et al.\  1997; Salazar et al.\ 1997; Stepanov et al.\ 2002; Heden  et al.\ 2003; Echt et al.\ 2005;  
Cavar et al.\ 2005; and Appendix G). The corresponding  steps of internal energy conversion following the absorption of a UV photon are  schematized in Fig.\ 2: 
1) a very fast  ($\le$20\,ps) transfer first to adjacent electronic levels and then cascade to the lowest electronic excited singlets S$_1$, S$_2$, S$_3$..., by internal conversion from electronic to vibration energy (IEC); 2)
very low yield (a few 10$^{-4}$) of radiative fluorescence from S$_{123}$ to S$_0$,
but fast (nanosecond), almost complete IEC transfer to the triplet state; long lifetime of this triplet state; followed mainly by radiationless transfer to the ground state S$_0$ with IEC conversion of the triplet electronic energy  to vibrations.  
3) Besides negligible phosphorescence, a significant decay mode of the triplet state should be delayed
fluorescence (`Poincar\'e fluorescence'; L\'eger et al.\ 1988a) with thermal return from T$_1$ to S$_{123}$ electronic states. This is discussed in Appendix G for various fullerene compounds. It seems that  the precise contribution of the Poincar\'e fluorescence to the cooling of  C$_{60}$ compounds  remains somewhat uncertain, albeit significant, in interstellar conditions. 
 4) Finally, in isolated molecules, the rest of the vibrational energy should be
emitted in the C$_{60}$ infrared modes. However, because of the total transition weakness of these modes, the overall time constant $\tau_{\rm IR}$ for such infrared gas-phase cooling should be two orders of magnitude longer than the time given by (L\'eger et al.\ (1989) for typical PAHs (§6.2)  and thus possibly exceed 100\,s.

\subsubsection{ C$_{60}$ ions}
The excited level structure of the C$_{60}$ ions, such as C$_{60}^+$
(and C$_{60}^-$, see Appendix F), is substantially different because their ground state
is not a closed shell. Because of this, they have more  allowed transitions in the visible and
even near-infrared range. However, their excited states remain more undetermined than those of C$_{60}$,
especially because of  the Jahn-Teller effect.

Good-quality optical-near-IR absorption spectra of C$_{60}^-$ and C$_{60}^+$ in
neon matrices were first  obtained and disentangled by Fulara et al.\ (1993a)
(see also Kato et al.\ 1991; Gasyna et al.\ 1992; Kern et al.\ 2014). Both ions show  strong
features around 1\,$\mu$m. Recently, the sophisticated experiment of
Campbell et al.\ (2015) has provided accurate data, devoid of matrix
effects, for the prominent double band at 9577.56 and 9632.76\,\AA\
of C$_{60}^+$ and its vibronic satellites (Fig.\ 3). While Campbell et al.\ cannot
measure accurate band strengths, improved f-values, for the corresponding
bands in an Ne matrix, are provided by Strelnikov, Kern and Kappes (2015), namely f\,=\,0.015$\pm$0.005 for the 966\,nm absorption and f\,=\,0.01$\pm$0.003 for the 958\,nm absorption. 
However, matrix effects could substantially affect line strengths, as possibly seen from the large difference between the line ratios measured by Strelnikov et al.\ in matrix and Campbell et al.\ in free  C$_{60}^+$ molecules, and we note that Walker et al.\ (2015) favour twice higher f-values.

The electronic energy levels and the visible spectrum of C$_{60}^+$
were theoretically analyzed by Bendale et al.\ (1992) (see Fig.\ 1). They found that
the ground state of the ion distorts from I$_{\rm h}$ symmetry preferentially to  D$_{\rm
5d}$ symmetry 
(see Campbell et al.\ 2015) (however, the energy of the ground level of the isomer with D$_{\rm
3d}$ symmetry is only slightly higher). Bendale et al.\ identified the strong
$\sim$1$\mu$m transition as connecting the ground state $^2$A$_{\rm 1u}$, corresponding to the state 7h$_g^{10}$4h$_u^{9}$  in the  I$_{\rm h}$ symmetry,  to one $^2$E$_{\rm 1g}$ excited state corresponding to the I$_{\rm h}$ state 7h$_g^{9}$ 4h$_u^{10}$ (Fig.\ 1).
It is possible that the splitting of 60\,cm$^{-1}$ between the 9577 and
9633\,\AA\ components is due to the Jahn-Teller effect, but it does not seem to have been determined yet.
All other visible features derived in Table 6 of  Bendale et al.\ (1992) (Fig.\ 1) are significantly weaker, in agreement with the absorption
spectrum of Fulara et al.\ (1993a).  The SAMO states of C$_{60}^+$
do not seem to have been  studied yet. However, they should
not appear in the visible range.

Most vibrational frequencies of C$_{60}^+$ and C$_{60}^-$ remain close to,
but often significantly different from those of C$_{60}$. However, Jahn-Teller effect may break degeneracies and introduce new active vibration modes (see e.g.\ Kern et al.\ 2013; 
Bern\'e, Mulas \& Joblin 2013).

Only a few low doublets of C$_{60}^+$ are connected to the ground
doublet  by an allowed transition (Table 6 of Bendale et
al.\ 1992). In the whole near-infrared and visible
range, the strongest transition by far fom the ground level corresponds to the  bands at  9577/9633\,\AA\  towards  the low lying excited doublet $^2$E$_{\rm 1g}$. This level could perhaps carry a
marginal (Poincar\'e) fluorescence (Appendix G).  

\subsection{C$_{70}$ and other fullerenes}

Basing the analysis of the optical properties of fullerenes  on
those of C$_{60}$ may be misleading because its high symmetry relative
to other fullerenes makes its
spectrum somewhat unique. 
The next most stable fullerene, C$_{70}$, with a lower symmetry, represents a more realistic model for the intricate
details of the electronic spectroscopy. While its strongest absorption transitions are also located in the UV, it shows a significantly
stronger absorption in the visible than {C$_{60}$. It seems that only
a part of its rich visible absorption is identified well, except
in the red where a weak allowed transition coexists with Herzberg-Teller
bands  (Orlandi \& Negri 2002; Scuseria 1991).

From the original discovery experiment (Kroto et al.\ 1985), it is well known that a whole series of fullerene cages C$_{\rm 2n}$ generally
form together with {C$_{60}$ in various experiments producing
fullerenes (see e.g.\ Dunk et al.\ 2012 and Zhang et al.\ 2013 for updated references). The
general prominence of even numbers of carbon atoms, N$_{\rm C}$, and
the more or less specific prominence of C$_{60}$, C$_{70}$, and then other `magic'
numbers`, such as C$_{50}$  and C$_{32}$, 
are explained by stability 
arguments, especially by minimizing the number of adjacent pentagons in the
cage structure (see e.g.\ D\'iaz-Tendero, et al.\ 2006). However, all
cages with even N$_{\rm C}$, from $\sim$28  to well
above 70, share a similar basic stability, since  differences in the energy required for C$_2$ loss 
remain relatively small (e.g.\ D\'iaz-Tendero, et al.\ 2006; Bern\'e et al.\ 2015a,
and references therein). This should be kept in mind for interstellar
fullerenes by considering all such values of N$_{\rm C}$, since it is possible  that the prominence of C$_{60}$ abundance might remain relatively limited in astrophysical contexts 
(§2.2 \& 2.3.4).

Such a similarity of various cages is reflected in the values of their
first and second ionization potentials (IP; e.g.\ in Fig.\ 2
of  D\'iaz-Tendero, et al.\ 2006). Compared to the value for C$_{60}$
(7.6\,eV), the IPs of other fullerenes are smaller by typically only
0.5-1.0\,eV, while the C$_{70}$ IP is practically the same as  C$_{60}$. One can 
expect similar resemblances in electron affinities, bond strengths, vibration frequencies, chemical properties, etc., for the whole
family of fullerene cages. However, the lower symmetry may have various consequences including increasing chemical activity and  the number of active IR modes and shortening the time constant for IR energy emission. 

More importantly, 
variations in the electron number with N$_{\rm C}$ introduces basic
differences in filling the HOMO shell and thus in the actual properties of the ground state and the optical spectra of various pure fullerenes. This might be important
when considering  DIB carriers, since they might have  smaller
LUMO-HOMO gaps and richer optical spectra than C$_{60}$. But their known optical absorption
spectra  generally do not display prominent features in the visible range (Koponen et al.\ 2008; Lan,
Kang \& Niu 2015).  Such calculated data seem to still be lacking for most of
their cations. But  for C$_{70}^+$, Fulara et al.\ (1993b) observed a relatively weak absorption system in the range of 7000-8000\,\AA\ with vibrational structure. At shorter visible wavelengths, it
remains difficult to infer the actual gas-phase spectrum of C$_{70}^+$  either  from
the spectrum studied in an oleum solvent (Cataldo et al.\ 2012, 
2013) or from CNDO/S calculations (Kato et al.\ 1991).

The very long triplet lifetime of  C$_{70}$ makes its cooling dominated by Poincar\'e fluorescence (Appendix G). It is not impossible that the situation be the same for other cages.

SAMO states of fullerenes other than C$_{60}$ do not
seem to have been addressed yet. However, one may infer that they should
be similar to those of  C$_{60}$ and its ions and not  contribute to the visible absorption of
neutrals and cations. Anions might be considered  in a similar way  to  C$_{60}^-$ (Appendix F).

For each value of N$_{\rm C}$, fullerene cages may exist in an impressive number of different isomers. However, activation barriers for transformations to the most stable forms -- albeit substantial, 7.14\,ev for C$_{60}$ (e.g.\ Dunk et al.\ 2012) -- remain about twice  smaller than the  activation energies to degrade fullerene cages by C$_2$ ejection (§2.3.5).  One may therefore expect that interstellar fullerene cages reorganize into their most stable isomer, for instance  by simultaneous absorption of two UV photons (§2.3.4), on time scales much shorter than the fullerene lifetime. In addition, accretion of C$^+$ should lower the reorganization activation barriers (Dunk et al.\ 2012 and §2.3.4 Note 2).

\subsection{Fullerene physics and chemistry in interstellar conditions}

One key property of fullerenes is the exceptional stability of their
carbon cage. Once formed, it is difficult to destroy the cage
or to even modify it by, for example, subtracting or adding a C$_2$
molecule or introducing heteroatoms in the cage network
or inside.  On the other hand, minor modifications that preserve
the cage size and composition are much easier, such as ionization, isomerization,
exohedral addition of atoms and other chemical reactions
including association with PAHs. Therefore,
it is clear that the total abundance of a fullerene, such as C$_{60}$
in its different ionization or chemical states, makes much more sense
than its abundance in a given state.

\subsubsection{Ionization state}

In interstellar conditions  ionization
and recombination  processes for  fullerenes are likely to be similar
to those of PAHs (e.g.\ Tielens 2005, 2013). Both the UV absorption
cross-section per carbon atom and the photoionization yield  should be
comparable for PAHs and fullerenes (Verstraete \& L\'eger 1992; Berkowitz
1999). A broad picture of the state of ionization of interstellar
C$_{60}$ is thus provided by that of, say, C$_{54}$H$_{18}$
PAHs (e.g.\ by Fig.\ 6.7 of Tielens 2005). However, the value of the
electron recombination rate remains uncertain  for both PAHs (see e.g.\
Montillaud et al.\ 2013) and fullerenes. As for small PAHs, in the normal diffuse interstellar medium, the most frequent charge states of
fullerenes should be singly ionized (cations) and then neutral (see
e.g.\ Le Page et al.\ 2003; Tielens 2005; Montillaud et al.\ 2013).
This is consistent with the detection of C$_{60}^+$ there (§2.4).
However, anions should not be overlooked, as for PAHs (e.g.\ Salama et
al.\ 1996), and dications may also be rarely present.

The electron attachment energy to fullerenes is substantial,E$_{\rm A}$\,=\,2.666$\pm$0.001\,eV 
for C$_{60}$ (Stochkel \& Andersen 2013; see also Huang et al.\ 2014). Despite contradictory early estimates,
it is now confirmed that the rate of electron attachment to fullerenes
 is high  (e.g.\ Viggiano
et al.\ 2010 and references therein).  Therefore, despite their large
photo-detachment cross-section, the abundance of fullerene anions,
especially C$_{60}^-$, may be substantial  in regions  that are well shielded  from
UV radiation, as with PAHs (e.g.\ Omont 1986; Salama et al.\ 1996).

The ionization potential of  C$_{60}^+$ is only 11.5\,eV (Dresselhaus
et al.\ 1996).  C$_{60}^+$ may thus be converted into C$_{60}^{2+}$
by interstellar far-UV photons just below 13.6\,eV. However, efficient
internal energy conversion between electronic levels, including super-excited auto-ionizing states, implies relatively small cross-sections  for photo-ionization close to threshold.    
 It results  in very low values for the photoionization yield, which often reaches unity only $\sim$9\,eV above threshold (see e.g.\ Jochims et al.\ 1996). 
Nevertheless,
the photo-ionization is more efficient than photo-destruction by C$_2$
ejection (e.g.\ Leach 2001), so that the abundance of C$_{60}^{2+}$ might
be significant in regions with very strong UV radiation. We note,
nevertheless, that 
the C$_{60}^+$ has been found  substantial
even, for example, in Orion sight lines when the UV photon intensity is high
(Galazutdinov et al.\ 2000; Misawa et al.\ 2009; and Fig.\ 4). This puts limits on a very easy formation of C$_{60}^{2+}$.

\subsubsection{Chemistry - Preliminary remarks}

Despite their overall chemical stability, fullerenes are also known
to be fairly reactive.  Chemical processes mostly imply addition
reactions that may be seen in a simplistic way as breaking one of their double C=C bonds (located only on the
hexagons). 
Fullerenes should thus react with various atoms, radicals,
ions, or other reactive molecular species found in the interstellar
medium. However, in most cases it is probable that the adducts thus
formed  would rapidly return to the initial, more stable state
 mostly under the action of UV photolysis. 
In addition to the possible durable association of some strongly bound
heteroatoms discussed in §3 \& 4, it seems that at least three classes 
of chemical processes with abundant species might play a significant role in the overall
balance of fullerene interstellar compounds: hydrogenation; accretion of C$^+$, and chemical
association of fullerenes with PAHs (§6.1).  Chemical reactions with other  species are generally less
important. But notable exceptions include destruction of fullerenes
in reactions with energetic ions such as He$^+$ (§2.3.5), possible formation of some EEHFs (§3 \& 4) and perhaps reaction with atomic oxygen.

Because of its overwhelming abundance reactions with highly reactive atomic
hydrogen are by far the most frequent. They lead mostly to H accretion and to the formation of hydrogenated fullerenes (fulleranes). The physics and potential interstellar properties of these key compounds are discussed in §5. 

Accretion of C$^+$ might be thought of as a key starting point for the transformation of interstellar 
fullerene compounds.  Its importance is discussed in Appendix A and §2.3.4. It seems that the reaction C$^+$ + C$_{60}$ $\rightarrow$ C$_{61}^+$ might proceed without an activation barrier at a rate close to the Langevin rate. However, it is probable that the binding energy of C$^+$ to C$_{60}$ remains modest, which should make C$_{61}^+$ relatively easy to destroy by  UV photolysis. 
However, C$_{61}$H$_2^+$ might be more stable. Also note that small carbon chains attached to fullerenes seem to be pretty stable in laboratory conditions of fullerene formation (Pellarin et al.\ 2002; Shvartsburg et al.\ 2000), while C$^+$ accretion might efficiently grow such chains in interstellar conditions.

One may also wonder about the possible importance of fullerenes for interstellar chemical composition and processes (see e.g.\ Millar 1992; Petrie \& Bohme 2000) and about whether they might  play a significant role in interstellar chemistry like PAHs (see e.g.\ Lepp et al.\ 1988 ). But, despite their reactivity, it is not expected that fullerenes play as important a role because of their  much smaller abundance than PAHs. This is certainly true for gas-phase fullerenes with the modest abundance found for C$_{60}^+$ (§2.4). It might perhaps be less obvious for fullerenes bound to PAHs if their abundance were much larger (§6).

\subsubsection{Formation}

Possible modes
of formation of astrophysical fullerenes have been discussed by various
authors (see e.g.\ Tielens 2008; Chuvilin et al.\ 2010, Cami et al.\
2011; Bern\'e and Tielens 2012; Cami 2014  and references therein; Zhen
et al.\ 2014; Bern\'e et al.\ 2015a).  The two most likely processes now seem
to be either 1) degradation of PAHs or carbon grains under the violent
action of UV radiation,  especially dehydrogenation of PAHs (see e.g.\
Montillaud et al.\ 2013; Zhen et al.\ 2014; Bern\'e et al.\ 2015a) or of other hydrogenated carbonaceous particles (see e.g.\ Duley \& Williams 2011; Micelotta et al.\ 2012);
or 2) shattering in carbon grain collisions in shocks. 

 The physical conditions in carbon vaporization from grain shattering 
might be more or less roughly similar to classical `bottom-up' 
formation of fullerenes in the laboratory (Smalley 1992; Dunk et al.\ 2012; Kr\"{a}tschmer et al.\ 1990), such as by laser ablation of graphite,  laser ablation of  HAC (Scott, Duley \& Pinho 1997), 
and arc discharge with graphite electrodes, or in nature (Rietmeijer 2006).  

Very recent studies that demonstrate the easy
formation of C$_{60}$ from fast electron irradiation of graphene (Chuvilin et
al.\ 2010) or large PAHs (Zhen et al.\ 2014) or that model the folding
of PAHs into fullerene cages (Bern\'e et al.\ 2015a), strongly support
relatively easy `top-down' formation of fullerenes in a variety of interstellar and
circumstellar environments.
The size distribution of interstellar fullerenes is of course a
key question. There is currently evidence for the existence only of C$_{60}$ and  C$_{60}^+$ in the
interstellar medium (and C$_{70}$ in one source) (§2.4). This is not surprising
given their well-known higher stability compared to other
fullerenes. The detailed model of Bern\'e et al.\ (2015a)  confirms this trend in the conditions of strong UV radiation of
reflection nebulae. There, about 60-70 C atoms is the typical expected
size of dominant dehydrogenated PAHs.

However, in weaker UV radiation, the transition to fully dehydrogenated
PAHs occurs for N$_{\rm C}$\,$<\,$60 (e.g.\ Montillaud et al.\ 2013;
Le Page et al.\ 2003; Bern\'e \& Tielens 2012). Such graphene fragments should be as unstable
as larger ones discussed by Bern\'e et al.\ (2015a) or Pietrucci \&
Andreoni (2014), among others.  In interstellar conditions \footnote{As suggested e.\ g.\ by Duley \& Williams (2011) and  Micelotta et al.\ (2012), similar processes could form fullerenes especially in planetary nebulae from carbonaceous nanoparticles that are more aliphatic than pure PAHs, resulting from the decomposition of hydrogenated amorphous carbon (HAC) (see also Merino et al.\ 2014)}, they should likewise
quickly fold and transform into fullerene cages, with small activation
energies, only a few eV. One may thus imagine that interstellar PAHs might give
birth to a whole series of fullerene cages of various sizes.

\subsubsection{Cage growth and decay}

As Bern\'e et al.\ (2015a) have recalled, fullerene cages with arbitrary
even number of carbon atoms  N$_{\rm C}$ should be fairly stable against
abstraction or adjunction of C$_2$ particles (see e.g.\ their Table 3
giving values of the internal energy, E$_i$\,$\sim$\,25-40\,eV, needed for
C$_2$ abstraction from  C$_{58}$ to  C$_{66}$).  Dissociation energies
for C$_2$ ejection from fullerene cages are not significantly less 
for N$_{\rm C}$\,$\sim$\,40-60 than for  N$_{\rm C}$\,$\sim$60-70
(except for C$_{60}$ and C$_{70}$; see e.g.\ Gluch et al.\ 2004;
D\'iaz-Tendero et al.\ 2006). Nevertheless, the internal energy  E$_i$
needed to eject C$_2$ should decrease for smaller N$_{\rm C}$.
The value of E$_i$ is thus expected to be significantly lower for N$_{\rm C}$\,$\sim$\,25-40 than for N$_{\rm
C}$\,$\ga$60.  However, even for E$_i$\,$\approx$\,15-25\,eV as expected
for N$_{\rm C}$\,$\sim$\,30-40, C$_2$ UV abstraction still  needs the
simultaneous absorption of at least two hard-UV photons. Therefore such a process
must be relatively rare,  not more often than about once every 10$^6$\,yr  for N$_{\rm C}$\,$\sim$\,30-50 in standard diffuse clouds (§2.3.4). But such rates could be enough to perhaps stand as a dominant destruction process of such cages. 
One may even expect a more rapid
cage destruction for N$_{\rm C}$\,$<$\,25. This contrasts with  the likelihood that UV
sputtering is negligible for N$_{\rm C}$\,\,$\ga$60 in diffuse clouds, as 
may be inferred from Bern\'e et al.\ (2013). 

Of course, one should compare these rates with possible competing
growth by C$^+$ accretion resulting from the very rapid accretion rate given by Eq.(A1), $\sim$10$^{-4}$\,yr$^{-1}$. 
The relatively weak binding of the external C could make  C$_{60}^+$-C unstable\footnote{Similar to the accretion of a carbon atom (Dunk et al.\ 2012), the accretion of C$^+$ should substantially lower the activation barriers for isomer reorganization of fullerene cages, possibly to values as low as $\sim$2.5\,eV. It is therefore possible that C$^+$ accretion, even of short duration, favours isomer stabilization by soft-UV photon absorption.}
for a redistributed intramolecular vibrational energy $\ga$10\,eV (§2.2), i.e.\ even perhaps by the absorption of a single hard-UV photon, with a very uncertain photolysis rate.
 One may therefore expect that the rate of growth of the cage by  C$^+$  accretion is also relatively slow. However, the comparison of various uncertain rates -- UV C$_2$-decay of C$_{\rm 2n}$ cages, C$^+$ accretion by  C$_{\rm 2n}$, and UV photolysis of  C$_{\rm 2n}$-C,  C$_{\rm 2n}$-CH$_2$  and  their cations (Appendix A) -- makes it very difficult to predict whether a slow growth or decay of the cages will take over as a function of their size and the UV intensity. Therefore, the resulting size distribution of fullerenes in diffuse clouds remains unclear.  Detailed modelling is mandatory to try to predict its behaviour even approximately.

It seems, however, unavoidable that a significant part of the
tail of the distribution of PAHs that are small enough to be fully
dehydrogenated will eventually transform into  fullerenes. Of
course the rates at which the low-mass tail of PAH distribution is
destroyed and repopulated are highly uncertain. Many intricate processes
should be considered, including UV C$_2$/C$_2$H$_2$ abstraction, C$^+$
accretion, and even sputtering by reaction with atomic oxygen.  
However, since fullerenes are significantly more stable than PAHs (§2.3.5), it seems 
reasonable to surmise that the total amount of 
fullerenes could perhaps reach several percent of that of PAHs, i.e.\ up to 
10$^{-3}$ at least of total interstellar carbon (see also §6). 
Also, a significant amount of small fullerenes or PAHs could also be produced in grain shattering (e.g.\ Scott et al.\ 1997). 
But all these processes are so uncertain that the conjecture of a substantial abundance of small fullerenes remains somewhat  hypothetical, although it is now confirmed that gas-phase C$_{60}^+$ alone contributes  several 10$^{-4}$ of interstellar carbon (§2.4).

 \begin{figure*}[htbp]
	 \begin{center}
 \includegraphics[scale=0.73]{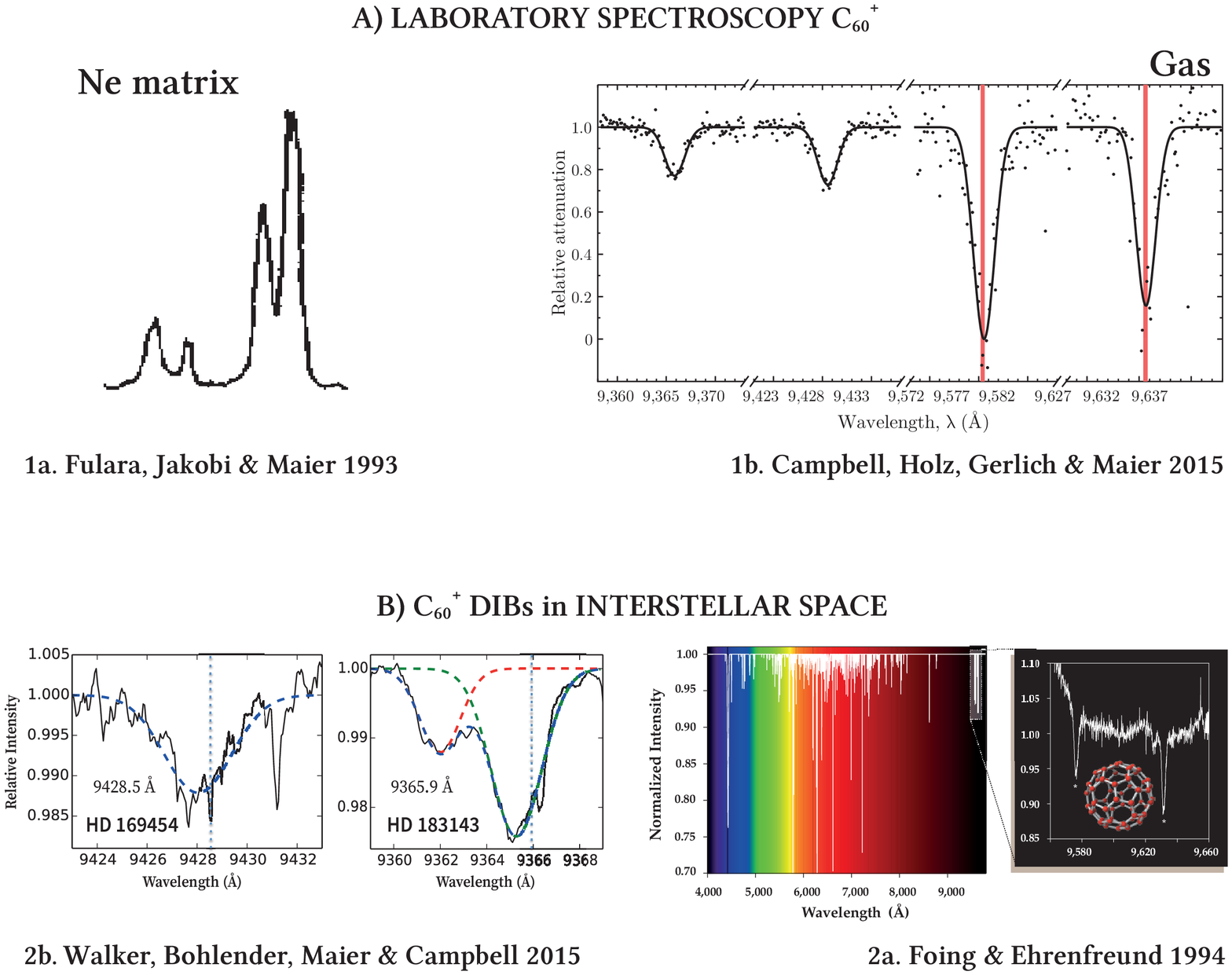}
 \caption{Successive key steps in identification of strong diffuse interstellar bands of C$_{60}^+$ at 9577.5\,\AA\ and 9632.7\,\AA\  (see also Maier 1994; Ehrenfreund \& Foing 1995, 2010, 2015; and http://www.kroto.info): 1a) High-quality laboratory spectroscopy of C$_{60}^+$ in a cold neon matrix by Fulara et al.\ (1993), showing two strong bands at 9583\,\AA\  and 9645\,\AA\ and two satellite bands. (Previous spectra by Kato et al.\ (1991) and Gasyna et al.\ (1992) were too perturbed  by matrix effects with respect to gas phase to allow DIB identification). 2a) Identification of two strong diffuse interstellar bands at 9577\,\AA\ and 9632\,\AA\ by Foing \& Ehrenfreund (1994) (confirmed by Foing \& Ehrenfreund 1997; Galazutdinov et al.\ 2000; Cox et al.\ 2014). The colour spectrum allows one to compare these two near-infrared DIBs with other strong visible DIBs (Ehrenfreund \& Foing 2015). 1b)  Laboratory spectroscopy of C$_{60}^+$ in gas phase at 5.8\,K by Campbell et al.\ (2015), confirming that the wavelengths of the two bands at 9577.5\,\AA\ and 9632.7\,\AA\ are exactly those of the two DIBs and that the wavelengths of the satellite vibronic bands are 9428.5\,\AA\ and 9365.9\,\AA .  2b) Identification of two weak interstellar bands by Walker et al.\ (2015) at these last  wavelengths.}
 \label{fig:h2o-e-level}
     \end{center}
 \end{figure*}

\subsubsection{Destruction}

Fullerene cages are known to be very resistant to various destruction
processes (e.g. Lifshitz 2000; Cataldo et al.\ 2009). Even in the harsh conditions of the ISM, it is believed
that fullerenes could be very resilient to the most frequent potentially
destructive processes such as very intense UV fields, shocks, PAH
collisions, X rays, or chemical reactions such as oxidation. Even if
the cage is substantially disrupted, it is possible that it might still
have a high probability of eventually returning quickly to some highly
stable fullerene compound of the same or different form.

All in all, despite curvature constraints, the fullerenes should be more stable than PAHs with
a comparable number of atoms, because they lack fragile C-H bonds
and dangling C atoms. As discussed above, the most efficient violent
destruction process, loss of a C$_2$ molecule, has a high dissociation
energy,  $\sim$11\,eV, in C$_{60}$,  $\sim$8-9\,eV for other cages, 
which requires a very high total internal energy, $\sim$\,25-40\,eV, 
 to achieve actual dissociation (Tables 2 and 3 of Bern\'e et al.\ 2015a).  
The UV photo-destruction of  C$_{\rm 2n}$ thus implies repeated simultaneous absorptions of several UV
photons. It is practically excluded in the normal interstellar medium for 2n\,$\ga$\,50 (however, see §2.3.4 for smaller cages), but it can be
achieved in exceptional conditions close to a star with strong UV emission
(Bern\'e et al.\ 2015a).  
 On the other hand, as seen above, repeated shrinking by UV C$_2$ loss could eventually destroy the smallest cages.

However, fullerenes will be eventually destroyed
by the same processes as PAHs, although less easily, mostly by successive
impact knockouts of C atoms or C$_2$ molecules. Micelotta
et al.\ 2014 estimate that the ejection of C from C$_{60}$ requires
an energy transferred during the collision that is twice as large as for
PAHs. Such dominant processes might include: 1)  mostly moderate shocks in the
diffuse interstellar medium, 2) cosmic rays especially in denser clouds, and
3) sputtering by electrons in the hot gas (see e.g.\ Micelotta et al.\
2010a,b, 2011, 2014; Dunk et al.\ 2012). Large fullerenes should thus
have lifetimes that are at least comparable to or longer than the estimate for PAHs of a few 10$^8$\,yr by  Micelotta et al.\ (2010a). 

Indeed, because of their stability, one of the most important sinks of
fullerene molecules of the interstellar gas could be their eventual
accretion onto dust grains.  On the one hand, sticking C$_{60}$ and other
fullerenes on pure graphenic surfaces, such as PAHs or even possibly
graphite, might seem difficult because of the spherical shape which may
prevent van der Waals binding energies as high as for PAHs which are flat, although charged particles
may provide more significant binding energies (e.g.\ Petrie \& Bohme
2000). 
But on the other hand, fullerenes display a much higher reactivity and may efficiently bind to carbonaceous grain surfaces similar to their addition reactions to PAHs (§6.1).
In addition,  previous strong binding of fullerenes to PAHs or PAH clusters
 might also 
mediate the incorporation of fullerenes into dust grains. Such fullerene
molecules deposited on grains may eventually be returned to the gas phase 
by rare violent desorption processes, such as grain-grain collisions,
exposure to cosmic rays, X rays, or strong UV (e.g. L\'eger et al.\ 1985). But it seems at least
equally probable that they are rather partially altered and more or
less definitely incorporated into larger carbon cluster structures in the grain.

Such physical and chemical properties of fullerenes in the ISM have been
repeatedly advocated since their discovery  in 1985 to predict
their significant presence in the ISM (see e.g.\ Kroto 1988, 1989;
L\'eger et al.\ 1988b; Kroto \& Jura 1992; Leach 2006; Cami 2014 and references
therein). This presence is now definitely proved by the astrophysical
detection of C$_{60}$, and C$_{60}$$^+$ in various sources\footnote{C$_{70}$ seems to only be confirmed in a single source, the planetary nebula Tc 1, Cami et al.\ (2010, 2011)}. 

\subsection{Interstellar abundances}

Long after the initial proposed identification of C$_{60}^+$ in the
ISM by Foing \& Ehrenfreud (1994)  taking advantage of the laboratory spectrum of Fulara et al.\ (1993a), the presence of C$_{60}$$^+$  has been
recently confirmed in diffuse interstellar clouds (Campbell et al.\ 2015)  
and the presence  of C$_{60}$ previously in various environments
as described  by Cami (2014) and Bern\'e et al.\ (2015b), among others: various types of carbon-rich planetary
nebulae (Cami et al.\ 2010; Garc{\'{\i}}a-Hern{\'a}ndez et al.\ 2010,
2011a,b, 2012; Gielen et al.\ 2011, Otsuka et al.\ 2013); a single pre-planetary
nebula (Zhang \& Kwok 2011); various interstellar environments (Rubin
et al.\ 2011; Peeters et al.\ 2012; Boersma et al.\ 2012),  including
reflection nebulae such as NGC\,7023 (Sellgren et al.\ 2007, 2010;
Bern\'e \& Tielens 2012) and young stellar objects (Roberts et al.\ 2012).

Except for the two strong DIBs close to 9600\,\AA\ of C$_{60}^+$
(Foing \& Ehrenfreud 1994; Campbell et al.\ 2015; and their vibronic satellites detected by  Walker et al.\ 2015; Fig.\ 3), all these fullerene
identifications are based on mid-IR spectral bands measured with the infrared spectrometer (IRS) of  {\it
Spitzer} Space Observatory. Amongst the four IR active bands of C$_{60}$,
those at 17.4\,$\mu$m and 18.9\,$\mu$m are easier to disentangle from
stronger PAH bands than those at 8.5 and 7.0\,$\mu$m. The identification
of C$_{60}^+$ was also similarly achieved in the mid-infrared (Bern\'e et
al.\ 2013). 

 \begin{figure}[htbp]
	 \begin{center}
 \includegraphics[scale=0.28]{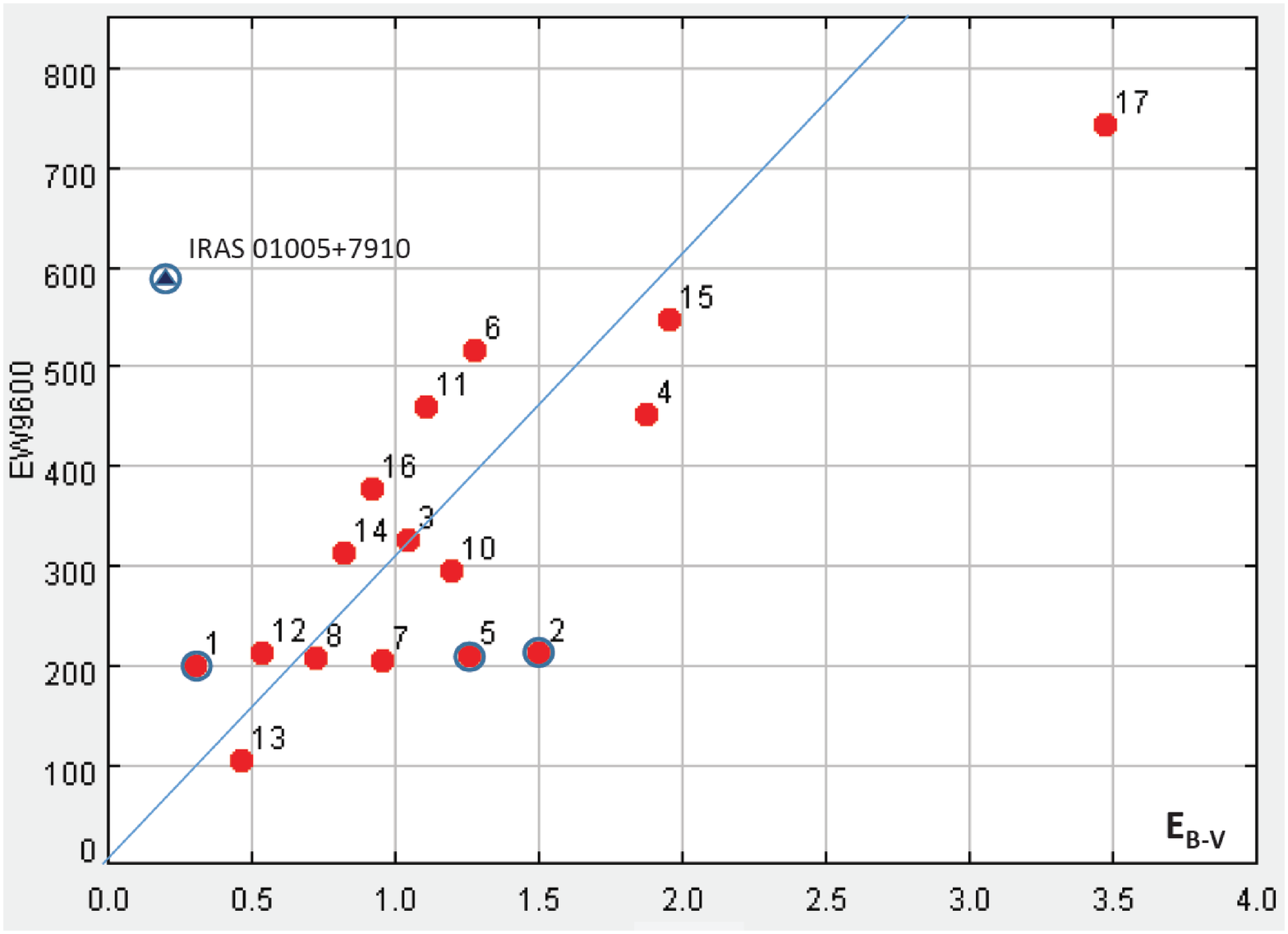}
 \caption{Correlation between the sum of the equivalent widths, EW$_{9600}$, of the two DIBs at 9577\,\AA\ and 9633\,\AA\ of C$_{60}^+$ with interstellar reddening, E$_{\rm B-V}$, for 16 sightlines from: 1) {\it   Galatzutdinov et al.\ (2000)}: 1.\ HD\,37022, 2.\ HD\,80077, 3.\ HD\,167971 4.\ HD\,168607, 5.\ HD\,169454 6.\
 HD\,183143, 7.\ HD\,186745, 8.\ HD\,190603, 10.\ HD\,194279, 11.\ HD\,195592, 12.\ HD\,198478, 13.\
 HD\,206165, 14.\ HD\,224055, 15.\ BD\,+40\,4220. 2) {\it  Cox et al.\  (2014)}: 16.\ HD\,161061, 17. 4U\,1907+09. 
The straight line represents the average value of the ratio EW9600/E$_{\rm B-V}$ = 300\,m\AA/mag, excluding the sightlines of \#1 HD\,37022 (Orion with strong UV) and \#2 \& \#5, HD\,80077 \& HD\,169454 (UV-shielded translucent clouds). 
The strong bands of C$_{60}^+$ detected by Iglesias-Groth \& Esposito (2013) in the direction of the only proto-planetary nebula IRAS 01005+7910, where IR bands of C$_{60}$ have been detected, are completely out of the correlation. This means that these bands should mostly form in the circumstellar shell. However, determination of the C$_{60}^+$ abundance is precluded there because of the lack of knowledge of the properties and the distribution of circumstellar dust.}
     \end{center}
 \end{figure}

With the confirmation of the C$_{60}^+$ DIB identification and the new
f-value provided in Ne matrix by Strelnikov et al.\ (2015), the abundance of
C$_{60}^+$ can now be relatively accurately estimated in interstellar diffuse clouds. We have used for that  13 typical sight lines of diffuse clouds where these DIBs have been detected well (Galazutdinov et al.\
2000; Cox et al.\ 2014a; see Fig.\ 4, excluding peculiar sight lines). 
The ratio of the sum of the equivalent widths
of the  two C$_{60}^+$ DIBs to reddening, EW9600/E(B-V), ranges from 210\,m\AA
/mag (HD186745) to 400\,m\AA
/mag (HD183143) with an average value of 300\,m\AA/mag. 
This is substantially lower than the value for HD\,183143, 1060\,m\AA
/mag, initially quoted by Foing and Ehrenfreund (1994), but very close to the more precise value, 451\,m\AA /mag,  given by Foing \& Ehrenfreund (1997). 

From Eq.(2), the average
abundance of C$_{60}^+$ (fraction of total interstellar carbon including dust, in gas-phase C$_{60}^+$) can be estimated as

\begin{equation}
{\rm <X_{C}(C_{60}^+)> =1.0 \times 10^{-3} \times (10^{-2}/f_{T})}
\end{equation}

where f$_{\rm T}$ is the sum of the oscillator strengths of the two
bands of  C$_{60}^+$. The new f-values measured in an Ne matrix by
Strelnikov et al.\ (2015) (see §2.1.2), yield
f$_{\rm T}$\,=\,0.025$\pm$0.008, hence ${\rm <X_{C}(C_{60}^+)>~~
\approx (4\pm2) \times 10^{-4}}$, if any matrix effect on f$_{\rm T}$ is negligible in this\footnote{Walker et al.\ (2015) favour a value of f$_{\rm T}$ twice larger which would yield ${\rm <X_{C}(C_{60}^+)>}$ twice smaller.}.

In the two dozen PNe where the C$_{60}$ mid-IR emission
bands are detected, its abundance is estimated in the range
0.05 $\times$ 10$^{-4}$--\,10$^{-2}$ of the available carbon  (Cami et
al.\ 2010, 2011; Garc{\'{\i}}a-Hern{\'a}ndez et al.\ 2010, 2011a,b,
2012; etc.).  However, these values might be affected by the assumptions
about the strength and the excitation of the infrared vibration modes (see e.g.\ Bernard-Salas et al.\ 2012). It should also be stressed that only $\sim$\,3\% of all PNe observed in the Milky Way with {\it
Spitzer}-IRS show evidence of the C$_{60}$ mid-IR bands (Otsuka et al.\
2014).
The abundance of both C$_{60}$ and  C$_{60}^+$ seems significantly
smaller in reflection nebulae such as NGC 7023 (Bern\'e \& Tielens
2012; Sellgren et al.\ 2010; Bern\'e et al.\  2013).  

The weakness or the absence of observed DIBs at the wavelength of a number of weak forbidden visible
transitions of C$_{60}$, as well as of the two broad allowed bands at
3980 and 4024\,\AA , seems to  put roughly similar constraints, $\sim$10$^{-4}$ of
interstellar carbon, on the abundance of  neutral C$_{60}$ in diffuse clouds
(see Appendix E).  This is consistent with the C$_{60}^+$ abundance quoted above, since  a significantly larger  abundance of C$_{60}^+$  than C$_{60}$ is expected in regions with strong UV, as for  PAH cations (e.g.\ Montillaud et al.\ 2013; Tielens 2005, 2008).

 C$_{70}$ has been identified  in one of the many sources where  C$_{60}$ has been found, the planetary nebula Tc 1. Its abundance there has been estimated to be comparable to   C$_{60}$ (Cami et al.\ 2010). However, the predominance of cooling by Poincar\'e fluorescence in C$_{70}$ (Appendix G) may have led to underestimating its abundance in all sources. Although Poincar\'e fluorescence is less important for C$_{60}$, its possible influence should also be carefully addressed for abundance determinations from IR line intensity and the interpretation of IR-line ratios.

\subsection{Outline of endohedral/exohedral fullerenes and
heterofullerenes of possible interstellar interest}

Fullerenes are known to form a rich variety of compounds by association
with many atoms that may be bound to the carbon cage in three main
ways. In endofullerenes the atoms (or molecules) are locked inside
the cage; such a configuration is specific to fullerenes when compared to PAHs, and it is 
remarkably stable. In addition, fullerenes share two other modes of
binding to heteroatoms with other carbon clusters, especially PAHs:
exohedral adsorption on the outer surface, and incorporation  as
heteroatoms  in the carbon network, replacing a C atom. Many exohedral
atoms only have a limited activation energy for desorption, at most a
few eV. Their expected interstellar stability is thus much less than for
endofullerenes, with a few exceptions (§3.3).  Heterofullerenes display
higher stabilities, albeit variously (§4).

All these compounds, endohedral exohedral hetero fullerenes (EEHFs), display the basic cage structure of fullerenes and most of its electronic
level structure. Generally little is changed in the vibrational modes
and the cage stability. But the presence of the
heteroatom may bring fundamental changes in chemical properties and
spectra, and  in the ionization potential, owing to the addition of
electrons to the cage by charge transfer (see e.g. Jura \& Kroto 1992;
Shinohara 2000; Dunk et al.\ 2014). In addition, covalent bonding with
the cage may strongly perturb and hybridize the orbitals, lowering the
symmetry and strengthening forbidden transitions of pure fullerenes.  

More specifically, charge exchange with associated atoms
may add one or several electrons in the LUMO orbital;  as analyzed e.g.\ by Leach
(2004), bonding to adducts may take one or two $\pi$ electrons out of the HOMO orbital, as also in C$_{60}^+$, without changing much the remaining
electronic structure of the cage. Such an unfilled HOMO shell allows new
transitions from lower shells. Both cases open the way to new optical
or IR transitions absent in C$_{60}$.
Associated metals also introduce additional levels in the DIB
energy range, corresponding to their own energy levels.  

As discussed in detail below, interstellar formation of EEHFs is often not obvious. However,  EEHFs might perhaps form somewhat normally in the
same interstellar processes together with pure fullerenes as long as
heteroatoms are present to take part in the formation process.
Interstellar cages of EEHFs are probably as resilient as those of pure fullerenes. Extraction of endohedral atoms also proves very difficult, but there are a few exceptions. But most exohedral compounds should be fragile to UV photolysis.

The next five sections summarize the most relevant basic properties of
potential interstellar EEHFs for those that may be relevant in the ISM. 

 \begin{figure}[htbp]
	 \begin{center}
 \includegraphics[scale=0.45]{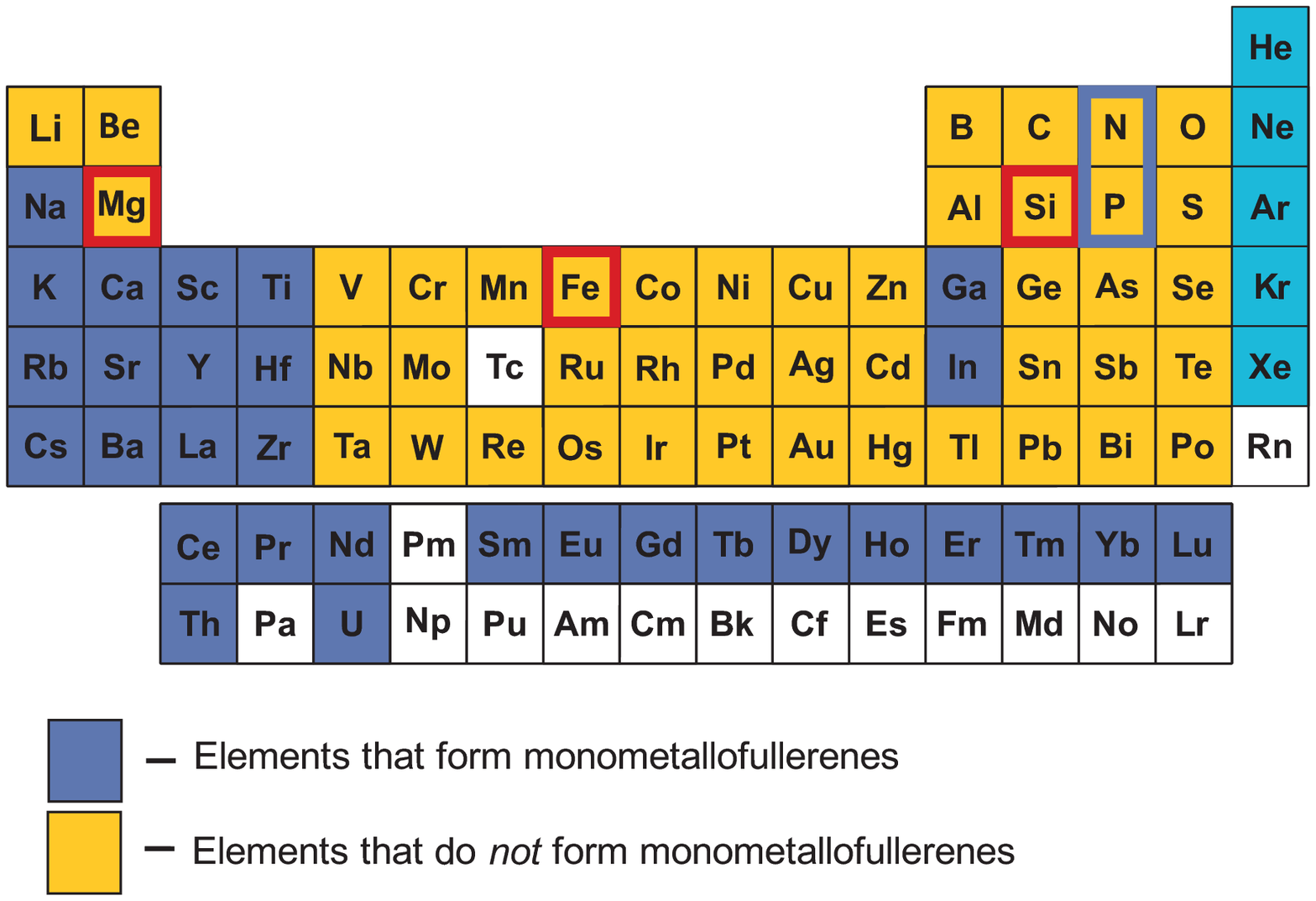}
 \caption{{(\it Adapted from Supplementary Fig.\ 5 of Dunk et al.\ 2014)}. Behaviour of the chemical elements for the formation of mono-atom endohedral compounds with C$_{60}$: {\it grey-blue squares}, elements that form mono-metallofullerenes in laser graphite ablation; {\it full yellow squares}, elements that do not form confirmed mono-atom endohedral compounds; {\it light blue squares}, rare gases that form mono-atom endohedral compounds; {\it yellow squares with grey-blue thick borders}, other non metals that form mono-atom endohedral compounds. In addition, a few abundant cosmic elements that do not form confirmed mono-atom endohedral compounds in  laser graphite ablation, are indicated by {\it yellow squares with red thick borders}}
     \end{center}
 \end{figure}

\section{Potential interstellar metallofullerenes}

\subsection{Basic properties of endofullerenes}

The ability to stably encapsulate various atoms and even molecules
inside a carbon cage was immediately recognized as an essential
property of fullerenes, with possible astrophysical implications
(e.g.\ Kroto \& Jura 1992).  Long  after that,
endohedral compounds (endofullerenes, EF), especially with metals
(metallofullerenes) are still a key component of the fullerene landscape
(e.g.\ Dresselhaus 1996; Shinohara 2000; Eletskii 2000; Akasak \& Nagase
2001; Guha \& Nakamoto 2005; Lu et al.\ 2012; Popov et al.\ 2013; Cong et
al.\ 2013; Dunk et al.\ 2013, 2014).  Mono-atom endohedral fullerenes are known for a substantial number of the elements, as seen in Fig.\
5 and Table B1.  Confirmed configurations include mainly metals, especially alkalis and some alkaline earths,
but also  non-metals such as N, P and rare gases.

But such endofullerenes have not been substantiated for most d-transition metals and a few other cosmically abundant elements (Mg, Si, Al), as well as the oxygen family. Indeed, as described  by Shinohara (2000)
and Dunk et al.\ (2014), for metals, the existence and properties of
endohedral compounds are closely related to the possibility of charge
transfer between the metal and the cage (§3.1.1).

Most  EFs of this confirmed list are formed relatively easily in
the various classical processes of formation of fullerenes, such as
laser graphite ablation or arc discharge, by mixing some compound of the
considered element with graphite (see e.g.\ a summary of these methods by Lu et al.\ 2012; Cong et al.\ 2013; Eletskii et al.\ 2000; Dunk et al.\
2013, 2014). In addition, in many cases, EFs may be formed in small
quantities just by bombarding fullerenes with atomic ions of moderate
energy (a few tens of eV). Such EF synthesis has been achieved for rare gases 
(He, Ne, Ar), alkalis (Li, Na, K), Ca, etc.  The threshold collisional energy of
implantation increases from $\sim$6\,eV for He$^+$ (Campbell et al.\
1998) to $\sim$50\,eV for Ar$^+$ or K$^+$ (Deng et al.\ 2003).

There are various indications that most EFs might resist destruction
practically as might empty fullerenes. One clue is precisely the
high value of the energy of maximum efficiency for ion implantation,
proving that below this value the probability of cage destruction or
opening remains  low. Another proof is the evidence that delayed
laser photoemission of electrons is similar for endohedral and empty
fullerenes, showing that both are capable of storing large amounts of
internal energy without dissociation (Beck et al.\ 1996; Clipston et al.\
2000).  Except for very small atoms and their ions
(H and He, but also N), opening the cage enough to allow the
atom or its ion to escape requires injecting a lot of energy, typically
a few tens of eV (see e.g.\ Wan et al.\ 1992; Campbell \& Rohmund 2000),
which is not very different from what is needed to extract a C$_2$ molecule
from the cage (discussed in §2.3.5).

\subsubsection{Charge distribution and structure of endofullerenes}

Electron charge transfer from the atom to
the cage is general for confirmed endometals whose formation is  favoured by it 
(Dunk et al.\ 2014). It typically implies one to three electrons depending on the
metal valence: one  for alkalis; two for alkaline earths like Ca, Ba; three for La (e.g.\ Shinohara 2000; Dunk et al.\ 2014;
but see below for Ca and Mg).  On the other hand, charge transfer is
practically non-existent for rare gases and N (e.g.\ Eletskii et al.\ 2000). It is
a key factor for the various properties of EFs: atom position, bonding
to the cage, internal motions, ionization potential, the structure of
excited electronic levels, the optical spectrum (§3.1.2), electron and
nuclear spin resonance; superconductivity, etc.

Charge transfer obviously depends mainly on the
ionization potential of the metal (IP) being small enough; e.g., very roughly, IP$_1$~-~e$^2$/r$_{\rm
F}$~$\la$~E$_{\rm A}$, where IP$_1$ is the first IP, r$_{\rm F}$ the
radius of the fullerene cage  and E$_{\rm A}$ the electron attachment
energy to the cage. For r$_{\rm F}$\,$\approx$\,3.5\,\AA\   and E$_{\rm
A}$\,$\approx$\,2.7\,eV, it requires IP$_1$\,$\la$\,6.8\,eV. This
immediately shows (see e.g.\ Table B1) that  charge transfer is not
possible for Mg, as well as for Fe and other d-transition metals. It may thus
explain why these metals do not (easily) form endohedral compounds.

When charge transfer fully operates for n electrons, neutral
metallofullerenes with alkalis and alkaline earths have thus a structure
close to M$^{n+}$@C$_{60}^{n-}$. But, as noted, this is not the case
for Mg. Even for Ca, if the transfer of the first 4s electron
proves complete, the case does not seem as clear for the second one  
because the
energies of the systems Ca$^{+}$@C$_{60}^{-}$ and Ca$^{2+}$@C$_{60}^{2-}$
are close, therefore the ground state is instead a hybrid state (e.g.\
Stener et al.\ 2002), even if the transfer of two electrons
dominates (Wang et al.\ 1993).

Because the extra-electrons occupy the LUMO orbital of the empty fullerene,
their ionization energy is reduced compared to the latter by an amount of the order of the LUMO-HOMO gap. For example, the ionization potential (IP) of
M@C$_{60}$ is lowered with respect to C$_{60}$ from 7.6\,eV to typically
$\sim$6.0-6.4\,eV for alkalis and $\sim$6-7\,eV for Ca@C$_{60}$ (Broclawik
\& Eiles 1998; Stener et al.\ 1999, 2002). On the other hand, the second IP, i.e.\   the IP
of (M@C$_{60})^+$, does not seem much smaller than that of  C$_{60}^+$ (11.5\,eV). 

Charge transfer induces a strong electrostatic interaction between the
positive ion and the negatively charged cage. This implies that the most
stable position for the metal ion is generally, but not always, not the centre of the cage, but a position closer to the cage, with some
 chemical bonding. However, for C$_{60}$
the possible position is always multiple with energy degeneracy due to
the initial  high symmetry. 
Such a configuration breaks the original symmetry and opens 
complex possibilities for splitting the energy levels and for the internal
motion of the encapsulated ion  (e.g.\ Hern\'andez-Rojas et al.\ 1996;
Zhang et al.\ 2008).

\subsubsection{Electronic energy levels of endofullerenes}

Endohedral fullerenes, like other EEHFs, are interesting candidates for DIB
carriers because they may have relatively strong electronic transitions
in the visible range thus differing from C$_{60}$ and other pure
fullerenes. But this is probably not the case for non-metal atoms located in the centre of
a C$_{60}$ cage such as He and Ne, as well as for H$_2$@C$_{60}$
(§7).

On the other hand, in metallofullerenes the optical spectrum may be
completely modified by charge exchange. In addition, the outer orbitals of the endohedral atom or ion are strongly  perturbed by the strong attractive potential of the cage (e.g. Connerade et al.\ 2000, Xu et al.\ 1996, Hasoglu et al.\ 2013). They may also be transformed by entering chemical bonding between the metal and the cage, and eventually hybridize with fullerene SAMO orbitals (§2.1.1). 

As quoted, charge
transfer may add n electrons to the cage C$_{\rm NC}$ giving the
same electronic structure as C$_{\rm NC}^{n-}$. One may thus expect,
e.g.\ for Na@C$_{60}$, a spectrum similar to C$_{60}^-$ for the lower
electronic transitions, especially in the infrared, but one should find
a richer spectrum at shorter wavelengths since the ionization potential
(IP) of Na@C$_{60}$ ($\sim$6-7\,eV) is much higher than the
attachment energy of C$_{60}^-$ (2.7\,eV). This allows the presence
of additional bound electronic levels, including SAMO ones.  In the same way, the spectrum
of metallofullerene cations, (M@C$_{60}$)$^+$, for metals of valence n, is similar to C$_{60}^{(n-1)-}$ --  e.g., C$_{60}$ for alkalis, and again C$_{60}^-$
for alkaline earths such as (Ca@C$_{60}$)$^+$, i.e.\ Ca$^{2+}$@C$_{60}^-$.

Analysis of the optical spectra of metallofullerenes thus presents
the same difficulties as do C$_{60}$ or C$_{60}^-$. It is therefore
not surprising that even theoretical studies of their energy levels
and excitation are lacking, incomplete or ambiguous, and we are
still waiting for laboratory spectra. The
accuracies of wavelength determinations are still far from being
useful in constraining possible associations of visible DIB features
with metallofullerenes. Estimates of line strengths are practically
always lacking.

The simplest cases, such as K@C$_{60}$ and Ca@C$_{60}$ and their cations
provide a good illustration of this situation (see Appendix C, where various calculations about Mg@C$_{60}$ are also reviewed despite the absence of confirmed experimental evidence).

The situation is  more confused for several other cosmically important
elements, such as  Si, Fe, and other d-transition metals for which the
existence and  a fortiori the structure of endohedral compounds are not
established (see e.g.\ Cong et al.\ 2013; Dunk et al.\ 2014). However,
there is evidence that at least Si may have  stable heterocompounds (§4.3).
The case of endohedral Fe is, of course, especially tantalizing because of the large Fe  abundance and the richness of its visible spectra (e.g.\ Simon \& Joblin 2009; Tang et al.\ 2006). But Fe@C$_{60}$ does not seem to have been isolated in the laboratory (see e.g.\ Basir \& Anderson 1999), while various forms of exohedral Fe are known as discussed below (§3.3.1). 

\subsection{Possible interstellar formation of metallofullerenes}

As quoted above,  some metals easily form endohedral compounds when they are present in fullerene formation processes (Dunk et al.\ 2013, 2014). Therefore, metals such as Na, K, Ca, which display significant abundances  (see Table B1), should be first candidates for interstellar endometals. This is especially true  for possible fullerene formation in interstellar grain collisions, whose carbon-vaporization physics might be somewhat similar to laser ablation. 

But one should keep in mind that the possible interstellar `top-down' fullerene formation from PAHs (§2.3.3) is completely different from `bottom-up' cage formation in laboratory. We therefore keep considering the possibility that other most abundant metals such as Fe, Mg, Si, Al, and Ni, might give interstellar endofullerenes while they are known not to normally do in laboratory conditions (Dunk et al.\ 2014 and Table B1). The only possibility for ensuring this seems to be  with a metal atom tightly bound to the edge of highly dehydrogenated PAHs (dPAHs) when they  begin the folding process to form the cage. However, evaluating the order of magnitude of the actual rates of endohedral formation appears as a difficult task by implying two key steps in the modelling:  metal attachment to the PAH, and non-ejection of the attached metal in the final outcome of the folding process. Both processes are very uncertain but may be roughly addressed as follows.

It seems that most metal ions could efficiently attach to more or less dehydrogenated neutral dPAHs. 
For order-of-magnitude estimates, we may tentatively assume a rate up to 10$^{-9}$cm$^{-3}$s$^{-1}$, although it is a significant fraction of the Langevin rate. Assuming n$_{\rm H}$\,=\,50\,cm$^{-3}$ and a total interstellar abundance of metals in the gas phase of 2$\times$10$^{-6}$ yields that a dPAH frequently attaches a metal about every 3$\times$10$^5$yr or so.  However, the duration of the time the metal remains stuck to the PAH skeleton crucially depends on the nature of its binding to the PAH. Only  the dangling bonds at the periphery of dPAHs should open the possibility of tight bonding, with binding energies possibly higher than the strongest hydrogen bonding to PAHs ($\sim$4.8\,eV, e.g.\ Montillaud et al.\ 2013). The actual existence of such strong metal-dPAH binding is not warranted. It needs to be explored further.  Nevertheless, a significant  probability of  finding some metals that are strongly bound to dPAHs, such as Fe, Ni, or Si,  seems possible since they are known to form strong bonding with  carbon shell structures (see e.g.\ Lee et al.\ 1997 for binding of Ni to the edge of carbon nanotubes and §4.3 \& 3.3.1 for strongly bound incorporation of Si and Fe into fullerene shells, respectively). But the question remains more uncertain for metals such as Mg, Ca, and Al, and a fortiori alkalis.

An additional difficulty is keeping the metal atom attached to the carbon shell, possibly inside, when the latter finally folds into a stable fullerene cage. This seems possible since the folding activation energy to be overcome by  UV heating impulse is estimated to be only  $\sim$4\,eV for N$_{\rm C}$\,$\sim$\,66 by Bern\'e et al.\ (2015a). 
When an endofullerene is eventually formed, one may expect that  it remains stable in the last steps when the cage reorganizes with eventual C$_2$ ejections. The same could be true for hetero Si (§4.3).

But there is no need to say that all these processes, hence their outcome, are extremely uncertain. Nevertheless it would be very useful to explore them with the power of modern simulations.

 \subsection{Strongly bound exohedral fullerenes}

\subsubsection{Fe and other metals}

As usual the case of iron fullerene compounds is complicated because of the propensity of iron to form complex bonding, especially with carbonaceous surfaces. Anyway, independently of the uncertain existence and properties of Fe@C$_{60}$,  iron reacts efficiently with the outer surface of fullerenes to form two main compounds, as for other d-transition metals  (Basir \& Anderson 1999). 

One of this compounds is an exohedral coordination complex weakly bound (1–3\,eV) to the surface. It forms with no activation barrier at low collision energy between Fe$^+$ and C$_{60}$. It seems similar to Fe binding to  PAHs with binding energy 0.6-2.5\,eV,  as calculated by Simon \& Joblin.\ (2007). However, such weakly bound Fe or Fe$^+$ cannot survive long on free fullerenes to interstellar UV photolysis.

The other complex, C$_{60}$Fe, or rather C$_{60}^+$Fe, is observed in C$_{60}$+Fe$^+$ collisions at  energies  $\ga$10\,eV. It thus has a substantial activation barrier to formation, $\sim$8-10\,eV, is chemically bound and remains stable for injected energy up to $\sim$20\,eV. Basir \& Anderson (1999) have proposed that this complex is a network-bound structure, with the metal atom probably sitting above the fullerene surface, chemically bound to two or more carbon atoms. Not much more information seems available about such  C$_{60}$Fe$^+$ or C$_{60}$Fe compounds,  which might be  stable in interstellar conditions, but should probably not form easily there. Nothing seems to be known either about their electronic levels and optical spectrum. One may expect that the strong binding with Fe and the distortion of the cage significantly modify the C$_{60}$ electronic level structure. More importantly, the large number of low excited levels of Fe and Fe$^+$ in the range $\sim$2--3\,eV must add a number of visible transitions with significant oscillator strengths, as found by Simon \& Joblin (2009) for Fe-PAH$^+$ weakly bound complexes.

Either way, when discussing possible structures for interstellar Fe bound to carbonaceous particles, one should keep in mind the strong catalysis power of Fe in forming complex particles (see e.g.\ Zhou et al.\ 2006), although significantly abundant carbon nanotubes seem excluded in the ISM. As stated above, the case of fullerene compounds with other less cosmically abundant d-transition metals seems very similar to Fe compounds (Basir \& Anderson 1999); e.g., various compounds of Ni have been recently theoretically addressed by Neyst et al.\ ( 2011).

\subsubsection{Oxygen compounds}

Oxygen is known to react efficiently with fullerenes; even  atmospheric O$_2$ is the main cause of their decomposition in normal conditions. Oxygen may thus form a long list of oxides with C$_{60}$, including hetero C$_{59}$O, possibly O@C$_{60}$, and many somewhat unstable exohedral mono- or poly-oxides (see e.g.\ Heymann \& Weisman 2006). 

Christian et al.\ (1992a) have reported  that  C$_{59}$O$^+$ significantly forms in collisions of O$^+$ with  C$_{60}$ and is amazingly stable for collision energies up to at least $\sim$30\,eV. 
 They suggest two possible structures for C$_{59}$O$^+$: either hetero inclusion of trivalent O$^+$ in the carbon cage network or endohedral CO@C$_{58}^+$. 
However, the DFT theory of Jiao et al.\ (2002) has found that the most stable  shape of C$_{59}$O should be an open fullerene cage, but C$_{59}$O seems to have never been synthesized (Vostrowsky \& Hirsch 2006). Therefore,
the question of its long survival in the interstellar medium remains open. 

The most stable exohedral oxide is the  epoxy C$_{60}$O with a bridging bond of O on one double bond C=C of  C$_{60}$ (Heymann \& Weisman 2006); however, it is very unstable  (Christian et al.\ 1992a). It should therefore be rapidly photolyzed by the interstellar UV radiation. 
 
Sulphur fullerene compounds present some similarities with oxygen (e.g.\ Jiao et al.\ 2002; Ren et al.\ 2004). It is even more difficult to assess their eventual interstellar importance.

\section{Potential interstellar  azafullerenes, silafullerenes and other heterofullerenes}

As reviewed by Vostrowsky \&  Hirsch (2006), many heteroelements are able to enter the carbon network of fullerene cages, replacing a carbon atom. Besides the most studied C$_{59}$N and other azafullerenes, the synthesis of C$_{59}$X has been reported for X\,=\,B, P, Si, O, Ni and other d-transition metals,  and many calculations have addressed them. We stress the possible interstellar interest of C$_{59}$Si and other silafullerenes.  For transition metals, O and S we refer to §3.3.2. 

\subsection{C$_{59}$N, C$_{59}$HN and other azafullerenes in the laboratory and models}
Nitrogen heterofullerenes (azafullerenes) are among the best-studied fullerene derivatives as a result of the  propensity of N to  easily replace C atoms in carbon compounds. C$_{59}$N (or rather its dimer (C$_{59}$N)$_2$) and its derivatives stand as the most studied  heterofullerenes. The high reactivity of this radical may generate a variety of stable azafullerenes including C$_{59}$HN and C$_{59}$N-PAH dyads. 

 All such species keep most properties of pure fullerenes, but besides enhancing the chemical activity,  the presence of the additional N electron on the cage, may significantly alter the electronic structure and its energy levels. 
Compared to C$_{60}$, the ionization potential is significantly decreased (from 7.6\,eV for C$_{60}$ to about 6\,eV for C$_{59}$N, see Xie et al.\ 2003), but the electron affinity remains similar.

As reviewed by Vostrowsky \& Hirsch (2006), although it is hard to produce in graphite laser ablation, C$_{59}$N forms more or less  easily in a variety of processes -- mostly concerning exohedral nitrogen derivatives of fullerenes, but also gas discharge, ion implantation, etc.  
The stability of its cage proves comparable to, albeit slightly less than C$_{60}$ (Clipston 2000 from delayed photoionization, Christian et al.\ 1992b from ion collisions,  etc). However, its destruction begins by losing a CN molecule rather than C$_2$. The radical reactivity of C$_{59}$N is such that only its dimer, (C$_{59}$N)$_2$, is stable in laboratory conditions and most experimental information about C$_{59}$N indeed comes from dimer studies.

In addition to (C$_{59}$N)$_2$, many adducts of C$_{59}$N are known.  
However, despite the rich chemistry of azafullerenes 
 it seems that, as for C$_{60}$, the only adducts that have a chance to reach significant interstellar abundances, are with hydrogen, carbon or PAHs. 
C$_{59}$HN is well studied both experimentally (e.g.\ Keshavarz-K et al.\ 1996; Buchachenko \& Breslavskaya 2005) and theoretically (e.g.\ Andreoni et al.\ 1996). The binding energy of H to C$_{59}$N and the ionization potential are estimated to be $\sim$3.4\,eV (Buchachenko \& Brestlavskaya 2005) and $\sim$6.1\,eV (Andreoni et al.), respectively.  The binding energy of many other C$_{59}$N adducts is probably similar. This includes C$_{59}$N--PAH dyads and 
their cations. 
Such dyads of  C$_{59}$N bound to coronene, coranulene, etc.\ can be  easily synthesized as shown by Hauke et al.\ (2004).

As the cage properties are mostly preserved,  the vibrational modes of azafullerenes do not prove very different from the corresponding fullerenes. A number of theoretical studies have provided models for the electronic energy levels  of basic azafullerenes,  
corroborated by spectral measurements.
All optical  
transitions of C$_{59}$N$^+$ remain very weak with f-values staying below  a few 10$^{-3}$  (Xie et al.\ 2004). There is a significant increase in the case of  C$_{59}$N with theoretical f-values reaching $\sim$10$^{-2}$  in several bands in the range  4000--8000\,\AA\ (Ren et al.\ 2000). Band  intensities prove even larger for C$_{59}$N adducts with a strong broad feature about 4400\,\AA\   (Hauke et al.\ 2006). This includes C$_{59}$HN and (C$_{59}$N)$_2$ (see Fig.\ 3a of Keshavarz-K et al.\ 1996, 
and  f-values estimates  by Ren et al.\ 2000).

\subsection{Interstellar C$_{59}$N and other azafullerenes}

C$_{59}$N and other derived azafullerenes appear to be interesting candidates for interstellar fullerenes because of the extraordinary stability of the C$_{59}$N cage and the related propensity of N atoms to replace C in carbon networks. It has been repeatedly suggested that interstellar PAHs could contain significant amounts of nitrogen in their graphene network, at the level of possibly $\sim$1-2\%, in order to explain the properties of their 6.2\,$\mu$m feature (Hudgins et al.\ 2005; Tielens 2008) (although this interpretation has been contested by Pino et al.\ 2008). If this is confirmed for PAHs, there is  a possibility of finding a similar situation for interstellar fullerenes. Even only 1\% of randomly distributed N-atoms would mean that there is one N-atom in 33\% of the 60-atom cages, to be compared with 55\% and 10\% for 0 and 2 N atoms, respectively. But a high abundance of azafullerenes, similar to nitrogen-substituted PAHs, is not warranted in the most likely formation processes of interstellar fullerenes  (§2.3.3).  N atoms migh tend to be ejected in the formation process of fullerenes from PAH because their binding is weaker than for carbon atoms. Despite the  nitrogen presence  expected in carbon grains (e.g.\ Jones 2013), azafullerene formation in grain-grain shattering might be inefficient because of the expected similarities of this process with graphite laser ablation, which produces little C$_{59}$N.
One very hypothetical way of forming C$_{59}$N could perhaps be UV-processing of exohedral nitrogen derivatives of fullerenes bound to PAHs or grains, if any.

In C$_{59}$HN  the binding energy of H to the cage in the vicinity of N, $\sim$3.4\,eV, is significantly higher than the direct binding energy of H to a C$_{60}$ cage,  $\sim$2.7\,eV (§2.3.1).  
One may infer that the ratio n(C$_{59}$HN$^+$)/n(C$_{59}$N$^+$) is probably greater than n(C$_{60}$H$^+$)/n(C$_{60}^+$);  
it could thus be significant. One might therefore expect a significant abundance of C$_{59}$HN$^+$ if C$_{59}$N$^+$ was abundant enough.

\subsection{C$_{59}$Si and other silafullerenes}

Besides azafullerenes, silicon heterofullerenes (silafullerenes) seem by far the most attractive case because of their stability that results from the similarity of binding properties of Si and C and of the large abundance of silicon. Fullerene cages may thus amazingly support the stable  replacement of up to $\sim$20 C atoms  by Si  (Ray et al.\ 1998; Vostrowsky \& Hirsch 2006). 
However, in the ISM  we will only consider C$_{59}$Si here (and possibly C$_{58}$Si$_2$).

The high stability of silafullerenes is  amazing. Although  the binding energy of Si is computed to be much smaller than that of C  (Billas et al.\ 1999; Marcos et al.\ 2003),
silafullerenes resist temperatures up to $\sim$4000\,K before dissociating, and the first dissociation product is generally C$_2$, seldom SiC.  This behaviour is explained well by molecular dynamics simulations (Marcos et al.\ 2003).

The computed  electronic level structure of C$_{59}$Si and other silafullerenes is substantially different from C$_{60}$. The LUMO-HOMO gaps are smaller. The  first visible allowed optical transitions appear then at longer wavelength;  however, they remain relatively  weak (Koponen et al.\ 2008; Lan, Kang \& Niu 2015) and no results seem available for cations. Similarly, the ionization potentials decrease by about 0.5-0.7\,eV per Si atom 
(Tenorio \& Robles 2000).

Substantial abundances of silafullerenes in the ISM are not excluded. Since grain shattering could create  physical conditions similar to graphite laser vaporization and Si is  abundant in carbon grains, shattering could efficiently produce C$_{\rm 2n-1}$Si and  C$_{\rm 2n}$Si$_2$ with various n, like  laser ablation of mixtures of graphite with a small Si proportion (e.g.\ Pellarin et al.\ 1997, 1999). Si atoms could also strongly bound to dehydrogenated PAHs and survive their folding into fullerene cages (§3.2)\footnote{Si is known to bound on fully hydrogenated PAHs with a modest adsorption energy $\sim$1.5\,eV (e.g.\ Joalland et al.\ 2009) and with a large rate of accretion for Si$^+$ (Dunbar et al.\ 1994). However, such exhohedral Si cannot survive long to interstellar UV photolysis.}.

\section{Interstellar fulleranes}

 \subsection{Fulleranes in the laboratory and models}
Hydrogen may be bound  in large quantities on the outer surface of fullerenes by opening their numerous double bonds (30 in  C$_{60}$) to form strong CH alkane bonds. This sort of more or less hydrogenated fullerenes, fulleranes, up to highly hydrogenated compounds, such as C$_{60}$H$_{18}$ and C$_{60}$H$_{36}$, have long been recognized as major members of the fullerene family.  Their cosmic importance has also been discussed for a long time (Webster 1992; Iglesias-Groth 2006 and references therein). Their presence has been claimed in meteorites (Becker \& Bunch 1997), but this experimental evidence has been contested (Hammond \& Zare 2008).

 The energetics of the binding of H atoms to a fullerene cage  is  known rather precisely both experimentally and theoretically (see e.g.\ Bettinger et al.\ 2002 and also Vehvilainen et al.\ 2011; Okamoto 2001). It is summarized in Table 1. The H binding energies for larger even or odd numbers of H atoms seem comparable to C$_{60}$H$_2$ and C$_{60}$H, respectively, probably  up to about C$_{60}$H$_{36}$.
The situation is similar for cations with significantly stronger binding  -- 2.9\,eV for H on  C$_{60}^+$ (Petrie et al.\ 1995).

  \begin{table}
      \caption[]{Estimated reaction  
and activation energies  (eV) for dissociation reactions of C$_{60}$H and  C$_{60}$H$_2$ (Bettinger et al.\  2002).}
         \label{tab:lines}
            \begin{tabular}{l l c c }
            \hline
            \noalign{\smallskip}
N  &  Reaction    & $\Delta$E & E$_{\rm ac}$ \\
            \noalign{\smallskip}
            \hline 
            \noalign{\smallskip}
1 &   C$_{60}$H $\rightarrow$ C$_{60}$ + H    & 1.9  &  1.9   \\
2 &   C$_{60}$H$_2$ $\rightarrow$ C$_{60}$H + H      & 3.3   &  $\approx$3.3-3.5   \\
3 &   C$_{60}$H$_2$ $\rightarrow$ C$_{60}$ + H$_2$      & 0.7  &  $\approx$4.0  \\
4 &   C$_{60}$H$_2$ + H  $\rightarrow$ C$_{60}$H + H$_2$  & -1.3    &  0.1  \\
            \noalign{\smallskip}
            \hline
           \end{tabular}
\end{table}

The visible spectrum of C$_{60}$H$_2$ was measured by Bensasson et al.\ (1995), together with that of C$_{60}$H$_4$. Both spectra display a relatively strong feature near 4350\,\AA , similarly to most other C$_{60}$ adducts (e.g.\ Smith et al.\ 1995; Kordatos et al.\ 2003; Wang et al.\ 2012). Not much seems known about the optical spectra of C$_{60}$H, C$_{60}$H$_2^+$ and C$_{60}$H$^+$. 

Infrared, visible and UV spectroscopy studies of highly hydrogenated fulleranes were performed in matrix, mostly for astrophysical implications, by Cataldo \& Iglesias-Groth (2009), Cataldo et al.\ (2014), Iglesias-Groth et al.\ (2012), and references therein. The main findings are 
strong  characteristic bands due to C--H stretching close to 3.5\,$\mu$m and the  washing out most of the optical and near-UV bands of C$_{60}$. 

The variable large number of H atoms in fulleranes destroys the unique symmetry properties of C$_{60}$.  This might   diminish their interest for DIBs, except  light fulleranes, such as C$_{60}$H,  C$_{60}$H$_2$, and their ions.

\subsection{Interstellar  fulleranes}

Fulleranes should form efficiently  in the ISM by H accretion, especially onto ionized fullerenes (Appendix D).  
Although the activation energies needed to extract H or H$_2$ from fulleranes are significant ($\sim$1.9-3.3\,eV for neutrals, Table 1), their values remain significantly smaller  than in PAHs (4.8\,eV). One therefore expects that fulleranes lose more easily their H atoms by UV photolysis than PAHs for similar numbers of carbon atoms N$_{\rm C}$. However, PAHs are known to remain practically completely hydrogenated for N$_{\rm C}$\,$\sim$\,60 in standard interstellar UV (e.g.\ Fig.\ 10 of Montillaud et al.\ 2013 and Fig.\ 10 of Le Page et al.\ 2001). As a result, the interstellar dissociation of fulleranes is not straightforward. It is qualitatively discussed in Appendix D.

The question is made difficult for various reasons including the need of an elaborate thermodynamic treatment of H-photolysis by a single hard-UV photon, the uncertainty about internal energy conversion, the various H binding modes,  the possible role of specific reactions of fulleranes with H, forming H$_2$ (\#4 in Table 1).  Although there is no doubt that UV photolysis should be very efficient for the first H atom of fulleranes with an odd number of H atoms, such as C$_{60}$H, the outcome is less clear for even H numbers, such as C$_{60}$H$_2$. Although an efficient dehydrogenation of C$_{60}$ fulleranes in diffuse clouds does not seem  impossible, it is concluded in Appendix D that the question should be considered as unsettled until an elaborate treatment is provided, similar to the methods used for PAHs, as in Le Page et al.\ (2001, 2003) or Montillaud et al.\ (2013). Therefore, the ratio of the abundances of fulleranes to fullerenes, such as C$_{60}$H$_2$/C$_{60}$ remains uncertain. However, the rate of UV H photolysis is obviously reduced when the number of H atoms is large, for larger fullerenes including C$_{70}$, in UV-shielded regions, and even more for fullerenes attached to PAHs or dust grains. It is on the other hand increased for smaller cages than C$_{60}$.

There are many other fullerene hydrocarbons,  such as  C$_{61}$H and C$_{61}$H$_2$, longer chains, and their cations. But, as discussed in §2.3.2, the UV photolysis of  C$_{61}$ might be too fast to allow significant abundances, unless  C$_{61}$H$_2$ (methanofullerene) is  more stable. Note that  C$_{61}$H$_2$ has a characteristic spectral feature at 432\,nm (Anthony et al.\ 2003 and §5.1) while the optical spectra of  C$_{61}$H$_2^+$ seems unknown.

\section{Associations of interstellar fullerenes with PAHs. }

As  pointed out in §1, fullerenes are known to form somewhat stable associations with PAHs. They deserve special consideration because of the large interstellar abundance of PAHs. 
Such associations could be  frequent, and  fullerenes might keep some of their specific properties when engaged in them  and more or less easily episodically return to the gas phase. 

\subsection{Formation, destruction and abundance}

Orders of magnitude of the percentage of interstellar fullerenes bound in such associations might be tentatively estimated when building on the work carried out for modelling PAH clustering (e.g.\ Montillaud \& Joblin 2014). However, as noted, the most stable and important fullerene-PAH associations are likely to be  chemically bound because of the fullerene reactivity, while slightly weaker van der Waals interactions probably dominate the binding of clusters of larger PAHs.
As shown, for instance,  by Dunk et al.\ (2013) (see also Garc{\'{\i}}a-Hern{\'a}ndez et al.\ 2013),  fullerene-PAH dyads form easily in the laboratory in reactions of PAHs such as coronene with C$_{60}$ (and C$_{70}$) in energetic conditions. Coronene then loses one or two H atoms and is bound to C$_{60}$ by one or two C-C bonds (see Fig.\ 2 of Dunk et al.).  C$_{60}$ easily reacts with linear PAHs (e.g. Petrie \& Bohm 2000; Garc{\'{\i}}a-Hern{\'a}ndez et al.\ 2013), but not or with difficulty with fully hydrogenated non-linear PAHs such as coronene (Petrie \& Bohm 2000; Dunk et al.\ 2013). As expected, Dunk et al.\ find that partially dehydrogenated coronene shows a higher reactivity. In interstellar conditions, we may thus expect that similar dyads form easily  in reactions of fullerenes with dehydrogenated PAHs, but more rarely with fully hydrogenated PAHs.
The following simple considerations show that it is not impossible that such associations might constitute an important reservoir for interstellar fullerene compounds. 

The frequency of close collisions between fullerenes and PAHs depends a lot on the charge state of the two particles. It is clear that it is negligible when the two particles have a charge with the same sign. 
One of the most favourable cases is a collision between a cation and a neutral particle. These collisions  should occur close to the Langevin rate, $\sim$10$^{-9}$ cm$^3$s$^{-1}$, in both cases not depending on which is the cation. The rate of close collisions when both PAH and fullerene are neutral is significantly smaller by a factor $\sim$3-5. 

It is proven that there is a sharp transition between almost full hydrogenation and dehydrogenation of interstellar PAHs depending on their number of carbon atoms  N$_{\rm C}$ and the interstellar UV radiation field (ISRF). In the model of Montillaud et al.\ (2013) their Figure 10 shows that, for N$_{\rm C}$ $\sim$ 60, this transition occurs for typical values $\sim$ 3--20 G$_0$ (where  G$_0$ is the Habing intensity, see Note 7). 

For crude estimates, we assume  that PAHs contain 10\% of interstellar carbon in PAHs and that they are half ionized  with  N$_{\rm C}$\,=\,80. 
This yields 3x10$^6$ yr for the order of magnitude of the interval between close collisions of C$_{60}$ or C$_{60}^+$ with PAHs in diffuse clouds. It is likely that there is on average a low but significant probability of cycloaddition reactions in such close collisions. It is thus possible that the rate of formation of  fullerene-PAH dyads be faster than  the rate of destruction of fullerenes -- about every 10$^9$\,yr  from Micelotta et al.\ (2010a, 2014), see §2.3.5. 

However, such dyads are certainly more fragile than fullerenes themselves, by just breaking their double C-C bridging bond. The exact binding energy seems unknown, but  
it should be slightly higher than for the C$_{60}$ or  C$_{60}^+$ dimers. If one adopts the DFT theoretical estimate, 0.6-1\,eV,  of Bihlmeier et al.\ (2005) and Wang et al.\ (2014) for these dimers, a  best guess for the binding energy of C$_{60}$ to a PAH  could be $\sim$2-3\,eV, but this value is highly uncertain.
The activation energy is even more uncertain.

If one tries to compare the rate of dissociation in the interstellar UV of such dyads with H atoms in PAHs, it is possible that the much larger number of carbon atoms compensates for the weaker bond, $\sim$3\,eV  vs 4.8\,eV.  
Therefore, it is possible that such dyads of fullerenes with heavy PAHs might survive fairly long.

On the other hand, as discussed in §2.3.3, fully dehydrogenated PAHs should be highly unstable and they possibly very quickly fold and transform into fullerene cages with N$_{\rm C}\,\sim$\,30-50. Information is lacking for estimating the reactivity of  such unstable smaller fullerenes cages with larger PAHs and with various fullerenes.

All in all, it is possible that such PAH-fullerene dyads and associations of fullerenes with PAH clusters contain a significant fraction of interstellar fullerenes (the associations to carbonaceous grains should also be considered as Cami et al.\ 2010 suggested). But the importance of such dyads still has to be confirmed.

\subsection{Properties}

A key question about such fullerene association with carbonaceous particles is whether the engaged fullerene might keep most of its main properties. 
In such dyads, the structure of the cage should not be much affected together with its  electronic levels, and also its vibration modes (Garc{\'{\i}}a-Hern{\'a}ndez et al.\ 2013).
However, one may expect the appearance of a few additional modes linked to the PAH-C$_{60}$ bond, some mixing of the vibration modes of the C$_{60}$ and PAH moities, the significant modification of the IR spectrum, and the  slight but significant changes of the wavelengths of electronic fullerene transitions. On the other hand,  one needs  to consider   the possibility of exchanging energy and charge with the PAH moiety and the modification of statistical properties related to the total number of  C atoms N$_{\rm C}$. As discussed below, hydrogenation and the stability of other adducts could be deeply modified, the fullerene ionization made more difficult and most of the IR emission transferred to the PAH bands.

As is obvious from the discussion above, the stability of PAH hydrogenation against photolysis should be enormously improved  with the large increase in N$_{\rm C}$  
and the high energy of the PAH C-H bonds (assuming that the intramolecular vibrational redistribution is effective), so that the PAH should recover its total hydrogenation immediately after the dyad formation.  However, even if the H-photolysis of fulleranes bound to PAHs is probably  inefficient, one should perhaps also consider reactions of atomic H  
with fulleranes bound to the PAH forming H$_2$ (such as reaction \#4 of Table 1).
If their very low activation energy ($\sim$0.1\,eV) could be overcome, their rate  
can be comparable to the rate 
of accretion of H onto fullerenes or fulleranes,  possibly reducing the average number of H atoms on the PAH-bound fullerene.
This question should be carefully addressed.

As the ionization potential of C$_{60}$ is significantly higher (7.6\,eV) than for relatively large PAHs (e.g.\ 6.36\,eV for ovalene, C$_{36}$H$_{16}$), one may expect that the ionization mostly stays on the PAH moiety in ionized C$_{60}$-PAH associations while the  C$_{60}$  moiety remains practically neutral. This is obviously capital for the expected fullerene optical spectrum. It may also significantly bear on its infrared emission since the vibrational  energy conversion between the two moieties is certainly  fast and the infrared bands of PAHs are much stronger than those of C$_{60}$. Adopting the values from Choi et al.\ (2000) for the IR band strengths of C$_{60}$ yields a total strength for its  four IR bands of 58\,km/mol. In comparison, from models of Langhoff (1996),  the total strengths of the infrared bands of ovalene (C$_{32}$H$_{14}$), neutral and cation, are  772 and 1629 km/mol, respectively. It turns out that only a small fraction of the UV energy absorbed by a C$_{60}$-PAH dyad should be emitted in the IR bands of C$_{60}$. 
As C$_{60}$ contributes about half the dyad UV absorption, one might underestimate the abundance of fullerenes locked in such dyads by an order of magnitude  when applying the usual assumption of complete conversion of the UV energy that it absorbs, into its IR emission bands.  

As other (6,6) adducts, C$_{60}$-PAH dyads are expected to display universal characteristic optical features absent in C$_{60}$ (§5.1): a relatively strong, sharp band close to 4330-4350\,\AA , and a  broader feature close to 7000\,\AA . This is confirmed for neutral Diels-Alder adducts such as C$_{60}$-anthracene (Udvardi et al.\ 1995). But no precise information seems to exist for the corresponding cation spectra.

C$_{59}$N and other azafullerenes form single-bond adducts  easily with many compounds, including PAHs (§4.2). Because it is likely that reaction rates and dissociation energies are similar to those of fullerenes with PAHs, the above discussion of various PAH-fullerene dyads should also  roughly apply to such PAH-azafullerene dyads.

 \section{Interstellar H$_2$@C$_{60}$?}

It has long been recognized that space inside fullerene cages is large enough to contain  molecules.  We limit our interstellar discussion to H$_2$@C$_{60}$. 

\subsection{Properties of H$_2$@C$_{60}$}

\subsubsection{H@C$_{60}$?}
H-atoms and protons are certainly among the atomic particles that can most easily go through the cage surface of fullerenes in both directions. The most recent theoretical discussion of the interaction energy between H and C$_{60}$ seems to be that of Vehvilainen et al.\ (2011), who summarizes previous works. The lowest value of the energy barrier preventing H atoms entering or getting out the cage is found to be  only about 2.5-3\,eV. Bound states may exist on both sides of this barrier. While H may chemisorb on the outer surface of the cage with a substantial binding energy, $\sim$2\,eV (§5),  the binding energy at the centre of the cage seems practically zero. Endohedral H@C$_{60}$ should thus have H confined by a relatively high energy barrier. However, H@C$_{60}$ seems to have never been confirmed in the laboratory. This looks consistent with its expected instability at room temperature. 

Similarly, H@C$_{60}^+$ does not seem to have been formally identified either. All the positive charge should stay on the cage, again with a neutral H atom inside, a binding energy close to zero at the centre of the cage, and  a significant lowering of the energy barrier with respect to H@C$_{60}$ (Ramachandran et al.\ 2008). The last point should make H@C$_{60}^+$ even easier to form and dissociate than H@C$_{60}$.

\subsubsection{H$_2$@C$_{60}$ in the laboratory}

Unlike elusive H@C$_{60}$, H$_2$@C$_{60}$ is a well-established stable endofullerene.
Its formation proves to not be easy, however (Komatsu et al.\ 2005), confirming the impenetrability of the fullerene cage to easy H$_2$ crossing. Such an efficient confinement warrants the stability of H$_2$@C$_{60}$ (Murata et al.\ 2006),  which might compare to C$_{60}$.

Detailed studies  illustrate the spectacular consequences of the confinement of the H$_2$ molecule (see e.g. the review by Vehvilainen et al.\ 2011), including, for example, infrared spectroscopy (Mamone et al.\ 2009, 2011; Ge et al.\ 2011), inelastic neutron scattering (Vehvilainen et al.\ 2011), and NMR studies (Murata et al.\ 2006; Turro et al.\ 2010).

The 2\,$\mu$m absorption spectroscopy of solid films of H$_2$@C$_{60}$  has fully confirmed theoretical predictions for its fine structure. The classical rotation structure of H$_2$  is replaced by a completely different set of quantum levels resulting from the coupled translational and rotational motion of the H$_2$ quantum rotor in the  C$_{60}$ cage.

In addition, confinement strongly enhances the intensity of the forbidden IR transitions similarly to the  effect observed for H$_2$  confined in a gas at  high pressure (e.g.\ Frommhold 1993). 

\subsection{Is there a possibility of finding  interstellar H$_2$@C$_{60}$ ?}

H$_2$@C$_{60}$  could have an ISM lifetime that is  not so much shorter than fullerene cages themselves because it is hard to imagine a way to take H$_2$ out of the cage without destroying the cage. Given the cosmic abundance of hydrogen, H$_2$@C$_{60}$ might be one of the fullerene compounds to be considered  in the interstellar medium. However, forming H$_2$@C$_{60}$ in interstellar conditions does not seem straightforward since there is no hope of directly injecting an H$_2$ molecule into the cage. One may nevertheless try to imagine a few processes that are eventually able to form it.

As stated in §7.1.1, it seems  easy to inject a proton H$^+$  into the cage, where it must immediately stabilize  into H by charge exchange with the cage. This accretion rate could equal a small but significant fraction $\eta$ of the large Langevin rate (k'$_{\rm L}$=2.0  10$^{-8}$\,cm$^3$s$^{-1}$), 
because of the small size of H$^+$ and the likely absence of a high activation barrier (Ramachandran et al.\ 2008). For  n(H)\,=\,50\,cm$^{-3}$ and a proton abundance of 3x10$^{-4}$, the rate of forming H@C$_{60}^+$ from C$_{60}$ + H$^+$ could reach  $\eta$ $\times$ 10$^{-2}$\,yr$^{-1}$. 
Similar to  photolysis discussions (Appendix D), it seems that, with an activation energy as low as $\sim$\,2.5\,eV, H@C$_{60}$ can hardly survive any absorption of a hard-UV photon.  The latter occurs every few years with standard UV radiation (Note 7). This does not leave much time for the arrival of a second H$^+$, forming H$_2$@C$_{60}$ at the rate $\eta$' $\times$ 10$^{-2}$\,yr$^{-1}$.

 All in all, this would yield a global formation rate of H$_2$@C$_{60}$ that is lower  than $\eta \eta$' $\times$ 10$^{-4}$\,yr$^{-1}$, i.e.\ perhaps $\sim$10$^{-8}$-10$^{-9}$\,yr$^{-1}$ in such very uncertain assumed conditions.  This could be just enough to yield that a  certain number of  C$_{60}$ cages could contain an H$_2$ molecule at the end of their lifetime estimated to be $\ga$ a few 10$^{8}$\,yr (§2.3.5).  But this result remains extremely fragile because of the many uncertainties of the various parameters. 

An alternativ and even more tentative less documented process to form H$_2$@C$_{60}$ might perhaps be violent perturbations of the cage, e.g.\ in shocks, in presence of  a large amount of hydrogen bound to the outer surface. As seen in laser formation of H$_2$@C$_{60}$ (Oksengorn 2003) or in simulations (e.g.\ Long et al.\ 2013 for a breathing-trap mechanism for H-encapsulation), such violent perturbations may temporarily distort or open the cage, with or without loss of C$_2$ molecules, eventually allowing exohedral H$_2$ to enter it. Appealing cases for such a process could be H-saturated fulleranes bound to PAHs (§6.2) or grains, or perhaps free fulleranes.

It seems that the presence of endohedral H$_2$ would not change  the basic properties of interstellar fullerenes much, making it difficult to reveal its presence, except thanks to specific spectroscopic features.
However, even if it were significant, the presence of endohedral H$_2$ should be difficult to confirm because the  abundance of  H$_2$@C$_{60}^+$, for instance, could hardly be expected to exceed or even reach 10$^{-9}$. There is thus practically no chance to ever detect  its  near-IR H$_2$ 2\,$\mu$m absorption. UV spectroscopy is more sensitive by orders of magnitude than absorption of weak IR lines. It can detect H$_2$ column densities lower than $\sim$10$^{14}$\,cm$^{-2}$ (e.g. Gillmon et al.\ 2006)). But even so,  the UV detection of H$_2$@C$_{60}$ seems out of reach. 

The presence of endohedral H$_2$, if any, in fullerene DIB carriers should, however, cause a very small wavelength shift of the DIB itself with respect to a H$_2$-less fullerene. One might thus expect weak satellites to fullerene DIBs. This could be the best way to confirm or deny the existence  of significant amounts of endohedral H$_2$. 
The expected value of such shifts should be addressed theoretically. A very rough idea of their expected order of magnitude might be obtained by considering the average polarizability per unit volume corresponding to an H$_2$ molecule within a C$_{60}$ fullerene cage. Using  the shift value measured for C$_{60}$ visible forbidden bands by Sassara et al.\ (1997), 
$\sim$35\,cm$^{-1}$, in a Ne matrix, for example, one may infer a rough order of magnitude of 9\,cm$^{-1}$ for the endohedral-H$_2$ shift  by scaling from the polarizability per unit volume of Ne. There is no hint of such weak satellites in the observed spectra of the 9600\,\AA\ C$_{60}^+$ DIBs. It thus appears that the presence of endohedral H$_2$ cannot exceed a few percent of the C$_{60}^+$ cages. But high telluric noise in this spectral range still restricts the sensitivity of this limit value.

\section{Fullerene compounds and DIBs}

In spite of so many question marks, such an inventory of potential properties of various interstellar fullerene compounds allows us to briefly discuss the possibility further that they could significantly  be carriers of DIBs. Even after the first success of the recent confirmation of the C$_{60}^+$ DIBs, the two main obvious difficulties for this purpose remain the uncertainties about the wavelengths and strengths of visible (and near-IR) bands of other fullerene compounds, and about their interstellar abundances.

\subsection{Line strengths, abundances and DIB strengths}

\noindent It  is useful to recall the general relation between the expected equivalent width of a DIB, the abundance of its carrier M and the band oscillator strength f. It may write (see e.g. Cami 2014)
\begin{equation}
{\rm EW}({\rm mA/mag)=10(\lambda/5500)^2(f/10^{-2})(X_{CM}/10^{-4})(60/N_{C})}
\end{equation}

\noindent where EW(mA/mag) is the equivalent width expressed in m\AA\ for 1\,mag of interstellar reddening E(B-V), and  X$_{{\rm CM}}$ denotes the fraction of interstellar carbon locked up in the considered type M of
`C$_{\rm NC}$' molecules (e.g.\ Ca@C$_{60}$ or C$_{59}$N). Note that X$_{{\rm CM}}$ in  Eq.(2) refers to the total carbon abundance in the interstellar medium assumed to be 
$\chi_{{\rm C}}$\,=\,N$_{\rm C}$/N$_{\rm H}$\,=\,3.9\,10$^{-4}$, while Cami (2014) refers to N$_{\rm C}$/N$_{\rm H}$\,=\,1.4\,10$^{-4}$, i.e.\ the carbon abundance in the interstellar gas. Our convention yields that the abundance of molecules M relative to H is $\chi_{{\rm M}}$\,=\,N$_{\rm M}$/N$_{\rm H}$\,=\,0.65\,10$^{-5}$\,X$_{{\rm CM}}$ $\times$ (60/N$_{\rm C}$).

Therefore, typical weak DIBs with EW\,$\sim$\,10\, m\AA /mag might be explained with a modest but significant oscillator strength f\,$\sim$\,10$^{-2}$, and a carbon fraction X$_{{\rm CM}}$\,$\sim$\,10$^{-4}$, comparable to standard abundances reported for interstellar C$_{60}$ and $\sim$4 times smaller than the average abundance  found for C$_{60}^+$ in diffuse clouds  (§2.4).

Note that for such low values of X$_{{\rm CM}}$\,$\sim$\,10$^{-4}$ (and N$_{\rm C}$=60), the abundances of endo- or heterofullerenes M remain very small, $\chi_{{\rm M}}$\,$\sim$\,10$^{-9}$.
When the fullerene compound includes a heteroatom, this abundance is also the abundance of the heteroatom locked up in M. Such low abundances are always smaller than cosmic abundances of all important elements  (Table B1). They are even smaller than depleted abundances in the cold diffuse ISM (Savage \& Sembach 1996 and Table B1) for most interesting elements such as Fe, Mg, Si, Na, K, Ca, etc.\ discussed in §3 \& §4. Therefore, the limitation for EEHFs reaching an abundance large enough to be considered as carriers of DIBs will practically always come from the total number of fullerene cages available, not from limited interstellar abundance of their associated heteroatom.

 \begin{figure}[htbp]
	 \begin{center}
 \includegraphics[scale=0.55]{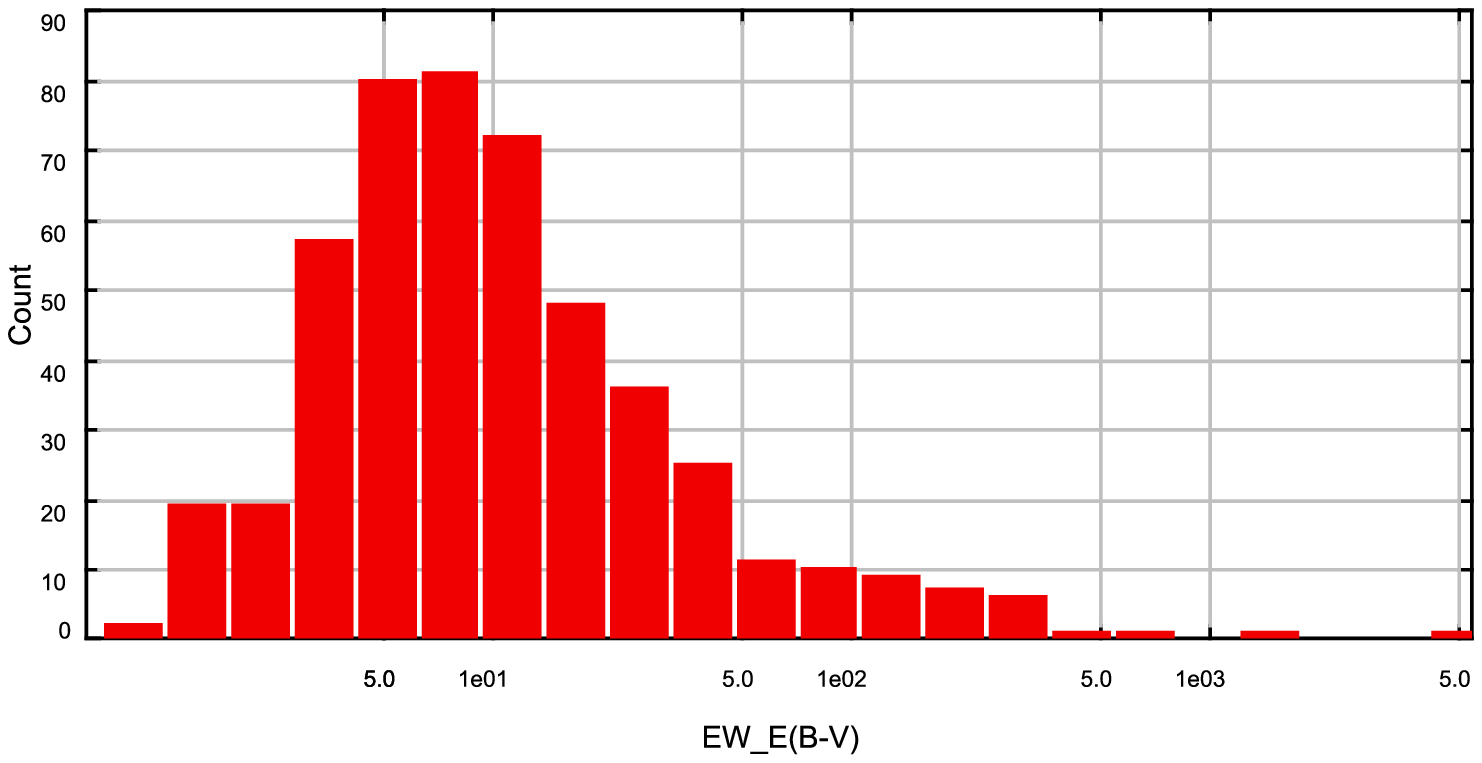}
 \caption{Distribution of the equivalent widths per unit reddening, EW/E$_{\rm B-V}$, of 489 diffuse interstellar bands observed in the line of sight of HD\,183143 with E$_{\rm B-V}$\,=\,1.27, by Hobbs et al.\ (2009).}
     \end{center}
 \end{figure}

\subsection{Which  fullerene compounds could be carriers of strong DIBs ?}
 
It is well known  that the vast majority of the $\sim$500 DIBs are relatively weak with EW/E(B-V)\,$<$\,50\,m\AA, see e.g.\ Fig.\ 6 for the best studied line of sight, HD\,183143, from Hobbs et al.\ (2009).  For this star with E(B-V)\,=\,1.27, apart from four very strong DIBs whose EW/E(B-V) spreads out between 350 and 4000\,{\rm mA/mag}, there are 22 strong bands with EW/E(B-V) between 100 and 300\,{\rm mA/mag}  and 19 slightly weaker ones between 50 and 100\,{\rm mA/mag} -- i.e. the total approaches 50 as noted by Salama et al.\ (1996). The strategy for identifying the DIB carriers should first address these strong bands, but  a proper assessment and understanding of the weak 
DIBs might also be fundamental (see §8.3). 

It is obvious from Eq.(3) that targeting strong DIBs with EW\,$\sim$\,100--300 {\rm mA/mag} would imply values of f or X$_{{\rm CM}}$ significantly larger than standard values of Eq.(3), i.e.\
\begin{equation}
{\rm X_{CM}~x~f} \approx (1-3)~{\rm x}~10^{-5}
\end{equation} 
\noindent with e.g.\ f\,$\sim$\,0.05--0.1 and X$_{{\rm CM}}$\,$\sim$\,(2--3)\,10$^{-4}$. As discussed in §3-6, it seems difficult to get oscillator strengths f that are much greater than  10$^{-2}$ for visible transitions connected to pure levels of the `C$_{60}$' cage. However, f\,$\ga$\,0.1 does not seem completely impossible for a very few  strong lines in the case of heteroatoms or adducts, but the difficulty should be to keep the carbon abundance X$_{{\rm CM}}$\,$\sim$\,10$^{-4}$ or more for such compounds (§3 \& §4).

Therefore, with pure carbon C$_{60}$-C$_{70}$ compounds, it seems impossible to meet the requirement of Eq.(3) for strong DIBs except in a few exceptional cases.
The C$_{60}^+$ DIBs at 9577 and 9633\,\AA\ are the most obvious examples. The search for other cases should focus on cations of other pure fullerenes and a few heteroatoms and adducts.  

Azafullerenes do not seem very promising  because of their lack of strong transitions and their questionable abundances. The case of fulleranes seems  more interesting, especially in molecular diffuse clouds and even perhaps as cations in typical atomic diffuse clouds (see §5.2 and Appendix D). 
As quoted in §3.1.2, alkali endofullerenes should have a few strong allowed transitions from the ground state in the optical range. 
But the possibility that these endometals might be DIB carriers relies entirely on their abundance being large enough as the result of an extremely efficient formation (Dunk et al.\ 2013). Their formation from PAH folding does not look very promising. Grain shattering could be more efficient, but the required abundances for strong DIBs seem difficult to  reach. 

Similar considerations apply to other C$_{60}$-C$_{70}$ EEHFs that are much more poorly known and more or less hypothetical, but perhaps they are more promising for a few of them.   C$_{60}$Fe$^+$ and C$_{59}$Si$^+$ should be considered first.  The optical spectrum  of C$_{59}$Si (and C$_{58}$Si$_2$) display a few significantly strong visible transitions (§4.3) and it is possible that some abundances of silafullerenes approach those of pure fullerenes.  The visible spectrum of C$_{60}$Fe$^+$ and other fullerene-Fe compounds should be richer (§3.3.1), but  the possibility that they could eventually carry strong DIBs remains highly hypothetical because of the required abundances. 

Besides the most stable fullerenes C$_{60}$-C$_{70}$, it might remain attractive to explore the field of smaller fullerene compounds (mostly cations), since folding dehydrogenated PAHs with N$_{\rm C}$\,$<$\,60 appears as the most promising way of forming interstellar fullerene cages (§2.3.3 \& §2.3.4). One should similarly explore the case of carbon chains or rings attached to fullerenes if they could prove abundant enough in the ISM, since they might combine the stability of fullerenes with the large f-values of carbon chains.

We finally recall that the 4430\,\AA\ DIB is stronger by an order of magnitude than most of the other strong DIBs that we have considered, and that the abundance-strength constraint for its carrier increases in proportion. It seems practically excluded that we can  meet this constraint by any of the single fullerene compounds we have reviewed. One  possibility with fullerenes could  be to involve a whole family of compounds if they could meet the tough constraints on wavelength, band strength, and abundance. 

\subsection{Fullerene compounds and weak DIBs}

Now that the detection of the C$_{60}^+$ strong DIBs at $\sim$9600\,\AA\ is confirmed, it is clear that a number of weaker DIBs -- with EW\,$\sim$\,5--30\,{\rm mA/mag} -- could have some C$_{60}$ compound as carrier. This first includes a few weaker C$_{60}^+$ allowed bands, as estimated in Table 6 of Bendale et al.\ (1992). Their  uncertain values yield expected equivalent widths for those bands that are smaller than the observed sum of the $\sim$9600\,\AA\ DIBs by factors 5 to 40. But the strength they computed for the $\sim$9600\,\AA\ DIBs is around five times larger than the actual strength of these bands.

Of course, first-priority efforts must further address the 3980\,\AA\ and 4024\,\AA\ broad bands of C$_{60}$, which could have been tentatively detected (see Sassara et al.\ 2001). One may also  expect 
more weak bands of other  C$_{60}$-C$_{70}$ compounds, including  C$_{70}$ and C$_{70}^+$, and 
some weak vibronic lines of  C$_{60}$ -- EW $\sim$5--10\,{\rm mA/mag} -- as addressed by Herbig (2000) and discussed in Appendix E. Similarly, more  weak forbidden vibronic bands of  C$_{60}^+$ could be present. 

Even C$_{60}^-$ and anions of other abundant fullerene compounds might be considered as possible carriers of `C$_2$' DIBs  in UV shielded clouds (§2.3.1  and Appendix F).

\subsection{Interstellar environment, DIBs and fullerenes}

Of course the properties of DIBs in some way reflect  the physical conditions and abundances of the interstellar environment where they form, which influence DIB carriers. Correlations among DIBs themselves  allow the definition of DIB families whose members behave similarly with respect to the  environment. 
It is well established that the UV interstellar radiation field 
seems to be the main factor intervening in the differentiation of the main DIB families. They are traditionally named `$\sigma$' and  `$\zeta$'  (discussed in many works:  
Krelowski \& Walker 1987; Krelowski et al.\ 1992; Krelowski 2014; and, e.g., Vos et al.\ 2011; Kos et al.\ 2013; Herbig 1993, 1995; Cami et al.\ 1997; Tielens 2013; Sonnentrucker 2014; Lan, M\'enard \& Zhu 2015; Bailey et al.\ 2015a,b) and  `C$_2$' (Thorburn et al.\ 2003; Ka{\'z}mierczak et al.\ 2010, 2014); and they 
correspond to UV non-shielded, partially-shielded and very-shielded sight lines, respectively. 

The most important effects of the radiation field on DIB carriers obviously concern their ionization state, although it may also induce photo-dissociation. The ionization (and perhaps hydrogenation) properties of various fullerenes might fit well with these characteristics. It seems especially likely that cations should be associated with  `$\sigma$' DIBs such as 5780\,\AA\ (e.g.\ Cami et al.\ 1997; Friedman et al.\ 2011; Vos et al.\ 2011; Farhang et al.\ 2015), including C$_{60}^+$.

Fullerene (or PAH) anions could be good candidates as carriers of `C$_2$' DIBs (Appendix F), but the lack of gas-phase laboratory spectra still precludes checking this conjecture.

 One  may also look for DIBs especially strong in regions with  strong UV and low density, which might perhaps be carried by dications. From the high latitude study of Lan, M\'enard \& Zhu (2015), corroborated by Ma\'is-Apell\'aniz et al.\ (2015), a possibly intriguing case seems the DIB at about 4885\,\AA . The average value given by Lan et al.\ for EW/E$_{\rm B-V}$ of this DIB is three times larger than for the classical sightline HD\,183143 (Hobbs et al.\ 2009); however, the reported band width is also much broader, which makes its strength more uncertain. In contrast, most other DIBs have a ratio of the high-latitude value to the HD\,183143 value comprised between 0.7 and 1.4. Figs.\ 13, 12 \& 8 of Lan et al.\ confirm that this 4885\,\AA\ DIB displays a strong anti-correlation with H$_2$ and a pronounced saturation at high extinction.

The extension of DIB studies to a large  number of sight lines from massive surveys at high Galactic latitude (van Loon et al.\ 2013; Kos et al. 2014; Lan, M\'enard \& Zhu 2015; Baron et al.\ 2015a,b) confirms  the general resilience of DIBs at high  latitude, amazingly even at latitudes higher than dust for DIBs such as 8120\,\AA\ (Kos et al.\ 2014). It might also show a possible peculiar behaviour of the prominent strongest 4430\,\AA\ DIB, in spite of the difficulty of accurately measuring its strength because of its large width; for example,  Lan et al.\ (2015), as well as Ma\'is-Apelll\'aniz et al.\ (2015), found an average value for EW/E$_{\rm B-V}$ of this DIB that is more than three times less than for the HD\,183143 sight line (Hobbs et al.\ 2009). This could confirm that its ratio to `$\sigma$' 5780\,\AA\ and even `$\zeta$' 5797\,\AA\ DIBs might  decrease at high Galactic latitude, as reported earlier by McIntosh \& Webster (1993), while it seems to be  a well confirmed `$\sigma$' DIB (see e.g.\ Fig.\ 13 of Lan et al.). One might thus infer a special fragility of its likely cation carriers. This could be confirmed by its early destruction in SN 2012ap (Milisavljevic et al.\ 2012). We also note  the extreme behaviour of the 5797\,\AA\ DIB in the sight line of SN 2014J in M82 (Welty et al.\ 2014).

DIB studies now extend well into the near-infrared range up to 1.8\,$\mu$m (Geballe et al.\ 2011; Zasowski et al.\ 2015; Cox et al.\ 2014a).  Correlations have already been found with visible DIBs. It would be interesting to improve these correlations  since it is possible to have a better chance there to find `perfect' correlation between two DIBs with the same carrier, one in the visible and the other in the infrared.  EEHFs should be good candidates for such correlations since many of them are expected to have strong near-infrared  transitions.

\subsection{Emission bands}

The amazing appearance in the Red Rectangle (RR) nebula of sharp emission features coinciding with four major DIBs (Schmidt et al.\ 1980) has always been recognized as a key clue for identifying of DIB carriers (see e.g.\ van Winckel et al.\ 2002; Sarre et al.\ 2006; van Winckel 2014). It is especially striking that these four DIBs (5797\AA , 5850\AA , 6379\AA\  and 6614\AA ) are exactly the prominent members of the best defined `DIB family' (`$\zeta$') whose carriers 
have a positive correlation with H$_2$ (§8.4 and e.g.\ Fig.\ 13 of Lan, M\'enard \& Zhu 2015) and  are thus enhanced in UV-shielded interstellar  `$\zeta$’ lines of
sight (see e.g. Vos et al. 2011). This might be consistent with milder UV conditions in the 
RR nebula than in the diffuse ISM, as also shown by the good correlation of the spatial emission of these bands with neutral Na (Kerr et al.\ 1999). Such conditions might be expected, but would need to be confirmed from models similar to that of Men'shchikov et al.\ (2002), yielding a lack of ionizing far-UV photons resulting from the reddened F-type stellar spectrum and the absence of an additional ionizing source, together with a  high gas density n\,$\ga10^4$\,cm$^{-3}$. But building such models would be difficult because of the complexity of the far-UV spectrum of the RR (Sitko et al.\ 2008).

The nature of the carriers of these bands stil need to be elucidated: why they are so uniquely important in the RR\footnote{The detection of a similar weak emission in the 6614\,\AA\ DIB from the interstellar medium of our Galaxy has very recently been announced by Burton Williams et al.\ (2015)} and what the emission mechanism is. Fullerene compounds may appear as good candidates, for the same arguments as given before for DIBs in general (but PAHs and other carbonaceous particles cannot be excluded). However, the origin of their abundance there should not be obvious. The peculiar abundances of the RR photosphere, devoid of refractory elements (Waelkens et al.\ 1992 and Table B.1), could point to the absence of these elements (such as Fe, Ca, Mg, etc.) in these emission DIB carriers. 

The most promising emission mechanism of these RR bands could be some fluorescence process as advocated by Witt (2014; and references therein). However, the fluorescence yield must be high, which implies tight conditions for the internal energy conversion. Either the spontaneous emission from the upper  state U of the DIB to the ground state of the same multiplicity  
(typical lifetime $\sim$\,10$^{-6}$s for f\,$\sim$\,a few 10$^{-2}$) is faster than internal conversion to other metastable states M, 
which finally leads to vibrational excitation and IR emission. Or  the energy may be returned from M to U through thermodynamic equilibrium, if the difference in energy between U and M is small enough (delayed or Poincar\'e fluorescence (PF), L\'eger et al.\ 1998a, see Appendix G). As pointed out by Witt (2014), such an efficient delayed fluorescence should be easier to achieve with small particles such as C$_{20}$. But it does not seem impossible for fullerenes if their level structure and internal conversion could meet these conditions (§2.1.1). As discussed in Appendix G, PF appears specially favourable for  C$_{70}$ compounds; it could also operate  for other cages, even perhaps for  C$_{60}$ compounds. 
The case of peculiar fullerene orbitals, such as those of endohedral atoms, adducts, or SAMOs, should be specially addressed. Since, the interaction of these orbitals with the bath of regular cage orbitals should be much weaker than for cage orbitals, one may expect that internal conversion is slowed down.

\subsection{Matching band profiles and other DIB properties?}

Following Snow (2014), among others, DIB carriers must comply with an impressive list of requirements (see §1). As regards fullerene compounds as candidates, we have already addressed a number of these requirements, such as size, stability, carbonaceous nature, ionization states, interstellar formation, required abundance, and line strengths. 
It is easily seen that additional properties  shared with PAHs, such as strong dependence on metallicity or absence of polarization (see e.g.\ Cox et al.\ 2007, 2011b), are also easily met by fullerene compounds.

Key information about DIB carriers is also clearly embodied in their wavelength distribution and band profiles (e.g.\ Sarre 2014). The wavelength domain where DIBs show up has always been recognized as highly characteristic (see e.g.\ Fig.\ 1 of Witt 2014) - concentration in the visible range, and more specifically high density in the range 5500-6500\,\AA , low density at $\sim$4000-5000\,\AA\ and  absence in the UV (but see Bhatt \& Cami 2015).
As for PAHs (and other carbonaceous particles), such a distribution is well accounted for if the DIBs correspond to the first allowed transitions of fullerenes between the highly symmetric ground state and the first singlet (or doublet for ions) excited state accessible through an electric dipole transition. 

Such a situation also fits  well with the widths observed for DIBs. Their full width at half maximum (FWHM) ranges from $\sim$\,0.5 to 2\,\AA\ ($\sim$\,1-5\,cm$^{-1}$) for most of them, with an extension to  $\sim$\,4 to 20\,\AA\ ($\sim$\,10-100\,cm$^{-1}$) for the broadest ones ($\sim$5\% of them at the most; see e.g.\ Hobbs et al.\ 2009; but see York et al.\ 2014 for confirmed ones).  The short lifetime of the excited state and rotational broadening have long been proposed as the main causes susceptible to explain such line widths. 

The Lorentz width connected to a lifetime $\tau$ may be written
\begin{equation}
 {\rm FWHM(cm^{-1})\,=\,1/2\pi c\tau = 5.3/\tau(ps)}
\end{equation}
The width of the broadest DIBs ($\sim$\,10-100\,cm$^{-1}$) might thus be explained by ultrafast sub-picosecond internal conversion (IC) of fullerenes (§2.1.1) or other carbonaceous particles such as PAHs (see e.g. Marciniak et al.\ 2015, Reddy et al.\ 2010). The 4430\,\AA\ DIB, however, seems the only one strong enough to allow the verification that the profile actually seems Lorentzian, as shown for this band by Snow et al.\ (2002). The width they report, 17\,\AA\ ($\sim$90\,cm$^{-1}$) yields $\tau$\,$\sim$\,0.06\,ps. 
Although very short, this is still within  the expected order of magnitude  for the IC lifetime of the main singlet levels of fullerenes such as C$_{60}$ (e.g.\ Stepanov et al.\ 2002, see §2.1.1 and Fig.\ 2). 

The rare confirmed very broad DIBs (York et al.\ 2014) 
might be explained in this way.  (Weak broad DIBs could be much more numerous but are not detectable, and other large molecules, such as PAHs, have similar IC features).  
The large width of UV bands, close to 10\,\AA\ (or more, Sassara et al.\ 2001) on top of a crowded absorption background, could be the main reason for the quasi absence of strong UV DIBs (Bhatt \& Cami 2015).

On the other hand, one has to explain why the majority of DIBs are too narrow to allow internal conversion times faster than $\sim$1\,ps. This might point  to more peculiar transitions that are not as well coupled to the main phonon bath of large molecules, i.e.\ to cage vibrations for fullerene compounds. Such a slower IC might be expected with the lowest excited levels of most fullerene compounds or with SAMO states or with transitions of EEHFs connected with their heteroatoms (§3 \& §4) or adducts.

The 9577/9633\,\AA\  bands  give tight constraints on IC in C$_{60}^+$. Their intrinsic FWHM, 2.4-2.7$\pm$0.2\,cm$^{-1}$, measured at 5.8\,K in the laboratory by Campbell et al.\ (2015), is interpreted as mainly due to internal conversion, indicating a lifetime of $\sim$2\,ps for the excited electronic state. The slightly larger width measured for the DIBs (3.05$\pm$0.05\,cm$^{-1}$ towards HD183143, Foing \& Ehrenfreund 1997) might include a small contribution from rotational broadening, but smaller than expected from the modelling of Edwards \& Leach (1993) and Foing \& Ehrenfreund (1997). Therefore, this result seems to point out to low rotational excitation in diffuse clouds, perhaps similar to the conclusions by Cami et al.\ (2004) for the 6614\,\AA\ DIB. 

The width of C$_{60}^+$ DIBs is close to the median (in cm$^{-1}$) of the distribution of DIB widths. These C$_{60}^+$ data  strengthen previous conclusions that there is no problem for explaining DIB widths with a combination of IC and rotational broadening for fullerene carriers.   

On the other hand, the most extreme broadening observed for some DIBs especially in the line of sight of He-36 (Dahlstrom et al.\ 2013) seems to require larger rotation constants that implies molecules smaller than $\sim$20 atoms (Oka et al.\ 2013). This seems to exclude fullerenes as carriers of these bands, unless perhaps one considers the very smallest cages or some other peculiar fine structure of fullerene lines. 

A few DIBs, such as 6614\,\AA , display a clear multicomponent structure that could contain crucial information on their carrier (see e.g.\ Sarre 2014 and references therein). The case of  the 6614\,\AA\ DIB  analyzed by Cami et al.\ (2004), might be hardly compatible with spherical tops such as fullerenes. Various  possible reasons for the structure
 of this DIB have been recently discussed by Marshall et al.\ (2015).

\section{Conclusion}

Despite the difficulties assessing the actual importance  of fullerene compounds, one may  confirm that they display  an impressive panel of  attractive features as DIB carriers.  Many of these compounds should be practically as stable in interstellar conditions as pure fullerenes such as C$_{60}$. They therefore stand  among the most stable interstellar compounds. 
Easy interstellar formation of cages of various sizes from dehydrogenated PAHs seems now likely, while other formation modes, such as grain shattering,  are possible. 
The properties of their excited electronic levels and fast internal conversion may explain the gross features of the DIB wavelength distribution. When combining a variety of all factors, their rich diversity of forms compare well with the number of weak DIBs. 
Compared to PAHs, their high symmetry, especially for  C$_{60}$ compounds, the well-defined hierarchy of cage and isomer stability, and the diversity of sizes, adducts, and heteroatom abundances and properties could be able to produce  the observed pattern of DIB strengths. 
The comparison of interstellar and laboratory gas-phase widths of 9600\,\AA\ bands of C$_{60}^+$ strengthens the propositions that other observed DIB widths and profiles might be eventually accounted for by the combination of fast internal conversion, rotation levels and fine Jahn-Teller structure of fullerene compounds. But it is possible that the extreme and variable broadening observed for some DIBs in peculiar sight lines, such as He-36, be incompatible with fullerene carriers.

As for C$_{60}$, fullerene compounds should exist in various ionization states, cations being dominant in standard diffuse interstellar clouds. These ionization properties, as well as those of PAHs, might eventually nicely explain the behaviour of the different DIB `families' in various UV environments. But this is still unclear. The same might be true for the strong emission DIBs  of the Red Rectangle (which are major members  of the `$\zeta$' low-UV family of classical absorption DIBs). 
One might imagine that this emission might be explained by some special properties of internal conversion and fluorescence such as Poincar\'e fluorescence.

At any rate, two major drawbacks affect fullerene compounds as DIB candidates. To start with, many varieties completely lack precise laboratory data, especially gas-phase visible spectra, which are mandatory for carrier identification. More fundamentally, practically all known cases seem to fail at the requirement for strong DIBs in the product of the oscillator strength and the carbon fraction in the DIB carrier, f$\times$X$_{\rm CM}>10^{-5}$ (Eq.3). Visible transitions with f larger than a few 10$^{-2}$ seem extremely rare in all fullerene compounds with known spectra because of selection rules and screening by cage $\pi$ electrons. Fullerenes  are less abundant than PAHs by one or two orders of magnitude, holding probably less than a few 10$^{-3}$ of interstellar carbon in total. It thus seems unlikely that the abundance of many individual fullerene molecules may exceed X$_{\rm CM}=10^{-4}$  of interstellar carbon. Identification of strong DIB carriers with fullerenes should thus eventually rely on exceptional cases, which remain mostly hypothetical.

The recently confirmed near-IR double DIB of  C$_{60}^+$ at 9577/9633\,\AA\  might be considered as the first outstanding example. Despite their peculiar wavelength range, these DIBs display all the standard properties of the `$\sigma$' DIB family, strengthening the proposition that the `$\sigma$' carriers should be cations. The substantial, but modest, abundance of C$_{60}^+$,  $4\pm2 \times 10^{-4}$ of interstellar carbon, provides a good scale for the abundances of other fullerene compounds, since there are good arguments that they could only be smaller than or  comparable to that of  C$_{60}^+$  at best. 

However, even if eventually other fullerene compounds reached high abundances, X$_{\rm CM}\sim\,(1-3)$\,10$^{-4}$, this would leave serious difficulties for allowing them to be the carriers of twenty or so strongest DIBs (§8.2), since it implies f\,$\ga$\,0.05-0.1 for these DIBs (Eq.3). Such f-values are practically absent in our survey of important fullerene compounds with known or estimated visual spectra (except perhaps a few endohedral metallofullerenes, and carbon-chain adducts). However,  the possible existence of transitions approaching these values might not be totally excluded  in  fullerene and fullerane cations with unknown spectra, such as  C$_{70}^+$, C$_{50}^+$, 
silafullerenes, C$_{60}$Fe$^+$ 
 and smaller analogues.  Azafullerene cations, as well as other very stable fullerene compounds of transition metals or Mg, if any, might also be tentatively considered for weaker DIBs. But it is unlikely that the abundances of these more or less hypothetical compounds may approach that of  C$_{60}^+$.

On the other hand, now that the detection of the 9600\,\AA\ C$_{60}^+$ strong DIBs is confirmed, it is clear that a number of {\it weaker DIBs} -- with EW\,$\sim$\,5--30\,{\rm mA/mag} -- could have some fullerene compound as carrier. This should  include: 
1) A few weaker  visible allowed bands of C$_{60}^+$; 
2) Some of the many identified weak forbidden vibronic lines of  C$_{60}$ that should give DIBs with  EW\,$\sim$5--10\,{\rm mA/mag},  as well as  weak vibronic bands of  C$_{60}^+$ to be identified;  
3) In the same strength range, similar bands of  various other fullerene compounds (see §8.3), 
but their number could be limited and remains very uncertain because the abundances and many f-values are unknown.  
4) Even C$_{60}^-$ and other anions, as possible carriers of `C$_2$' DIBs.
However, the crowding of such weak DIBs and the complexity of matrix distortion of fullerene spectra make gas-phase spectroscopy mandatory for  identifying these fullerene DIBs. 
A few fullerene dications could also be considered (§8.4). 

To summarize, the confirmation of the 9600\,\AA\ C$_{60}^+$ strong DIBs strengthens the importance of interstellar fullerenes in two main ways: 

1) The substantial abundance of C$_{60}^+$  in diffuse clouds shows that fullerenes are significant compounds of the interstellar medium (but much less abundant than PAHs). From this abundance, one may expect that a number of other fullerenes (with different  sizes and various atoms in hetero, endohedral or exohedral compounds) might eventually prove significantly abundant. But identifying their importance and estimating their abundance remain awfully difficult because of the complexity of their physics and of interstellar processes and of the lack of spectroscopic data for their identification. 

2) This confirmation of the first DIB identification also strengthens the attraction of various fullerene compounds as DIB carrier candidates. One might expect that many weak fullerene DIBs will be identified as soon as sensitive gas-phase spectroscopy is available for major fullerene compounds, such as  C$_{60}$ and  C$_{70}$ and their ions. But the possibility that other fullerene compounds are the carriers of other {\ strongest} DIBs remains subject to very constraining conditions about their abundance, which should approach that of C$_{60}^+$, and large strengths of their visible bands, which seem rare among known compounds. Fullerenes remain key DIB candidates, but it is also possible that other carriers, such as PAHs or carbon chains, ares dominant. 

Such an uncertainty about the status of fullerenes as DIB carriers, even after the confirmation of C$_{60}^+$ DIBs, may seem frustrating. However, it is just an illustration of the
enormous complexity of the issue of the DIB carriers and the fact that solving this problem will require long-term collective organized efforts and the close cooperation of astronomers, molecular physicists, astrochemists, and spectroscopists, as mentioned by Tielens (2014) and Cox \& Cami (2014), among others.

Identification of DIBs, even if weak, appears the best way to trace interstellar fullerene compounds and their abundances. More generally, the importance of interstellar fullerenes obviously justifies major efforts for improving the knowledge of their properties that are relevant to interstellar conditions. Refined quantum calculations of various compounds remain essential for disentangling complex structures and their energy levels and tracing possible strong  transitions in the visible or near-infrared range. They should be extended to simulate the formation, destruction, and evolution of these compounds  in interstellar conditions. However, once key compounds and transitions have been identified, the most crucial step is accurate laboratory spectroscopy. Matrix spectroscopy may remain useful, but gas-phase spectroscopy appears to be  ultimately mandatory. The tour-de-force experiment of Campbell et al.\ (2015) shows the level of experimental complexity required. Although it should be difficult to extend to neutral fullerenes, it should open the way to systematic studies of the most important ionized interstellar fullerene compounds.

\bigskip

\begin{acknowledgements}

I thank Christine Joblin and Olivier Bern\'e for numerous comments and discussions and for sharing insights and unpublished work.

I am very deeply indebted to Sydney Leach for many discussions and for sharing his interest and insights in fullerene physics, for careful reading and improving the manuscript, and for also sharing unpublished work.

I am  indebted to Matt Lehnert for his detailed, careful reading and for improving the manuscript. 

I especially thank Dmitry Strelnikov for his careful reading of the manuscript, important comments and suggestions, and for providing me in advance with their unpublished data on C$_{60}^+$ spectroscopy.

I would like to thank Giacomo Mulas, Elisabetta Micelottta, Roland Gredel, Fernand Spiegelman, Aude Simon, Nick Cox, Jose Cernicharo, Jeronimo Bernard-Salas, Steven Federman, Patrick Boiss\'e, Françoise Remacle, Benoit Mignolet, Michel Broyer, Thomas Pino, Adolf Witt, Brice M\'enard, Ting-wen Lan, Chao Liu and Xiaowei Liu for many discussions and comments on various aspects of the manusript, as well as Chentao Yang for his nice help.  

I especially thank Madeleine Roux-Merlin for her precious help in finding many references.

I thank Joli Adams for her many and detailed suggestions for correcting the english.

I thank  the anonymous referee very much for the very helpful suggestions for shortening and improving the manuscript.

\end{acknowledgements}

\bigskip

\appendix

\section{Interstellar accretion of  C$^+$ to C$_{60}$}

From their measurement of the C$^+$ + C$_{60}$
$\rightarrow$ C$_{61}^+$ accretion process, Christian et al.\ (1992c) suggested
that there is no activation energy, though they cannot exclude a small
($<$\,0.5\,eV) barrier. Indeed, a reaction barrier seems unlikely since, e.g., no reaction barrier was found by Strelnikov \& Kr\"{a}tschmer (2010) for C$_3$ reacting with C$_{60}$, and C and C$^+$  
should be even more reactive than C$_3$. However, as quoted by Bern\'e et al.\ (2015a), the actual
rate of C$^+$ accretion, k$_{\rm C+}$, is apparently unknown. A possible guess is that it equals a
 significant fraction $\eta$ of the Langevin rate k$_{\rm L}$,
with perhaps $\eta$ at least $\sim$\,0.1. 
 For a reduced mass m (reduced atomic mass $\mu$) and a particle of polarizability $\alpha$, k$_{\rm L}$\,=\,2$\pi$e\,($\alpha$/m)$^{0.5}$\,=\,2.34\,x\,10$^{-9}$\,[$\alpha$(\AA$^3)$/$\mu$]$^{0.5}$. Therefore, for C$^+$ and the polarizability of C$_{60}$, $\alpha$$\approx$80x\AA$^3$, k$_{\rm L}$\,=\,6\,x\,10$^{-9}$\,cm$^3$s$^{-1}$. This yields, for the rate of C$^+$ accretion, 
$\tau^{-1}_{\rm C+}$={\rm k$_{C+}n_{C+}$},

 \begin{eqnarray}
 {\rm \tau^{-1}_{C+} \approx (\eta/0.1)~(n_H/50)~(N_C/60)~x~0.4~10^{-11}~s^{-1}}
   \end{eqnarray}

Although fullerenes with unpaired numbers of C are much less stable than C$_{\rm 2n}$, 
the binding energy of C$_{60}$-C remains substantial, possibly  $\sim$4\,eV,  as Slanina \& Lee 1994  found using the AM1 quantum-chemical method, or $\sim$4.9\,eV found from DFT by Pellarin et al.\ (2002). It could even be smaller ($\sim$3\,eV, Mulas, priv.\ comm.) or  larger (Strelnikov, priv.\ comm.). A low value, $\sim$3\,eV,  is consistent with the finding by Christian et al.\ (1992c) that the formation of C$_{61}^+$ in collisions of C$^+$ with C$_{60}$ is efficient up to a collision energy approaching 10\,eV, but that C$_{61}^+$ is negated  when it exceeds 10\,eV.
 
Such a mild  binding should  make   C$_{60}$-C$^+$ (and perhaps C$_{60}$-C) relatively easy to destroy by  UV photolysis as in  the case of smaller-PAH hydrogenation, if their internal conversion of energy is comparable to PAHs. 
However, it is not clear whether this photolysis should  prevent some cage or adduct growth by further C$^+$ accretion, or reactions of C$_{60}$-C$^+$ with H (§2.3.4).

Molecules, such as C$_{60}$-C$^+$, C$_{60}$-C and their possible derivatives, should be much more stable if  C$_{60}$ is bound to a PAH (§6.2).

\section{Abundances of interstellar elements}

See Table B.1.

  \begin{table}
      \caption[]{Abundances for elements of interest for endohedral fullerenes and heterofullerenes.}
         \label{tab:lines}
            \begin{tabular}{l c c c c c c}
            \hline
            \noalign{\smallskip}
Z       & Elem & Sol$^a$ & IS$^a$ & RedRec$^b$ &  Ion.p$^d$ & Endo$^c$ \\
            \noalign{\smallskip}
            \hline 
            \noalign{\smallskip}
  01    &  H   & 0    &  0    &   0 & 13.6   &  no \\
  02    &  He  & -1.1 & -1.1  &   -1.01   & 24.6   & yes  \\
  03    &  Li  & -8.7 & -10.3 &      & 5.4   & yes   \\
  06    &  C   & -3.5 & -3.8  &   -3.34   & 11.2    & no?  \\
  07    &  N   & -4.1 & -4.1  &   -4.05  & 14.5   &  yes   \\
  08    &  O   & -3.2 & -3.5  &    -3.26 &   13.6   & no   \\
  10    &  Ne  & -4.1 & -4.1  &      & 21.6     & yes   \\
  11    &  Na  & -5.7 & -6.6  &      & 5.1    &  yes  \\
  12    &  Mg  & -4.4 & -6.0  &   -6.49   & 7.6    & no  \\
  13    &  Al  & -5.6 & -7.9  &      & 6.0    & no  \\
  14    &  Si  & -4.4 & -5.8  & ($^e$)  & 8.2    & no  \\
  15    &  P   & -6.4 & -6.9  &      &  10.5   &  yes  \\
  16    &  S   & -4.7 &  4.6  & -5.0  & 10.4   &  no    \\
  18    &  Ar  & -5.2 & -5.2  &      & 15.8   &  yes   \\
  19    &  K   & -6.9 & -8.0  &      & 4.3   &  yes  \\
  20    &  Ca  & -5.7 & -9.4  &-8.75 & 6.1 & yes   \\
  21    &  Sc  & -8.6 &  -8.5 &    & 6.6   &  yes   \\
  22    &  Ti  & -7.0 & -10.1 &      & 6.8   &  yes    \\
  23    &  V   & -8.0 & -9.9  &      & 6.7   &  no  \\
  24    &  Cr  & -6.3 & -8.6  &      & 6.8   & no   \\
  25    &  Mn  & -6.4 & -7.9  &      & 7.4   & no   \\
  26    &  Fe  & -4.5 & -6.8  &  -7.82  & 7.9   & no   \\
  27    &  Co  & -7.1 & -9.9  &      & 7.9   & no   \\
  28    &  Ni  & -5.7 & -8.5  &      & 7.6   & no   \\
  29    &  Cu  & -7.7  &  -8.4  &      & 7.7   & no   \\
  30    &  Zn  & -7.3 & -8.0  &      & 9.4    &  no \\
            \noalign{\smallskip}
            \hline
           \end{tabular}
\begin{list}{}{}
\item[] $^a$Solar and typical cool gas-phase interstellar abundances, mostly from Jenkins et al.\ (2009) and  Savage \& Sembach (1996, Table 5, $\zeta$ Oph Cool), respectively.
\item[] $^b$Photospheric abundances in the Red Rectangle from Waelkens et al.\ (1992, Table 2, Model 1)
\item[] $^c$Confirmed endohedral compound (Dunk et al.\ 2014; Cong et al.\ 2013; Fig.\ 5)
\item[] $^d$First ionization potential (eV)
\item[] $^e$The  abundance of Si in the Red Rectangle seems unknown, but the total gas abundance of Si in the outer layers of the reference carbon circumstellar envelope IRC+10216 is -5.7  (Ag{\'u}ndez et al.\ 2012)

\end{list}
\end{table}

\section{Energy levels of C$_{60}$ endo alkalis and alkaline earths}

A complete computation of
the first excited levels of K@C$_{60}$ and Ca@C$_{60}$ and their cations was performed by Chang et al.\
(1991). It should provide reasonable patterns of the level structure, but,
as noted by Broklawik \& Eilmes (1998), such Hartree-Fock calculations,
which neglect electron correlations, greatly underestimate ionization
potentials (often by $\sim$2\,eV). Similarly, they cannot be trusted for
the energy of excited states either. Nevertheless, from the values
computed by Chang et al., one may infer that each EF among K@C$_{60}$,
Ca@C$_{60}$, and Ca@C$_{60}^+$ should have at least one
allowed absorption transition in the visible or near-infrared range.

It is important to compute the oscillator strengths of these
transitions. Most of them are between cage levels. The strength of
corresponding transitions of C$_{60}^-$, which are  weak (Appendix F), could probably
provide a guess of their order of magnitude for the first transitions close to 1\,$\mu$m, but this is questionable at shorter wavelengths. For the transitions related
to the atomic levels, one should remember that the fullerene cage is
a practically perfect Faraday cage for its interior (e.g.\ Connerade \& Solovyov
2005), which may considerably reduce their line strengths.

As quoted, the impossibility of charge transfer may explain the difficulty of forming Mg@C$_{60}$, contrary to Ca@C$_{60}$. 
Nevertheless, the case of Mg@C$_{60}$ has been taken as an example by several theoretical studies (e.g.\ Broclawik \& Eiles 1998; Stener et al.\ 2002; Lyras \& Bachau 2005). Indeed the possibility  that it could form in the peculiar interstellar conditions cannot be excluded. The results of these studies seem to show that  Mg@C$_{60}$ might display several transitions  in the visible, including between atomic levels. However, it is probable that the latter are substantially quenched by cage screening.

\section{Interstellar dehydrogenation of fulleranes}

The leading processes determining the abundance ratio of interstellar fulleranes to their parent fullerene (e.g.\ C$_{60}$) obviously are hydrogen photolysis by typical interstellar UV field\footnote{From e.g. Fig.\ 21 of Vos et al.\ (2011),  it is seen that most DIBs form in diffuse clouds with UV energy density 
$\sim$1-10\,G$_0$, where G$_0$\,=\,1\,habing\,=\,5.29 $\times$ 10$^{-14}$erg\,cm$^{-3}$\,=\,0.59 `Draine field'.} and the reactions  listed in Table 1 with hydrogen or the reverse reactions (Bettinger et al.\ 2002), or rather the reactions of H with the same fullerene compounds but ionized. 

The rate of formation of C$_{60}$H$^+$ from C$_{60}^+$  through  the reaction C$_{60}^+$ + H exceeds  5x10$^{-9}$\{n(H)/50\,cm$^{-3}$\}s$^{-1}$ (Petrie et al.\ 1995). Due to its small activation energy, $\la$2\,eV (Table 1), C$_{60}$H should be dissociated in practically every absorption of a hard UV photon (see e.g.\ Fig.\ 10 of Le Page et al.\ (2001) for PAHH$^+$ with similar H-binding energy), i.e.\  with a rate $\sim$1\,yr$^{-1}$\,=\,3$\times$10$^{-8}$s$^{-1}$ (see e.g. Tielens 2005). Assuming that the abundance ratios C$_{60}$H/C$_{60}$ or C$_{60}$H$^+$/C$_{60}^+$ roughly equal the ratio of these two rates, both ratios  might thus be $\sim$0.1-0.2, with a large uncertainty. 

The case  of C$_{60}$H$_2$ and other even-H fulleranes seems  more uncertain since their photo-dissociation is much more difficult because of their higher H binding and activation energies, $\ga$3.3\,eV (Table 1). Because of the high rate expected for the accretion of H onto C$_{60}$H$^+$ (and even onto C$_{60}$H), two-photon dissociation processes may be neglected (§2.3.4), and the photo-dissociation must be achieved by the absorption of a single hard UV-photon. As for PAHs, an elaborate modelling is needed to estimate the photo-dissociation rate of C$_{60}$H$_2$ rate even qualitatively. 

However, one could try to explore this photolysis by scaling from the modelling of PAH dehydrogenation in diffuse clouds (e.g.\ Le Page et al.\ 2001, 2003; Bern\'e and Tielens 2012; Montillaud et al.\ 2013 and references therein). For PAHs in diffuse clouds, all authors agree that the limit for full dehydrogenation lies for a number of C atom N$_{\rm C}$\,$\sim$\,30-40, but with a substantial uncertainty of several orders of magnitude for the average photo-dissociation rate (see e.g.\ Fig.\ 3 of Montillaud et al.\ 2013); for example, Fig.\ 10 of Le Page et al.\ (2001) shows that in their model the dehydrogenation limit is N$_{\rm C}$\,$\approx$\,30 for  PAHs (activation energy E$_{\rm ac}$\,=\,4.8\,eV) and N$_{\rm C}$\,$\approx$\,80 for a PAHH$^+$ with E$_{\rm ac}$\,=\,2.9\,eV.  It is thus expected that this limit for E$_{\rm ac}$\,=\,3.3\,eV lies just about N$_{\rm C}$\,$\approx$\,60. Even if other models could push this limit above N$_{\rm C}$\,=\,60, several additional specific factors may contribute to make this dehydrogenation limit even more uncertain for fulleranes than for PAHs. The internal energy conversion may indeed proceed differently. Several key rates and activation energies are unknown for the reactions of Table 1 and for their reverses, implying  various neutral and ionized fullerene compounds. 
The additional  destruction of C$_{60}$H$_2$ in the reaction  C$_{60}$H$_2$ + H  $\rightarrow$ C$_{60}$H + H$_2$  could be non-negligible in some circumstances despite the non-zero activation energy ($\sim$0.1\,eV).
 
We also stress that the degree of hydrogenation of interstellar fullerenes should critically depend on the UV intensity, the density of atomic hydrogen, and the size of the fullerene cage. The photo-dissociation rates are also expected to significantly depend on the number of H atoms of the fullerane for a given N$_{\rm C}$. 
Therefore, although an efficient dehydrogenation of C$_{60}$-fulleranes in diffuse clouds does not appear to be  impossible, the question should be considered as unsettled until an elaborate treatment is provided.

However,  the likelihood of finding fulleranes should be increased in UV-shielded molecular diffuse clouds, as well as in fullerenes attached to PAHs or carbon grains.

\section{DIBs and forbidden transitions of C$_{60}$}

Surprisingly,  the richest optical spectra of fullerenes have been obtained in the laboratory for forbidden transitions of  C$_{60}$ that cover a good part of the visible range.
Various techniques have been used, including  absorption spectroscopy in cold molecular beams (Haufler et al.\ 1991; Hansen et al.\ 1997) or He droplets (Close et al.\ 1997), combined absorption and fluorescence spectroscopy in rare gas matrices (Sassara et al.\ 1996a, 1997; Chergui 2000), phosphorescence (Sassara et al.\ 1996b),  fluorescence and phosphorescence from individual C$_{60}$ molecules excited by local electron tunneling (Cavar et al.\ 2005).

For identifying weak DIBs, the most interesting laboratory data are obviously absorption spectroscopy of Haufler et al.\ (1991),  Hansen et al.\ (1997),  Close et al.\ (1997), and Sassara et al.\ (1996a).
 Herzberg-Teller coupling with vibration states allows weak absorption in the forbidden transitions from the ground state  h$_u^{10}$ $^1$A$_g$ to the first singlet excited states h$_u^{9}$  t$_{1u}$ $^1$T$_{1g}$,  $^1$T$_{2g}$, and $^1$G$_{g}$.  However,  the very rich vibronic structure is difficult to disentangle because these electronic excited levels are almost completely degenerate within $\sim$100\,cm$^{-1}$ ($\sim$0.01\,eV) and eventually mixed by the coupling with the rare-gas matrix (Sassara at al.\ 1997; Chergui 2000; Orlandi \& Negri 2002; Cavar et al.\ 2005). Significant f-values have been computed by Sassara et al.\ (1997) for these transitions. Half a dozen of them approach 0.01, and a similar number lie in the range 0.001-0.003. 

As the abundance of  C$_{60}$, X$_{\rm C60}$, is expected to be a factor of a few units lower than that of  C$_{60}^+$ in diffuse clouds (§2.4), it should be close to 10$^{-4}$ of interstellar carbon. Therefore, the expected equivalent width of the corresponding DIBs (Eq.5) may approach $\sim$10\,m\AA/mag for the strongest ones, while others, which are two or three times weaker, should be barely detectable.

Of course, securely identifying weak DIBs at this level, close to the confusion limit, is a real challenge, which requires perfect spectroscopic data. But, despite the beautiful work performed in the past 25 years for the sophisticated spectroscopy of visible forbidden bands of  C$_{60}$, it seems that it has not yet fully reached the demanding requirements for definitely identifying them as DIBs, for various reasons. The most detailed and complete work was carried out by Sassara et al. (1996a, 1997) for their measurement in absorption and fluorescence, while further vibronic assignments and strength theoretical estimates are discussed by Orlandi \& Negri (2002). But these results are plagued by intricate matrix  effects. The case of  the strong DIBs of C$_{60}^+$ has shown how such effects may be difficult to predict (Campbell et al.\ 2015; Strelnikov et al.\ 2015). The situation is expected to be even more complicated for the first excited levels of C$_{60}$ because of the quasi-degeneracy of these levels, $^1$T$_{1g}$,  $^1$T$_{2g}$, and $^1$G$_{g}$. Checks with gas phase data are thus mandatory. But they are available only for a limited wavelength range, 5900-6300\AA\ (Haufler et al.\ 1991; Hansen et al.\ 1997) (as well as the He droplet data of Close et al.\ 1997).  

On this basis, Herbig (2000) was nevertheless able to tentatively identify a possible match of the strongest C$_{60}$ feature of Haufler et al.\ with a well-confirmed DIB at 6071.2\,\AA . 
However, as quoted by Herbig, this identification cannot be considered as secure, especially for such a weak DIB (EW\,=\,6.9\,m\AA /mag in HD183143, Hobbs et al.\ 2009). In addition, both Close et al.\ (1997) and Hansen et al.\ (1997) have said that there could be a problem in the wavelength recording of the preliminary spectrum shown in their Fig.\ 2 by Haufler et al. The main problem seems to be in the blue part of the spectrum. However, even for the central, strongest feature close to 6070\,\AA , Hansen et al.\ report a wavelength blue-shifted by $\sim$2\,\AA . Such a shift is enough to jeopardize identification with the DIB  at 6071.7\,\AA .  But it could make the association plausible with another nearby stronger, broad DIB at 6068.4\,\AA\ (EW\,=\,13\,m\AA /mag in HD183143). It is difficult to say which association, if any, has the best chance of being real. However, both DIBs would yield about 10$^{-4}$ of interstellar carbon for the order of magnitude of the C$_{60}$ abundance, if the f-value approaches 10$^{-2}$. 

To summarize, current molecular-beam data of C$_{60}$ are compatible with at least a match of the strongest band at  6070-6071\,\AA\ with a confirmed DIB of $\sim$10\,m\AA/mag. 
This would yield an abundance of C$_{60}$ $\sim$10$^{-4}$ of interstellar carbon in the sight line of HD183143, which seems consistent with the abundance of $\sim$5x10$^{-4}$ for C$_{60}$ (§ 2.4). But it is obvious that such a detection of a single band of C$_{60}$ cannot be considered as secure by any means.

\section{Could C$_{60}^-$ be carrier of `C$_2$' DIBs?}

As quoted in §2.3.1, the C$_{60}^-$ anion may include a significant percentage of C$_{60}$ cages in UV-shielded regions. It might thus be a good candidate for weak DIBs in such sight lines, such as the so-called `C$_2$' DIBs. Indeed, two of the three excited electronic energy levels of  C$_{60}^-$, $^2$T$_{1g}$ and $^2$A$_g$, are connected  to the ground level $^2$T$_{1u}$ by `allowed' optical transitions (Klaiman et al.\ 2013, 2014).

The first corresponds to a well-documented band in the very near infrared close to 1\,$\mu$m (Kato et al.\ 1991; Fulara et al.\ 1993a; Bolskar et al.\ 1995; Tomita et al.\ 2005; Hands et al.\ 2008; Strelnikov et al.\ 2015; and references therein). While most of these experiments were performed with C$_{60}^-$ in Ne matrix or in solution, the 300\,K storage ring spectrum of Tomita et al.\ provides an accurate value of the wavelength of the main band at 10659\,\AA .  Its strength is very comparable to the nearby band of C$_{60}^+$ with an oscillator strength of 0.022$\pm$0.005 (Strelnikov et al.\ 2015). This wavelength does not correspond to any of the NIR DIBs known in this wavelength range (Hamano et al.\ 2015, Cox et al.\ 2014a). Indeed, most of the observed sight lines have `$\sigma$' characteristics with  strong UV radiation where C$_{60}^-$ is not expected to be present. But no hint of absorption is  even seen  in the very reddened `$\zeta$' line of sight of 4U\,1907+19 with the very good sensitivity  of  Cox et al.\ (see their Fig.\ 6). One may infer the rough limit EW\,$<$\,10\,m\AA/mag, and Eq.(2) yields X$_{\rm C}$(C$_{60}^-$)\,$<$\,0.5 10$^{-4}$ in this sight line, and significantly lower values in other ($\sigma$-type) sight  lines, such as HD\,183143.

The other allowed transition, $^2$A$_g$-$^2$T$_{1u}$, of C$_{60}^-$ is expected in a very different (blue) spectral range. While the three other electronic levels of C$_{60}^-$ may be interpreted as the adjunction of one electron on the cage in one of the available first excited states of C$_{60}$, the A$_g$ level is very different. The wave 
function of the additional electron is very diffuse with a large extension far outside the central C$_{60}$ core. According to various estimates (Klaiman et al.\ 2013, 2014; Voora et al.\ 2013), its energy is very close to but  below the electron detachment limit (within $\sim$100-150\,meV). Together with the accurately determined value of the electron attachment energy, E$_{\rm A}$\,=\,2.666$\pm$0.001\,eV (Stochkel \& Andersen 2013), this yields the range $\sim$2.5-2.9\,eV for the energy of the transition  $^2$A$_g$-$^2$T$_{1u}$, i.e.\ $\sim$4300-5000\,\AA\ for its wavelength. This is the range where about half of all known  `C$_2$' DIBs are concentrated (Thorburn et al.\ 2003). The oscillator strength of this   $^2$A$_g$-$^2$T$_{1u}$ transition, if real, seems unknown. However, even a modest f-value of a few 10$^{-2}$ might be enough to account for  observed  `C$_2$' DIBs, as in the case of 4U\,1907+19;  considering one of the strongest `C$_2$' DIBs, at 4964.0\,\AA , the detection of which  is clear (but not reported)  in the spectrum of Fig.\ 1 of Cox et al.\ (2005), one may infer an equivalent width EW\,$\sim$\,10 \,m\AA/mag. The limit derived above for the abundance of  C$_{60}^-$ on this sight line, X$_{\rm C}$(C$_{60}^-$)\,$<$\,0.5 10$^{-4}$ of interstellar carbon, and Eq.(5) would then yield f\,$>$\,2 10$^{-2}$ if this DIB was due to  C$_{60}^-$. Of course, the half dozen confirmed `C$_2$' DIBs in this range may have completely different origin than C$_{60}^-$. However, one may also consider that more than one DIB might thus be explained if the C$_{60}^-$ band was split by some fine structure in the excited (or even the ground) level, and other fullerene anions might also contribute if abundant enough.

\section{Delayed (Poincar\'e) fluorescence in fullerenes}
Delayed  fluorescence (or more precisely  thermally activated delayed fluorescence, TADF)  is a well-known cooling process for large molecules by evacuating the (vibration) energy of a state $m$ with slow cooling through radiation from an upper electronic level $e$. Its rate remains generally low because of the factor exp(-$\Delta$E/kT$_{\rm v}$), where $\Delta$E\,$>>$\,kT${\rm _v}$ is the difference between $e$ and $m$ energies, and T${\rm _v}$ is the inner (vibration) temperature of the molecule. Its possible importance in the case of PAHs was stressed by L\'eger et al.\ (1988a) under the name of `Poincar\'e fluorescence' (PF). 

Its properties are significantly different for neutrals and ions. Neutral molecules, or more precisely those with an even number of electrons, often have a spin-singlet ground state, whereas most excited states exist in both forms, singlet and triplet. As shown e.g.\ by Salazar et al.\ (1997),  Bachilo et al.\ (2000), Baleiz\~ao \&  Berberan-Santos (2007), and references therein, several reasons may contribute to the importance of  PF in fullerenes: 1) very efficient intersystem conversion (ISC) channelling ~100\% of the energy of absorbed UV photons into the lowest triplet level T$_1$ (Fig.\ 2); 2) the very long lifetime of this level; 3) small  energy splitting,  $\Delta$E$_{\rm ST}$\,=\,E$_{\rm S1}$-E$_{\rm T1}$, typically $\sim$0.3\,eV for C$_{60}$ or C$_{70}$ compounds.

PF  dominates the cooling of  C$_{70}$ even for very low T${\rm _v}$ and vibration energy E$_{\rm v}$ (250-300\,K, a few eV) because of the very long triplet lifetime -- tens of ms -- (e.g.\ Bachilo et al.\ 2000;  Baleiz\~ao et al.\ 2011). C$_{70}$ derivatives, such as C$_{70}$H$_2$, also display PF, but at lower rates. PF is even significantly modified in the $^{13}$C isotopologue of C$_{70}$, confirming that key energy conversion rates may be extremely sensitive to actual energy-level patterns.  

PF has similarly been demonstrated for C$_{60}$ and derivatives, such as C$_{61}$H$_2$, but with lower rates because of the shorter triplet lifetime (e.g.\ Salazar et al.\ 1997; Anthony et al.\ 2003; Baleiz\~ao et al.\ 2011, and references therein). Its very well-behaved properties allow its use e.g.\ for accurate determination of the $\Delta$E$_{\rm ST}$ gap of C$_{61}$H$_2$  (Anthony et al.\ 2003). It is also probably involved in the observed exponential variation in the triplet lifetime with larger  E$_{\rm v}$ following the absorption of UV photons (Etheridge et al.\ 1995; Heden et al.\ 2003; Deng et al.\ 2003; Etch et al.\ 2005).

All in all, it seems obvious that PF should be considered as potentially important in the energy conversion processes of interstellar fullerenes. It may have led to underestimate the abundance of C$_{70}$ (§2.4). It should be eventually considered for the PF process proposed to explain emission DIBs (Witt 2014 and § 8.5). 

However, the actual importance of PF in the cooling of interstellar C$_{60}$ remains less obvious because of the more efficient IEC from the triplet to the ground state. If PF probably dominates at high E$_{\rm v}$\,$\ga$\,5-7\,eV, the case should be investigated in detail at lower energy, which corresponds to the bulk of the cooling. 

On the other hand, in the case of {\it singly ionized fullerenes}, the internal energy conversion (IEC) and radiative cooling may be deeply modified by the presence of a low doublet (although it is possible that higher metastable quadruplets play a similar role to triplets of neutrals). For example, both C$_{60}^+$ or C$_{60}^-$ exhibit a doublet D$_1$ about 1\,eV above the ground doublet D$_0$, with an allowed D$_1$-D$_0$ transition. The existence of this doublet may accelerate the IEC down to D$_0$, as proved for C$_{60}^-$  (Erhler et al.\ 2006; Andersen et al.\ 2001).  It would leave all the energy stored in vibrations for a very long time -- up to $\sim$10-100\,s if IR cooling played alone, §2.1.1. But PF could be much faster as proved in the case of C$_{60}^-$ for  T${\rm _v}$\,$\sim$\,1000-1500\,K (Andersen et al.\ 1996, 2001). For lower  typical values of  T${\rm _v}$\,$\sim$\,500-1000\,K (E$_{\rm v}$\,$\ga$\,3-10\,eV) following the absorption of a single UV photon, the transition between IR and PF cooling may arise in the range of E$_{\rm v}$\,$\sim$3-7\,eV, depending on the exact value of the energy of D$_1$. PF should thus be important for interstellar C$_{60}^-$ and might even be significant for C$_{60}^+$.

\bigskip

{\bf References}

Ag{\'u}ndez, M., Fonfr{\'{\i}}a, J.~P., Cernicharo, J., et al.\ 2012, \aap, 543, A48 

Akasak, T.\ \& Nakase, 2001, Endofullerenes: A New Family of Carbon Clusters, Ed. T. Akasak \& S. Nagase,
Kluwer Academic, Dordrecht, Netherlands

Andersen, J.~U.,  Gottrup, C., Hansen, K., Hvelplund, P.\ \& Larsson, M.~O.\ 2001, Eur.\ Phys.\ J.\ D, 17, 189

Andersen, J.~U.,  et al.\ 1996, Phys.\ Rev.\ Lett., 77, 3991 

Andreoni, W.\ et al.\ 1996, J.\ Am.\ Chem.\ Soc., 118, 11335

Andrews, H., Boersma, C., Werner, M.~W., et al.\ 2015, \apj, 807, 99 

Anthony, S.~M., Bachilo, S.~M., \&  Weisman, R.~B.\ 2003, J.\ Phys.\ Chem.\ A, 107, 10674

Arbogast, J.~W.\ \& Foote, C.~S.\ 1991, J.\ Am.\ Chem.\ Soc., 113, 8886

Bachilo, S.~M., Benedetto, A.~F., Weisman, R.~B.\ et al.\  2000, J.\ Phys.\ Chem.\ A, 104, 11265

Baffreau, J.\ et al.\  2008, Chem.\ Eur.\ J., 14, 4974

Bailey, M., van Loon, J.~T., Sarre, P.~J., \& Beckman, J.~E.\ 2015a, \mnras, 454, 4013 

Bailey, M., van Loon, J.~T., Farhang, A., et al.\ 2015b,  \aap, 585, A12  

Baleiz\~ao, C.\ \&  Berberan-Santos, M.~N, 2007, J.\ Chem.\ Phys., 126, 204510

Baleiz\~ao, C.\ \&  Berberan-Santos, M.~N, 2011, ChemPhysChem, 12, 1247

Baron, D., Poznanski, D., Watson, D., Yao, Y., \& Prochaska, J.~X.\ 2015a, \mnras, 447, 545 

Baron, D., Poznanski, D., Watson, D., et al.\ 2015b, \mnras, 451, 332 

Basir, Y.~S.\ \& Anderson, S.~L.\ 1999, International Journal of Mass Spectrometry 185/186/187, 603

Beck, R.~D., Weis, P., Rockenberger, J., \& Kappes, M.~M.\ 1996, Surf.\ Rev.\ Lett., 3, 771.

Becker, L., \& Bunch, T.E.\ 1997, Meteoritics \& Planetary Science 32, 479

Berkowitz, J., 1999, Chem.\ Phys.\ 111, 1446 

Bendale, R.~D., Stanton, J.~F., \& Zerner, M.~C.\ 1992, Chem.\ Phys.\ Lett., 194, 467

Bensasson, R~V., et al.\ 1995, Chem.\ Phys.\ Lett., 245, 566

Bernard-Salas, J., Cami, J., Peeters, E., et al.\ 2012, \apj, 757, 41 

Bern{\'e}, O., \& Tielens, A.~G.~G.~M.\ 2012, Proceedings of the National Academy of Science, 109, 401 

Bern{\'e}, O., Mulas, G., \& Joblin, C.\ 2013, \aap, 550, L4 

Bern{\'e}, O., Montillaud, J., \& Joblin, C.\ 2015a, \aap, 577, A133 

Bern{\'e}, O., Montillaud, J., Mulas, G., \& Joblin, C.\ 2015b, arXiv:1510.01642 

Bettinger, H.~F.\ et al.\ 2002, Chemical Physics Letters 360, 509

Bhatt, N.~H., \& Cami, J.\ 2015, \apjs, 216, 22 

Bihlmeier, A.,  Samson, C.~C.~M., \& Klopper, W.\ 2005, Chem.\ Phys.\ Chem.\ 6, 2625

Billas, I.~M.~L., et al.\   1999, J.\ Chem.\ Phys., 111, 6787

Boersma, C., Rubin,  R.~H., \& Allamandola, L.~J.\ 2012, \apj, 753, 168 

Bolskar, R.~D., et al.\ 1995, Chem.\ Phys.\ Lett., 247, 57

Broclawik, E.\ \& Eilmes, A.\ 1998, J.\ Chem.\ Phys., 108, 1 

Buchachenko, A.~L.\ \& Breslavskaya, N.N.\  2005, Russian Chemical Bulletin, International Edition, 54, 51

Burton Williams, T., Sarre, P., Marshall, C., Spekkens, K., \& Kuzio de Naray, R.\ 2015, IAU General Assembly, 22, 55619

Cami, J., Sonnentrucker, P., Ehrenfreund, P., \& Foing, B.~H.\ 1997, \aap, 326, 822 

Cami, J., Salama, F., Jim{\'e}nez-Vicente, J., Galazutdinov, G.~A., \& Kre{\l}owski, J.\ 2004, \apjl, 611, L113 

Cami, J., Bernard-Salas, J., Peeters, E., \& Malek, S.~E.\ 2010, Science, 329, 1180 

Cami, J., Bernard-Salas,  J., Peeters, E., \& Malek, S.~E.\ 2011, IAU Symposium, 280, 216 

Cami, J.\ 2014, IAU Symposium, 297, 370 

Cami, J., \& Cox, N.~L.~J.\ 2014, IAU Symposium, 297

Campbell, E. E. B., Ehlich, R.. Heusler, G., et al., 1998, Chem. Phys., 239, 299.

Campbell, E. E. B., \& Rohmund, F.\ 2000, Rep.\ Prog.\ Phys.\ 63, 1061

Campbell, E.~K., Holz, M., Gerlich, D., \& Maier, J.~P.\ 2015, \nat, 523, 322 

Cataldo, F., Strazzulla, G., \& Iglesias-Groth, S.\ 2009, \mnras, 394, 615 

Cataldo, F., \& Iglesias-Groth, S.\ 2009, \mnras, 400, 291 

Cataldo, F., Iglesias-Groth, S., \& Manchado, A.,\ 2012, Fullerenes Nanotubes and Carbon Nanostructures, 20, 656

Cataldo, F., Iglesias-Groth, S., \& Hafez, Y.\ 2013, Eur.\ Chem.\ Bull., 2, 1013

Cataldo, F., Garc{\'{\i}}a-Hern{\'a}ndez, D.~A., Manchado, A., \& Iglesias-Groth, S.\ 2014, IAU Symposium, 297, 294

Cavar, E.\ et al.\ 2005, Phys. Rev. Lett.\ 95, 196102

Chakraborty, H.~S.\ et al.\ 2008, Phys.\ Rev.\ A, 78, 013201 

Chancey, C.~C.\ \& O'Brien, C.~M.\ 1997. The Jahn-Teller effect in C$_{60}$ and other icosahedral complexes. Princeton University Press.

Chang, A.~H.~H., Ermler, W.~C. \& Pitzer, R.~M.\ 1991, Journal of Chemical Physics, 94, 5004

Chergui, M. 2000, Low Temperature Physics, 26, 632

Choi, C.\ et al.\ 2000, J.\ Phys.\ Chem.\ A, 104, 102

Christian, J.F.\ et al.\ 1992a, Chem.\ Phys.\ Lett., 199, 373

Christian, J.F.\ et al.\ 1992b, J.\ Phys.\ Chem., 96, 10597

Christian, J.F.\ et al.\ 1992c, J.\ Phys.\ Chem., 96, 3574

Chuvilin, A.\ et al.\ 2010, Nature Chemistry, 2, 450

Clipston, N.~L.\ 2000, J.\ Phys.\ Chem.\ A, 104, 9171

Close, J.D., Federmann, F., Hoffmann, K., Quaas, N.\ 1997, Chem.\ Phys.\ Lett., 276, 393

Cong, H.\ et al. 2013, Coordination Chemistry Reviews 257, 2880

Connerade, J.~P., Dolmatov, V.~K.\ \& P A Lakshmi, P.~A.\ 2000, J.\ Phys.\ B 33, 251

Connerade, J.~P.\  \& Solovyov, A.~V.\ 2005, J.\ Phys.\ B 38, 807

Cordiner, M.~A.\ 2014, IAU Symposium, 297, 41 

Cox, N.~L.~J., Kaper, L., Foing, B.~H., \& Ehrenfreund, P.\ 2005, \aap, 438, 187 

Cox, N.~L.~J.\ 2006, Ph.D.~Thesis, 

Cox, N.~L.~J., Boudin, N., Foing, B.~H., et al.\ 2007, \aap, 465, 899 

Cox, N.~L.~J.\ 2011a, in PAHs and the Universe: A Symposium to Celebrate the 25th Anniversary of the PAH Hypothesis, EAS Publication Series Vol.\ 46, 349-354, eds Joblin, C.\ \& Tielens, A.~G.~G.~M.

Cox, N.~L.~J., Ehrenfreund, P., Foing, B.~H., et al.\ 2011b, \aap, 531, A25 

Cox, N.~L.~J., Cami, J., Kaper, L., et al.\ 2014a, \aap, 569, AA117 

Cox, N.~L.~J., \& Cami, J.\ 2014b, IAU Symposium, 297, 412 

Cox, N.~L.~J.\ 2015, arXiv:1504.03281 

Crawford, M.~K., Tielens, A.~G.~G.~M., \& Allamandola, L. J.\ 1985, ApJ, 293, L45

Dahlstrom, J., York, D.~G., Welty, D.~E., et al.\ 2013, \apj, 773, 41 

Deng, R. Clegg, A. \& Echt, O.\ 2003, International Journal of Mass Spectrometry, 223–224, 695 

D\'iaz-Tendero, et al.\ 2006, International Journal of Mass Spectrometry, 252, 133

Dresselhaus, M.S., Dresselhaus, G.\ \&  Eklund, P.C.\  1996, Science of Fullerenes and Carbon Nanotubes, Academic press, San Diego

 Duley, W.~W., \& Williams, D.~A.\ 2011, \apjl, 737, L44 

Dunbar, R.~C.\ et al.\ 1994, J.\ Am.\ Chem.\ Soc., 116, 2466

Dunk, P.~W., et al.\  2012, Nat. Commun., 3, 855.

Dunk, P.~W., Adjizian,  J.-J., Kaiser, N.~K., et al.\ 2013, Proceedings of the National Academy of  Science, 110, 18081 

Dunk, P.~W., et al.\  2014, Nat. Commun., 5, 5844

Echt, O., Yao, S., and Deng, R.\ \& Hansen, K.\ 2005, J.\ Phys.\ Chem.\ A 108, 6944

Edwards, S.~A., \& Leach, S.\ 1993, \aap, 272, 533 

Ehrenfreund, P., \& Foing, B.~H.\ 1995, Planet.\ Space Sci., 43, 1183 

Ehrenfreund, P., \& Foing, B.~H.\ 2010, Science, 329, 1159 

Ehrenfreund, P., \& Foing, B.\ 2015, \nat, 523, 296 

Erhler, O.~T.\ et al.\ 2006, J.\ Chem.\  Phys., 125, 074312 

Etheridge, H.~T., Averitt, R.~D., Halas, N.~J.\ \& Weisman, R.~B.\ 1995, J.\ Phys.\ Chem., 99, 11306

Farhang, A., Khosroshahi, H.~G., Javadi, A., et al.\ 2015, \apj, 800, 64 

Feng, M., Zhao, J.\ \& Petek, H.\ 2008, Science, New Series, 320, 359

Foing, B.~H., \& Ehrenfreund, P.\ 1994, \nat, 369, 296

Foing, B.~H., \& Ehrenfreund, P.\ 1997, \aap, 317, L59 

Frommhold, L., Collisison-Induced Absorption in Gases, Cambridge Monographs on Atomic, Molecular, and Chemical Physics, Vol.\ 2 (Cambridge University Press, Cambridge, England, 1993)

Fulara, J., Lessen, D.,  Freivogel, P., \& Maier, J.~P.\ 1993a, \nat, 366, 439 

Fulara, J., Jakobi, M., \& Maier, J.~P.\ 1993b, Chem.\ Phys.\ Lett., 206, 203

Galazutdinov, G.~A., Kre{\l}owski, J., Musaev, F.~A., Ehrenfreund, P., \& Foing, B.~H.\ 2000, \mnras, 317, 750 

Garc{\'{\i}}a-Hern{\'a}ndez, D.~A., Manchado, A., Garc{\'{\i}}a-Lario, P.,  et al.\ 2010, \apjl, 724, L39 

Garc{\'{\i}}a-Hern{\'a}ndez, D.~A., Kameswara Rao, N., \& Lambert, D.~L.\ 2011a, \apj, 729, 126  

Garc{\'{\i}}a-Hern{\'a}ndez, D.~A., Iglesias-Groth, S., Acosta-Pulido,  J.~A., et al.\ 2011b, \apjl, 737, L30 

Garc{\'{\i}}a-Hern{\'a}ndez, D.~A., Kameswara Rao, N., \& Lambert, D.~L.\ 2012, \apjl, 759, L21 

Garc{\'{\i}}a-Hern{\'a}ndez, D.~A., Cataldo, F., \& Manchado, A.\ 2013, \mnras, 434, 415 

Gasyna, Z., Andrews, L.\ \& Schatz, P.\,N.\ 1992, J.\ Phys.\ Chem., 96, 1525

Ge, M.\ et al. 2011, Journal of Chemical Physics, 134, 054507

Geballe, T.~R., Najarro, F., Figer, D.~F., Schlegelmilch, B.~W., \& de La Fuente, D.\ 2011, \nat, 479, 200 

Geballe, T.~R., Najarro, F., de la Fuente, D., et al.\ 2014, IAU Symposium, 303, 75 

Gielen, C., Cami, J., Bouwman, J., Peeters, E., \& Min, M.\ 2011, \aap, 536, A54 

Gillmon, K., Shull, J.~M., Tumlinson, J., \& Danforth, C.\ 2006, \apj, 636, 891 

Gluch, K.\ et al.\ 2004, J.\ Chem.\ Phys.\, 121, 2137

Gredel, R., Carpentier, Y., Rouill{\'e}, G., et al.\ 2011, \aap, 530, A26 

Guha, S. \& Nakamoto, K.\ 2005, Coordination Chemistry Reviews, 249, 1111

Hamano, S., Kobayashi, N., Kondo, S., et al.\ 2015, \apj, 800, 137

Hammond, M.~R.\ \& Zare, R.~N.\ 2008, Geochimica et Cosmochimica Acta, 72, 5521

Hands, I.~D.\ et al.\ 2008, Phys.\ Rev.\ B, 77, 115445

Hansen, K., et al.\ 1997, Z.\ Phys.\ D, 42, 153

Hasoglu, M.~F.\ et al.\ 2013, Phys.\ Rev.\ A, 87, 013409

Haufler, R.~E.\ et al.\ 1991, J.\ Chem.\ Phys.\, 95, 2197 

Hauke, F.\ et al.\ 2004, Chem.\ Commun.,  766

Hauke, F.\ et al.\ 2006, Chem.\ Eur.\ J.\ 12, 4813

Hed\'en, M.\ et al.\ 2003, J.\ Chem.\ Phys., 118, 7161 

Heger, M.~L.\ 1922, Lick Observatory Bulletin, 10, 141 

Herbig, G.~H.\ 1975, \apj, 196, 129

Herbig, G.~H.\ 1993, \apj, 407, 142 

Herbig, G.~H.\ 1995, \araa, 33, 19 

Herbig, G.~H.\ 2000, \apj, 542, 334 

Hern\'andez-Rojas, J.\ et al.\ 1996, J.\ Chem.\ Phys., 104, 1179

Heymann, D.\ \& Weisman, R.~W.\ 2006, C. R. Chimie 9, 1107

Hobbs, L.~M., York, D.~G., Snow, T.~P., et al.\ 2008, \apj, 680, 1256 

Hobbs, L.~M., York, D.~G., Thorburn, J.~A., et al.\ 2009, \apj, 705, 32 

Huang, D.~L.\ et al.\ 2014, J.\ Chem.\ Phys., 140, 224315

Hudgins, D.~M., Bauschlicher, C.~W., Jr., \& Allamandola, L.~J.\ 2005, \apj, 632, 316

Iglesias-Groth, S.\ 2006, \mnras, 368, 1925 

Iglesias-Groth, S., Garc{\'{\i}}a-Hern{\'a}ndez, D.~A., Cataldo, F., \& Manchado, A.\ 2012, \mnras, 423, 2868 

Iglesias-Groth, S., \& Esposito, M.\ 2013, \apjl, 776, L2 

Jenkins, E.~B.\ 2009, \apj, 700, 1299 

Jenniskens, P., \& Desert, F.-X.\ 1994, \aaps, 106, 39 

Jiao, H.\ et al.\ 2002, Phys.\ Chem.\ Chem.\ Phys., 2002, 4, 4916

Jin, C.\ et al.\ 2009, Phys.\ Rev.\ Lett.\ 102, 205501 

Joalland, B., Simon, A., Marsden, C.~J., \& Joblin, C.\ 2009, \aap, 494, 969 

Joblin, C., D'Hendecourt, L., Leger, A., \& Maillard, J.~P.\ 1990, \nat, 346, 729

Jochims, H.~W., Baumgaertel, H., \& Leach, S.\ 1996, \aap, 314, 1003 

Johansson, J.O.\ \& Campbell, E.~E.~B.\ 2013, Chem. Soc. Rev., 2013, 42, 5661

Jones, A.~P.\ 2013, \aap, 555, A39

Jones, A.~P.\ 2014, \planss, 100, 26 

Jones, A.\ 2015, Highlights of Astronomy, 16, 707 

Kato, T.\ et al.\ 1991, Chem. Phys.\ Lett., 180, 446

Ka{\'z}mierczak, M., Schmidt, M.~R., Bondar, A., \& Kre{\l}owski, J.\ 2010, \mnras, 402, 2548 

Ka{\'z}mierczak, M., Schmidt, M., Weselak, T., Galazutdinov, G., \& Kre{\l}owski, J.\ 2014, IAU Symposium, 297, 121 

Kern, B., et al.\ 2013, J.\ Phys.\ Chem.\ A, 117, 8251

Kern, B., Strelnikov, D., Weis, P., B\"ottcher, A., \& Kappes, M.~M.\ 2014, J.\ Phys.\ Chem.\ Lett., 5, 457

Kerr, T.~H., Hurst, M.~E., Miles, J.~R., \& Sarre, P.~J.\ 1999, \mnras, 303, 446 

Keshavarz-K, et al.\ 1997, Nature, 383, 147

Klaiman, S., Gromov, E.~V.\ \& Cederbaum, L.~S. 2013, J.\ Phys.\ Chem.\ Lett., 4, 3319
  
Klaiman, S.\ et al.\ 2014, Phys.\ Chem.\ Chem.\ Phys., 16, 13287

Komatsu, K.\ et al. 2005, Science, 307, 238

Koponen, L., Puska, M.~J.\ \& Nieminen, R.\~M.\ 2008, J.\ Chem.\ Phys., 128, 154307

Kordatos, K., Da Ros, T., Prato, M., Bensasson, R.V.\ \& Leach, S.\ 2003, Chemical Physics 293, 263

Kos, J., Zwitter, T., Grebel, E.~K., et al.\ 2013, \apj, 778, 86 

Kos, J., Zwitter, T., Wyse, R., et al.\ 2014, Science, 345, 791 

 Kr\"{a}tschmer, W., Lamb, L.D., Fostiropoulos, K., \& Huffman, D.~R.\ 1990, Nature 347, 354 

Krelowski, J., \& Walker, G.~A.~H.\ 1987, \apj, 312, 860 

Kre{\l}owski, J., Snow, T.~P., Seab, C.~G., \& Papaj, J.\ 1992, \mnras, 258, 693 

Kre{\l}owski, J.\ 2014, IAU Symposium, 297, 23 

Kroto, H.~W., Heath, J.~R., Obrien, S.~C., Curl, R.~F., \& Smalley, R.~E.\ 1985, Nature, 318, 16

Kroto, H.~W.\ 1987, NATO Advanced Science Institutes (ASI) Series C, 191, 197. A. L\'eger et al.\ eds

Kroto, H.\ 1988, Science, 242, 1139 

Kroto, H.~W.\ 1989, Annales de Physique, 14, 169 

Kroto, H.~W., \& Jura, M.\ 1992, \aap, 263, 275 

Lan, Y., Kang, H.\ \& Niu, T.\ 2015, Eur.\ Phys.\ J.\ D, 69, 69
 
Lan, T.-W., M{\'e}nard, B., \& Zhu, G.\ 2015, \mnras, 452, 3629 

Langhoff, S.~R.\ 1996, J.\ Phys.\ Chem., 100, 2819

Leach, S., Vervloet, M.,  Despr{\`e}s, A., et al.\ 1992, Chemical Physics, 160, 451

Leach, S.\ 2001, Can.\ J.\ Phys., 79, 501

Leach, S.\ 2004, Israel Journal of Chemistry, 44, 193

Leach, S.\ 2006, {\it Fullerenes in Space : The hunting of the Snark ?}, unpublished manuscript, personal communication (2015)

Lee, Y.~H.\ Kim, S.~G.\ \& Tománek, D.\ 1997, Phys.\ Rev.\ Lettt., 78, 2393

L\'eger, A.\ \& D'Hendecourt, L.\ 1985, \aap, 146, L81

L\'eger, A., Jura, M., \& Omont, A.\ 1985, \aap, 144, 147 

L\'eger, A., D'Hendecourt, L., \& Boissel, P.\ 1988a, Physical Review Letters, 60, 921 

L\'eger, A., D'Hendecourt, L., Verstraete, L., \& Schmidt, W.\ 1988b, \aap, 203, 145 

L\'eger, A., D'Hendecourt, L., \& Defourneau, D.\ 1989, \aap, 216, 148  

Le Page, V., Snow,  T.~P., \& Bierbaum, V.~M.\ 2003, \apj, 584, 316 

Le Page, V., Snow, T.~P., \& Bierbaum, V.~M.\ 2001, \apjs, 132, 233 

Lepp, S., Dalgarno, A., van Dishoeck, E.~F., \& Black, J.~H.\ 1988, \apj, 329, 418 

Lifshitz, C.\ 2000, International Journal of Mass Spectrometry, 198, 1

Long, Z.\ et al.\ 2013, Chemical Physics Letters 583, 114

Lu, X., Feng, L., Akasaka, T., et al.\ 2012, Chem. Soc. Reviews, 41, 7723

Lyras, A.\ \& Bachau, H.\ 2005, J.\ Phys.\ B: At.\ Mol.\ Opt.\ Phys., 38, 1119

Maier, J.~P.\ 1994, \nat, 370, 423 

Maier, J.~P., Walker, G.~A.~H., \& Bohlender, D.~A.\ 2004, \apj, 602, 286 

Ma{\'{\i}}z Apell{\'a}niz, J., Barb{\'a}, R.~H., Sota, A., \& Sim{\'o}n-D{\'{\i}}az, S.\ 2015, \aap, 583, A132 

Mamone, S.\ et al.\ 2009,  Journal of Chemical Physics 130, 081103 

Mamone, S.\ et al.\ 2011, Coordination Chemistry Reviews 255, 938

Marciniak, A., Despr\'e, V., Barillot, T.\ et al.\  2015,  Nature communications, 6, 7909

Marcos, P~A.\ et al.\ 2003, J.\ Chem.\ Phys., 119, 1127

Marshall, C.~C.~M., Kre{\l}owski, J., \& Sarre, P.~J.\ 2015, \mnras, 453, 3912 

McCall, B.~J., Thorburn, J., Hobbs, L.~M., Oka, T., \& York, D.~G.\ 2001, \apjl, 559, L49 

McIntosh, A., \& Webster, A.\ 1993, \mnras, 261, L13 

Men'shchikov, A.~B., Schertl, D., Tuthill, P.~G., Weigelt, G., \& Yungelson, L.~R.\ 2002, \aap, 393, 867 

Merino, P., Svec, M., Martinez, J.~I.\ et al.\ 2014, Nature communications, 5, 3054

Merrill, P.~W.\ 1934, \pasp, 46, 206 

Merrill, P.~W.\ 1936, \pasp, 48, 179 

Micelotta, E.~R., Jones, A.~P., \& Tielens, A.~G.~G.~M.\ 2010a, \aap, 510, A36 

Micelotta, E.~R., Jones, A.~P., \& Tielens, A.~G.~G.~M.\ 2010b, \aap, 510, A37 

Micelotta, E.~R., Jones, A.~P., \& Tielens, A.~G.~G.~M.\ 2011, \aap, 526, A52 

Micelotta, E.~R., Jones, A.~P., Cami, J., Peters, E.\ Bernard-Salas, J.\ \&  Fanchini, G.\ 2012, \apj, 761, 35 

Micelotta, E.~R.,  Cami, J., Peeters, E., et al.\ 2014, IAU Symposium, 297, 339 

Milisavljevic, D., Margutti, R., Crabtree, K.~N., et al.\ 2014, \apjl, 782, L5 

Millar, T.~J.\ 1992, \mnras, 259, 35P 

Misawa, T., Gandhi, P., Hida, A., Tamagawa, T., \& Yamaguchi, T.\ 2009, \apj, 700, 1988

Montillaud, J., \& Joblin, C.\ 2014, \aap, 567, A45 

Montillaud, J., Joblin, C., \& Toublanc, D.\ 2013, \aap, 552, A15 

Munari, U., Tomasella, L., Fiorucci, M., et al.\ 2008, \aap, 488, 969 

Murata, M.\ 2006, J.\ Am.\ Chem.\ Soc.\ 128, 8025

Neyts, E.\ et al.\ 2011, Carbon 49, 1013

Nossal, J.\ et al.\ 2001, Eur.\ J.\ Org.\ Chem., 4167-4180

Oka, T., Welty, D.~E., Johnson, S., et al.\ 2013, \apj, 773, 42; 2014, \apj, 793, 68 

Okamoto, Y.\ 2001, J. Phys. Chem. A 2001, 105, 7634

Oksengorn, B.\ 2003, C. R. Chimie 6, 467

Omont, A.\ 1986, \aap, 164, 159 

Orlandi, G.\  \& Negri, F.\ 2002,  Photochem. Photobiol. Sci., 1, 289

Otsuka, M., Kemper, F.,  Hyung, S., et al.\ 2013, \apj, 764, 77 

Otsuka, M., Kemper, F.,  Cami, J., Peeters, E., \& Bernard-Salas, J.\ 2014, \mnras, 437, 2577 
 
Peeters, E., Tielens,  A.~G.~G.~M., Allamandola, L.~J., \& Wolfire, M.~G.\ 2012, \apj, 747, 44 

Pellarin, M., Ray, C., M\'elinon, P.\ et al.\ 1997 Chem.\ Phys.\ Lett., 277, 96

Pellarin, M., Ray, C., Lerm\'e, J.\ et al.\ 1999, Eur.\ Phys.\ J.\ D 9, 49

Pellarin, M.\ et al.\ 2002, J. Chem.\ Phys., 117, 3088 

Pennington, C.~H.\ \& Stenger, V.~A.\ 1996, Rev.\ Modern Phys., 68, 855

Petrie, S.\ et al.\ 1995, International Journal of Mass Spectrometry and Ion Processes 145, 79 

Petrie, S., \& Bohme, D.~K.\ 2000, \apj, 540, 869 

Pietrucci, F., \& Andreoni, W.\ 2014, Journal of Chemical Theory and Computation, 10, 913

Pino, T., Dartois, E., Cao, A.-T., et al.\ 2008, \aap, 490, 665 

Popov, A~.A., Yang, S.\ \& Dunsch, L.\ 2013, Chemical Reviews, 113, 5989

Puspitarini, L., Lallement, R., Babusiaux, C., et al.\ 2015, \aap, 573, AA35 

Ramachandran, C.~N.\ et al.\ 2008, Chemical Physics Letters 461, 87

Rawlings, M.~G., Adamson, A.~J., \& Kerr, T.~H.\ 2014, \apj, 796, 58 

Ray, C.\ et al.\ 1998, Phys.\ Rev.\ Lett., 80, 5365

Reddy, V.S.\ et al.\ 2010, Phys.\ Rev.\ Lett., 104, 111102

Ren, A.\ et al.\ 2000, International Journal of Quantum Chemistry, 78, 422

Ren, X.\ et al.\ 2004, Journal of Molecular Structure (Theochem), 686, 43

Rietmeijer, F.~J.~M.\ 2006,  Natural Fullerenes and Related Structures of Elemental Carbon. Springer

Roberts, K.~R.~G.,  Smith, K.~T., \& Sarre, P.~J.\ 2012, \mnras, 421, 3277 

Rubin, R.~H., Simpson,  J.~P., O'Dell, C.~R., et al.\ 2011, \mnras, 410, 1320 

R\"udel, A.\ et al.\ 2002, Phys.\ Rev.\ Lett., 89, 125503

Salama, F., Bakes, E.~L.~O., Allamandola, L.~J., \& Tielens, A.~G.~G.~M.\ 1996, \apj, 458, 621 

Salama, F., Galazutdinov, G.~A., Kre{\l}owski, J., et al.\ 2011, \apj, 728, 154 

Salama, F., \& Ehrenfreund, P.\ 2014, IAU Symposium, 297, 364 

Salazar, F.~A., Fedorov, A.\ \& Berberan-Santos, M~N.\ 1997, Chem.\ Phys.\ Lett.,  271, 361

Sarre, P.~J.\ 2006, Journal of Molecular Spectroscopy, 238, 1 

Sarre, P.~J.\ 2008, IAU Symposium, 251, 49 

Sarre, P.~J.\ 2014, IAU Symposium, 297, 34 

Sassara, A., Zerza, G. \& Chergui, M.\ 1996a, J. Phys. B: At. Mol. Opt. Phys., 29, 4997

Sassara, A., Zerza, G. \& Chergui, M.\  1996b, Chemical Physics Letters 261, 213

Sassara, A., Zerza, G., Chergui, M., Negri, F.\ \& Orlandi, G.\ 1997, J. Chem. Phys., 107, 8731

Sassara, A., Zerza, G., Chergui, M., \& Leach, S.\ 2001, \apjs, 135, 263 

Savage, B.~D., \& Sembach, K.~R.\ 1996, \araa, 34, 279 

Schmidt, G.~D., Cohen, M., \& Margon, B.\ 1980, \apjl, 239, L133 

Scott, A., Duley, W.~W., \& Pinho, G.~P.\ 1997, \apjl, 489, L193 

Scuseria, G.~E.\ 1991, Chem.\ Phys.\ Lett., 180, 451

Sellgren, K., Uchida,  K.~I., \& Werner, M.~W.\ 2007, \apj, 659, 1338 

Sellgren, K., Werner,  M.~W., Ingalls, J.~G., et al.\ 2010, \apjl, 722, L54 

Shinohara, H.\ 2000, Rep. Prog. Phys. 63, 843

Shvartsburg, A.~A., Hudgins, R.~R., Dugourd, P., Gutierrez, R., Frauenheim, T.\ \& Jarrold, M.~F.\ 2000, Phys.\ Rev.\ Lett., 84, 2421 

Simon, A., \& Joblin, C.\ 2007, Journal of Physical Chemistry A, 111, 9745

Simon, A., \& Joblin, C.\ 2009, Journal of Physical Chemistry A, 113, 4878 

Sitko, M.~L., Bernstein, L.~S., \& Glinski, R.~J.\ 2008, \apj, 680, 1426 

Slanina, Z., \& Lee, S.~L.\ 1994, Journal of Molecular Structure: {THEOCHEM},
304, 173

Smalley, R.~E.\^ 1992, Acc. Chem. Res, 25, 98

Smith, A~B.\  III, et al.\ 1995, J.\ Am.\ Chem.\ Soc., 117, 5492

Snow, T.~P.\ 1995, in Tielens, A.~G.~G.~M., \& Snow, T.~P., The Diffuse Interstellar Bands, {\it Astrophysics and Space Science Library}, 202, Kluwer

Snow, T.~P.\ 2001, Spectrochimica Acta, 57, 615 

Snow, T.~P., Zukowski, D., \& Massey, P.\ 2002, \apj, 578, 877

Snow, T.~P., \& McCall, B.~J.\ 2006, \araa, 44, 367 

Snow, T.~P.\ 2014, IAU Symposium, 297, 3 

Sonnentrucker, P.\ 2014, IAU Symposium, 297, 13 

Steglich, M., Bouwman, J., Huisken, F., \& Henning, T.\ 2011, \apj, 742, 2 

Stener, M.\ et al.\ 1999, Chem.\  Phys.\ Lett., 309,  129

Stener, M.\ et al.\ 2002, J.\ Phys.\ B: At.\ Mol.\ Opt.\ Phys., 35, 1421

Stepanov, A.~G.,  Portella-Oberli, M. ~T., Sassara, A. \& Chergui, M.\ 2002, Chem.\ Phys.\ Lett.,  358, 516

Stochkel, K.\ \& Andersen, J.U.\ 2013, J.\ Chem.\ Phys.\ 139, 164304

Strelnikov, D.\ \& Kr\"{a}tschmer, W.\ 2010, Carbon, 48, 1702

Strelnikov, D., Kern, B., \& Kappes, M.~M.\ 2015, \aap, 584, A55 

Tang, C.\ et al.\ 2006, Chinese Journal of Chemistry, 24, 1133

Tenorio, F.\ \& Robles 2000, J\ Int.\ J.\ Quantum Chem., 80, 220

Thaddeus, P.\ 1995, in Tielens, A.~G.~G.~M., \& Snow, T.~P.\ 1995, The Diffuse Interstellar Bands, {\it Astrophysics and Space Science Library}, 202, Kluwer

Thorburn, J.~A., Hobbs, L.~M., McCall, B.~J., et al.\ 2003, \apj, 584, 339 

Tielens, A.~G.~G.~M., \& Snow, T.~P.\ 1995, The Diffuse Interstellar Bands, {\it Astrophysics and Space Science Library}, 202, Kluwer 

Tielens, A.~G.~G.~M.\ 2005,  The Physics and Chemistry of the Interstellar Medium, by  A.~G.~G.~M.~Tielens, ISBN 0521826349.~Cambridge, UK: Cambridge  University Press,  2005

Tielens, A.~G.~G.~M.\ 2008, \araa, 46, 289 

Tielens, A.~G.~G.~M.\ 2013, Reviews of Modern Physics, 85, 1021 

Tielens, A.~G.~G.~M.\ 2014,  IAU Symposium, 297, 399 

Tomita, S.\ et al.\ 2005, Phys.\ Rev.\ Lett., 94, 053002 

Tulej, M., Kirkwood, D.~A., Pachkov, M., \& Maier, J.~P.\ 1998, \apjl, 506, L69 

Turro, N.J.\ et al.\ 2010, Accounts of Chemical Research 43, 335

Udvardi, L., Surja, P.~R., Kfirti, J.\ \& Pekker, S.\ 1995, Synthetic Metals, 70, 1377

Van der Zwet, G.~P., \& Allamandola, L.~J.\ 1985, \aap, 146, 76

van Loon, J.~T., Bailey, M., Tatton, B.~L., et al.\ 2013, \aap, 550, A108 

Van Winckel, H.\ 2014, IAU Symposium, 297, 180 

Van Winckel, H., Cohen, M., \& Gull, T.~R.\ 2002, \aap, 390, 147 

Vehvilainen, T.~T., et al.\ 2011, Physical Review B, 84, 085447

Verstraete, L., \& L\'eger, A.\ 1992, \aap, 266, 513 

Viggiano, A.~A.\ et al.\ 2010, J.\ Chem.\ Phys., 132, 194307 

Voora, D.~K., et al.\ 2013, J.\ Phys.\ Chem.\ Lett., 2013, 849

Vos, D.~A.~I., Cox, N.~L.~J., Kaper, L., Spaans, M., \& Ehrenfreund, P.\ 2011, \aap, 533, A129 

Vostrowsky, O.\ \& Hirsch, A.\ 2006, Chem.\ Rev.\ 106, 5191

Waelkens, C., Van Winckel, H., Trams, N.~R., \& Waters, L.~B.~F.~M.\ 1992, \aap, 256, L15 

Walker, G., Bohlender, D., Maier, J., \& Campbell, E.\ 2015, \apjl, 812, L8  

Wan, Z., Christian, J.~F., \& Anderson, S.~L.\ 1992, Phys.\ Rev.\ Lett.\, 69, 1352

Wang, L.~S., et al.\ 1993, Chem.\ Phys.\ Lett., 207, 354 

Wang, H., et al. 2012,  J.\ Phys.\ Chem.\ A, 116, 255

Wang, Y., Zettergren, H., Rousseau, P.\ et al.\ 2014, Phys.\ Rev.\ A, 89,  062708

Webster, A.\ 1992, \mnras, 257, 463 

Welty, D.~E., Ritchey, A.~M., Dahlstrom, J.~A., \& York, D.~G.\ 2014, \apj, 792, 106 

Westin, E.\ \& Ros\'en 1993, Zeitschrift f\"{u}r Physik D: Atoms, Molecules and Clusters, 26, S276 

Weisman, R.B.\ 1999, Optical Studies of Fullerene Triplet States. In {\it Optical and Electronic Properties of Fullerenes and Fullerene-Based Materials}; Shinar, J., et al.\  Eds.; Marcel Dekker: New York, 1999; p 83.

Witt, A.~N.\ 2014, IAU Symposium, 297, 173 

Xie, R.\ 2003, Phys.\ Rev.\ Lett., 90, 206602

Xie, R.\ 2004, Journal of Chemical Physics 120, 5133

Xu, Y.~B., Tan, M.~Q.\ \& Becker, U.\ 1996, Phys.\ Rev.\ Lett., 76, 3538

Yi, J.Y.\ \& Bernholc, J.\ 2005, Chem.\ Phys.\ Lett., 403, 359

York, B., Sonnentrucker, P., Hobbs, L.~M., et al.\ 2014, IAU Symposium, 297, 138 

Yuan, H.~B., \& Liu, X.~W.\ 2012, \mnras, 425, 1763 

Zack, L.~N., \& Maier, J.~P.\ 2014, The Diffuse Interstellar Bands, 297, 237 

Zasowski, G., M{\'e}nard, B., Bizyaev, D., et al.\ 2015, \apj, 798, 35 

Zhang, M.\ et al.\ 2008, J.\ Phys.\ Chem.\ A, 112, 5478

Zhang, Y., \& Kwok, S.\ 2011, \apj, 730, 126 


Zhang, J., Bowles, F.~L., Bearden, D.~W., et al.\ 2013, Nature Chemistry, 5, 880

Zhen, J., Castellanos, P., Paardekooper, D.~M., Linnartz, H., \& Tielens, A.~G.~G.~M.\ 2014, \apjl, 797, LL30

Zhou, Z.\ et al.\ 2006, \apjl, 638, L105

\end{document}